\pdfoutput=1
\documentclass[11pt]{article}

\usepackage[margin=1in]{geometry}
\usepackage{amsmath, amssymb, amsthm}
\usepackage{graphicx}
\usepackage{booktabs}
\usepackage{hyperref}
\usepackage{xcolor}
\usepackage{algorithm}
\usepackage{algpseudocode}
\usepackage{caption}
\usepackage[T1]{fontenc}
\usepackage{lmodern} 
\usepackage[expansion=false]{microtype}
\usepackage{enumitem}

\hypersetup{colorlinks=true, linkcolor=blue!50!black, citecolor=blue!50!black, urlcolor=blue!50!black,
  pdftitle={FP8 is All You Need (Part 2): Full-FP64 3-D FFT on FP8 Tensor Cores},
  pdfauthor={Satoshi Matsuoka},
  pdfsubject={FP64 FFT emulation on FP8 tensor cores: Ozaki-Bailey kernel, exact CRT reconstruction, codesign floors},
  pdfkeywords={FP8, FFT, Ozaki scheme, CRT, Kulisch, tensor cores, GPU, B300, Rubin, roofline}}

\newcommand{\fp}[1]{\texttt{fp#1}}
\newcommand{\bint}[1]{\texttt{int#1}}
\newcommand{\OII}{Ozaki-II}
\newcommand{\OBFFT}{Ozaki-Bailey FFT}
\usepackage{multirow}
\newcommand{\Pfpeight}{P_{\text{FP8-TC}}}
\newcommand{\Pfpsixteen}{P_{\text{FP16-TC}}}
\newcommand{\Bmem}{B_{\mathrm{HBM}}}
\newcommand{\Pnat}{P_\mathrm{fp64}}
\newcommand{\Plow}{P_\mathrm{low}}
\newcommand{\etaopt}{\eta_{\mathrm{opt}}}

\title{\Large\bfseries FP8 is All You Need (Part 2):\\[2pt]
Full-FP64 3-D FFT on FP8-Generation Tensor Cores\\[3pt]
\large\mdseries The Integer-Epilogue Wall and the Minimal Hardware
That Would Remove It}

\author{Satoshi Matsuoka\thanks{Correspondence: \texttt{matsu@acm.org}}\\[4pt]
\small Director, RIKEN Center for Computational Science (R-CCS), Kobe, Hyogo, Japan}

\date{Revision 10 --- September 3, 2026}

\begin{document}
\maketitle

\begin{abstract}
\noindent
Blackwell Ultra (B300) cuts FP64 vector throughput by ${\sim}30\times$
while multiplying FP8 tensor throughput. Following the recovery of
\fp{64} GEMM through Ozaki Scheme~II on \fp{8} tensor
cores~\cite{ozaki2025ii,uchino2026fp8} and the Tensor--Memory
Equilibrium model of the companion
papers~\cite{matsuoka2026tme,matsuoka2026ozaki25}, this paper asks
whether the fifth canonical HPC primitive---the full-\fp{64} $1024^3$
3-D FFT---can be carried by the same substrate, and answers with a
complete design and a precise account of where it stops. The design is
a Bailey six-step transform whose kernel contains no \fp{64}
arithmetic at all: the DFT GEMMs run on \fp{8} tensor cores through
Ozaki~II with the twiddle multiplication fused into the second-stage
operands, the complex arithmetic is carried in the residue domain by
Karatsuba combines, and the Chinese-remainder reconstruction is exact,
its bulk a small GEMM on the \fp{16} tensor path (one canonical residue
per operand, exact because every column sum stays below $2^{24}$) and
its remainder a Kulisch fixed-point accumulation with a two-sided
modulo-$M$ lift, so that the only rounding anywhere is the final
conversion. The modulus set is sized by an explicit capacity budget
($\log_2 M = 117.9$ against a $114$-bit requirement), and every
constant is machine-generated and verified bit-exactly by a
verification suite that also guards the rendered manuscript. The
central finding is that on such silicon the binding resource is not
floating point but a per-output \emph{integer epilogue}---the
deconstruction residual, canonicalisation, the combines, the Kulisch
deposits, the lift and the conversion---whose bandwidth-parity floor
has the closed form $(c_{\text{epi}}/8)\,\Bmem$ with
$c_{\text{epi}} \approx 203$--$281$ instructions per output. On B300
this holds the transform at $63$--$87$~ms against a $12.9$~ms memory
roof, $4.9$--$6.7\times$ short: at most $1.3$--$1.9\times$ faster than
the collapsed native path at peak integer issue, and possibly no faster
at realised issue efficiency. No software route we evaluate reaches the
roof on any part, and on Rubin, whose native \fp{64} already sits at
its own roof, emulation loses by $8$--$11\times$ and should not be run;
an \fp{32} variant meets the same wall at half precision, which
identifies the cause as reconstructing one scalar at a time rather than
\fp{64} itself. Because the floor is closed-form, each of its terms
names the hardware that would remove it, and the paper derives a
minimal co-design ladder: restore the INT8 tensor core that B200
already shipped ($4.5$~POPS against a $3.8$-POPS break-even) and give it
a \emph{position-weighted} cross-column accumulation primitive derived
from the two-limb \bint{8} layout (six INT64 group accumulators; the
exact $131$-bit signed magnitude needs a $132$-bit two's-complement
register, so INT128 is four bits short); add the load-path
deconstruction datapath the companion papers ask for as Option~C,
together with two multiword-carry ISA idioms; and, as the one moderate
ask, reduce modulo $m_k$ at the MMA output. The minor-hardware package
reaches $16.0$--$23.5$~ms, conditional on the accumulation primitive,
and the final rung $12.9$--$15.0$~ms---the roof at the optimistic end
of the epilogue interval---with no widening of the scalar INT32 pipe;
at Rubin-class bandwidth the break-even rises to $6.9$~POPS and the
epilogue becomes relatively harder, which is exactly why it merits a
named piece of silicon rather than reactive SIMT provisioning. All
figures are optimistic service floors projected from operation counts
and vendor peaks, not measurements, with sensitivity intervals stated
per term, the \fp{8} tensor budget stated conditionally on its plane
layout, and no area claims.
\end{abstract}

\section{Introduction}

The Tensor--Memory Equilibrium (TME) analysis of \cite{matsuoka2026tme}
established that Ozaki Scheme~II~\cite{ozaki2025ii} restores
\fp{64}-equivalent throughput to bandwidth-limited workloads on the
\fp{64}-collapsed B300 and Rubin GPUs, provided the kernel admits
register-level fusion of the residue decomposition (the $\beta\to 1$
discipline). Four of the five canonical primitives treated in that
line of work---dense GEMM, batched GEMV, structured stencils, and
Sparse Matrix--Vector multiply---share the property that the
\OII\ overhead parameters satisfy an amortisation condition under which
the kernel's wall time is bounded by $\max(\alpha W/\Plow,\,\beta
Q/\Bmem)$ and both conversion terms contribute negligibly---at the
engineered-$c_q$ bound; at the shipping $c_q$, Part~1 finds the
lowest-intensity of them (SpMV, single-RHS GEMV) conversion-bound and
interim-routed to the native pipe. Since the
snapshot on which this paper was first drafted, the TME model has been
extended (Part~1, from its Revision~32~\cite{matsuoka2026tme}, and the Ozaki 2.5
companion~\cite{matsuoka2026ozaki25}) from three overhead parameters to
\emph{four}: $\alpha$ (compute multiplier), $\beta$ (bandwidth
multiplier), $\gamma$ (per-output Garner reconstruction latency), and
$c_q$ (per-input residue \emph{deconstruction} cost)---the input-side
dual of $\gamma$, identified as a first-order omission by the NVIDIA
technical review of Bayraktar, Gunnels and Caday~\cite{bayraktar2026tme}.
The emulated wall time is now bounded by a \emph{three-way} service
maximum
$\max(\alpha W/\Plow,\,\beta Q/\Bmem,\,c_q r\varphi\,n_{\text{in}}/P_{\text{int}})
+ \gamma\, n_{\text{out}}$ (\S\ref{sec:tme}). For the four primitives
above, $\gamma$, $c_q$, and the tile-granularity charge all amortise;
the FFT is the primitive where they do not, which is what makes it the
fifth and the subject of this paper.

The thesis of this paper can be stated in three sentences, and we
state them at the outset rather than arriving at them: \emph{a
full-\fp{64} 3-D FFT needs no \fp{64} arithmetic anywhere in its
kernel---the Bailey GEMMs are emulated by \OII\ on \fp{8} tensor
cores, the twiddle Hadamard is fused into the second-stage DFT
operands, the Karatsuba combines run in the residue domain, the CRT
Phase~A rides the FP16 path of the same tensor cores, the reduction
is exact fixed-point with a two-sided modulo-$M$ lift, and the final
scaling is an exponent adjustment. What full-\fp{64} FFT does need,
and what this paper quantifies, is an integer SIMT pipe large enough
to carry the per-element deconstruction and per-output epilogue that
feed and drain the tensor cores---on B300 that pipe is the binding
floor, at $4.9$--$6.7\times$ the memory roof. Because the epilogue
floor is closed-form, it names its own hardware escape, and the
co-design ladder of \S\ref{sec:inttc} takes it to the roof.} The design is derived
directly from the cost model in \S\ref{sec:design}; the sequence of
weaker alternatives---recursive Garner, naive \fp{64} summation,
integer-hosted deposits---through which the result was historically
found is retained in \S\ref{sec:noint32}, compressed into an ablation
study, because seeing each alternative fail is what shows the
substrate assignment to be forced within the alternatives evaluated,
rather than one choice among many.

The companion paper~\cite{matsuoka2026tme} defers the
three-dimensional Fast Fourier Transform to the present analysis
(its \S7(b) states ``FFT. Handled in detail by the companion
paper'') and incorporates the wall-time and floor results
derived here into its end-to-end kernel-coverage audit. The
present paper develops those FFT results, and the two papers
together complete the \emph{analytical} canonical-kernel coverage of
the post-FP64 stack. The constructive findings sharpen the
message of \cite{matsuoka2026tme}: Ozaki-II emulation, combined
with the right classical numerical technique, removes every \fp{64}
dependence from the FFT even on GPUs whose FP64 vector pipe has been
deeply collapsed---and relocates the binding constraint to the
integer SIMT pipe, whose sizing becomes the FFT-specific procurement
question.

\paragraph{Contributions.} The paper leads with its destination and
derives it directly; the algorithmic lineage is retained, but as
evidence, not as the route. Eight results:
\begin{enumerate}[topsep=2pt, leftmargin=2em, itemsep=2pt]
\item \textbf{The direct design} (\S\ref{sec:design}, the headline,
constructive). A full-\fp{64} 3-D FFT with \emph{no \fp{64}
arithmetic anywhere}: Bailey GEMMs emulated by \OII\ on \fp{8} tensor
cores; the inter-stage twiddle Hadamard fused into the second-stage
DFT operand matrices; Karatsuba combines in the residue domain; CRT
Phase~A carried \emph{exactly on the FP16 path of the same tensor
cores} (one canonical residue per operand, no limb split; forced
there by the INT8-tensor collapse, item~2); an \emph{exact} Kulisch
fixed-point reduction~\cite{kulisch1976,kulisch1986} with a
two-sided modulo-$M$ centered lift, machine-generated from the
Rev-34 modulus set and unit-tested (\texttt{verify.py});
power-of-two exponent scaling.
Projected wall time (memory and tensor terms are lower bounds; the epilogue term is an estimate): $\approx 63$--$87$~ms for $1024^3$ at
full \fp{64} on B300 against a $12.9$~ms memory roof---\emph{at
most} $1.3$--$1.9\times$ the $116$~ms native path if attained.
\item \textbf{The integer-epilogue wall, and the second collapse}
(\S\ref{sec:obfft}, \S\ref{sec:deconstruction}). Blackwell Ultra
and Rubin have collapsed INT8 \emph{tensor} throughput ($167$/$250$
TOPS dense against $5$/$17.5$~PF FP8) exactly as B300 collapsed
\fp{64}---so every exact-arithmetic component must ride the
FP8/FP16 tensor path or the thin SIMT pipe. After migrating
everything migratable, an irreducible integer-SIMT
epilogue remains---deconstruction residual, canonicalisation,
combines, signed deposits, lift, conversion---totalling
$c_{\text{epi}} \approx 203$--$281$ instructions per output,
i.e.\ $\approx 25$--$35\,\Bmem$
against B300's $5.2\,\Bmem$ integer pipe. \emph{The FFT's binding
resource on shipping silicon is the integer pipe}, and no substrate
reassignment among those evaluated removes it.
\item \textbf{The software/hardware boundary, stated per part}
(\S\ref{sec:swhw}). Every construction in this paper runs on
shipping silicon; only the co-design rungs H1--H3 need new hardware.
Software alone takes B300 to $63$--$87$~ms at peak integer issue
($1.3$--$1.9\times$ its $116$~ms native path, $90$--$174$~ms at
realised issue efficiency; the modelled H0 optimisation would buy
$58$--$78$) and \emph{does not reach the memory roof
on either part}. On Rubin it takes the transform to $40$--$56$~ms
against a $4.9$~ms native path---emulation \emph{loses} there by
$8$--$11\times$, because Rubin's \fp{64} is still at parity with its
own roof. Hardware is required for any B300 result better than
$\approx 2\times$ native, and for roof attainment at all.
\item \textbf{The recommended co-design ladder, built on one
restoration} (\S\ref{sec:inttc}, in the manner of Ozaki 2.5: the
closed-form floor names its escapes). H0: pseudo-Mersenne moduli,
software, free. H1: restore the INT8 tensor core B200 shipped
(break-even $3.8$~POPS \eqref{eq:inttc-breakeven}; B200's $4.5$ is
$1.18\times$ enough) and give it split \emph{position-weighted}
accumulation derived from the two-limb layout---six INT64 group
accumulators; a two's-complement register for the exact 131-bit
signed magnitude needs 132 bits, so INT128 is 4 bits short. H2: the
load-path deconstruction datapath plus multiword-carry and
wide-\texttt{cvt} ISA idioms---both small. \emph{The package up to
H2 takes B300 from $63$--$87$~ms to $16.0$--$23.5$~ms against the
$12.9$~ms roof ($1.2$--$1.8\times$), conditional on the
position-weighted accumulation primitive that H1 also names.} H3
(moderate: modular reduction at the MMA output) reaches the roof at
the optimistic end of the epilogue interval ($12.9$--$15.0$~ms),
with no widening of the scalar INT32 pipe. A pure-SIMT alternative
(\S\ref{sec:codesign}) is priced and dominated.
\item \textbf{The ladder at Rubin-class bandwidth}
(\S\ref{sec:rubin22}). At 22~TB/s the Phase-A INT8-TC break-even
scales to $6.9$~POPS \eqref{eq:rubin-inttc}, and the up-to-H2
package lands $10.3$--$15.0$~ms against a $4.69$~ms roof
($2.2$--$3.2\times$; $1.4$--$2.1\times$ with H3): bandwidth grew
$2.75\times$ but the scalar pipe only $1.56\times$, so the epilogue
floor $(c_{\text{epi}}/8)\Bmem$ \emph{rises} with bandwidth. On
announced Rubin, native \fp{64} at $4.9$~ms makes emulation moot;
the target is a future Rubin-bandwidth part with B300's \fp{64}.
\item \textbf{Closed-form codesign floors, split by instruction
class} (\S\ref{sec:parity}). Native \fp{64} at $1.56\,\Bmem$
(B300 misses by $\sim 9\times$); FP8-TC at $444\,\Bmem$ and FP16-TC
at $96\,\Bmem$ (B300 tensor total $13.1$~ms---at the roof);
INT32-SIMT epilogue at $25$--$35\,\Bmem$ (B300 misses by
$4.9$--$6.7\times$). A prior formulation merged the tensor classes into
one floor and costed an INT8 MMA at the FP8 peak; both defects were
repaired throughout (\S\ref{sec:changelog}).
\item \textbf{The lineage as ablations} (\S\ref{sec:noint32}).
Recursive Garner ($\sim 111$~ms, Barrett double-count now fixed);
direct CRT with naive \fp{64} sum ($\sim 278$~ms); INT8-tensor and
dp4a Phase-A hostings ($\sim 85$/$46$~ms, dead on arrival); the
tensor-hosted deposit variant (qualitative only---its digitisation
cost was uncosted and its numbers are withdrawn). The
width-parameterised Kulisch fallback stays exact on any integer
width $\geq 8$~bits; cross-width performance awaits an ISA ledger.
\item \textbf{Parallel FP32 analysis} (\S\ref{sec:fp32}, corrected
twice). The native-FP32 parity floor is $3.125\,\Bmem$; at current
published specs the headroom is $6.4\times$ (H100), $3.0\times$
(B200), $3.33\times$ (B300), and $1.9\times$ (Rubin, at its
published 130~TF): FP32
FFT is adequately but not lavishly provisioned, and thins as
bandwidth grows.
\end{enumerate}
The argument is thus a single derivation: the cost
model locates the binding resource (2), the design minimises it (1),
the software/hardware boundary states what shipping silicon can do
about it (3), the floors quantify it (6), the lineage verifies no
evaluated alternative does better (7)---and the co-design ladder
(4), extended to Rubin bandwidth in (5), is what discharges it.

\begin{table}[h]
\centering
\caption{What is proved, what is modelled, and what is proposed. Each
claim in this paper belongs to exactly one row; the reader should
apply the corresponding confidence.}
\label{tab:confidence}
\small
\begin{tabular}{p{0.21\linewidth}p{0.44\linewidth}p{0.26\linewidth}}
\toprule
tier & contents & basis \\
\midrule
\textbf{proved arithmetic} & CRT capacity budget
  \eqref{eq:capacity}; \fp{16}-carrier exactness; biased Kulisch
  deposits; two-sided lift; exact positional bounds
  ($4360M{-}1$, $8724M{-}1$) & bit-exact unit tests
  (\texttt{verify.py}, 34 tests) \\
\textbf{modelled software} & epilogue instruction counts
  $[203,281]$; overlap and $\beta\to 1$; deconstruction residual
  $c_q^{\text{res}}$; instruction-mix $\varepsilon_{\text{mix}}$;
  the H0 factor & sensitivity intervals; no hardware measurement
  exists \\
\textbf{hardware proposals} & positional wide accumulation
  \eqref{eq:h1-group}; load-path converter; multiword-carry ISA;
  modular MMA-output reduction & dataflow stated; no area or power
  model \\
\bottomrule
\end{tabular}
\end{table} Implementing and measuring the construction---above
all the epilogue instruction counts that set the wall, now carried
explicitly as sensitivity parameters---is the immediate next step
(\S\ref{sec:future}).

\paragraph{Status of the results: a model, not a measurement.}
We state plainly at the outset what these results are and are not. The
proposed Ozaki-Bailey-Kulisch FFT has \emph{not} been implemented on
the target hardware; the wall-time figures (the headline
$\approx 63$--$87$~ms projection included) are derived from operation
counts and advertised peak rates,
not from measurement, and the per-output epilogue instruction counts
are stated as explicit sensitivity intervals rather than settled
constants. Every central ingredient remains unvalidated on
real silicon: the \fp{8} residue decomposition and the emulated matrix
operations; the fusion of decomposition, Bailey transforms,
reconstruction, and transpositions; the tensor-core formulation of
direct CRT; the six-word Kulisch accumulator with its carry
propagation, modulo-$M$ lift, and final conversion; and---most
consequentially---the assumed overlap of memory traffic, tensor work,
and integer-vector work, together with the achievable occupancy and
instruction throughput. Every figure of this kind in the paper is a
\emph{service-maximum lower bound on wall time} (equivalently, an
upper bound on attainable performance): a real kernel can only run
\emph{slower} than the stated number, never faster, and sequential
dependencies between GEMM stages, reconstruction, and the next
decomposition mean that sustaining several pipes near peak
simultaneously---which the bound assumes---is itself an unverified
scheduling claim. Which fraction of the bound a real kernel attains
is exactly what the measurement programme of \S\ref{sec:future} is
designed to determine. Like Part~1, this is a codesign argument about
what the hardware makes possible, not a report of delivered
performance.

\paragraph{On ``precision'' and ``accuracy''.}
We use these terms in their standard, distinct senses, and the
distinction matters for what is and is not claimed. \emph{Precision} is
a property of the number format---the \fp{64} format carries a 53-bit
mantissa---whereas \emph{accuracy} is a property of a computed result,
namely its closeness to the exact value. The Kulisch Phase~B is
\emph{exact} in a precise and provable sense: the accumulation
introduces no rounding error whatsoever, with a single correctly-rounded
integer$\to$\fp{64} conversion at readout (this is a property of the
fixed-point accumulator, independent of any hardware claim). The output
is therefore an ordinary \fp{64} number---it carries the full 53-bit
mantissa, no more and no less---with a \emph{worst-case} rounding
bound strictly stronger than that of naive \fp{64} summation, which
can lose bits to the $S=15$ intermediate roundings the exact
accumulator avoids. (A naive sum can be exact for favourable inputs;
the claim is about worst-case bounds for the CRT reduction step, not
a universal per-input dominance, and it covers the reduction only:
whole-transform accuracy additionally involves the scaling
quantisation, the residue arithmetic across six Bailey GEMMs, and
the stage cascade, whose end-to-end validation is future work,
\S\ref{sec:future}.) When we write ``full \fp{64}'' we mean exactly
this: an output of full \fp{64} mantissa width whose reduction error
is bounded by the stated final rounding(s). We do \emph{not} claim
more-than-\fp{64} precision (the result is an \fp{64} value), and we
reserve ``\fp{64}-equivalent'' for the throughput of the emulated path
(its rate matches what a native \fp{64} unit would deliver), never for
its accuracy. We have aimed to use this vocabulary consistently
throughout; the Phase~B accuracy analysis is in
Appendix~\ref{app:garner-detail}.

\paragraph{Novelty and prior art.}
The Bailey six-step decomposition~\cite{bailey1989} of FFTs is
classical; its application to tensor-core hardware was first
explored by \cite{durrani2021tensorfft} at reduced (\fp{16})
precision. The \OII\ scheme is due to Ozaki, Uchino and
Imamura~\cite{ozaki2025ii}, with the \fp{8} variant~\cite{uchino2026fp8},
FP64-emulation measurements on FP8 tensor cores~\cite{mukunoki2025},
and complex-CRT extension~\cite{uchino2025complex} from Q1-Q2 2026.
Concurrently with our work, Kawakami and
Takahashi~\cite{kawakami2026bluestein} have applied the Ozaki
scheme to FFT via a fundamentally different path: Bluestein's
algorithm reduces the DFT to a cyclic convolution, to which Ozaki
splitting is applied, with the split component convolutions
computed exactly via number-theoretic transforms (NTTs) on a CPU.
The two works are complementary (Bluestein--NTT for CPUs,
Bailey--tensor-core for GPUs); see \S\ref{para:kawakami} for a
detailed comparison.
Mantissa slicing for forward CRT is a standard technique in computer
algebra~\cite{bernstein2008fast}, but to the author's knowledge has
not previously been applied to tensor-core acceleration of \OII\
reconstruction. Kulisch's complete-arithmetic
accumulator~\cite{kulisch1976,kulisch1986} is also classical, but its
use as a software escape route for FP64-collapsed GPU architectures
appears to be new. The Phase~A/Phase~B direct-CRT decomposition, the
specific bandwidth-parity floors, and the codesign rule
are novel to this work. The input-side \emph{deconstruction} cost that
this revision charges on the integer pipe is not our own discovery: it
was identified as a first-order omission of the emulated-execution model
by the NVIDIA technical review of Bayraktar, Gunnels and
Caday~\cite{bayraktar2026tme} and developed into the four-parameter TME
model and the $c_q$ ladder by Part~1 (Revision~32 onward)~\cite{matsuoka2026tme} and
the Ozaki 2.5 companion~\cite{matsuoka2026ozaki25}; what is new here is
its instantiation for the Bailey-FFT kernel and the resulting combined
integer sub-floor. The width-parameterised Kulisch accumulator
(deliberately trading word width for lane count, building on classical
reduced-radix carry-save arithmetic~\cite{gueron2021avx512,collange2015repro,uguen2017kulisch})
and the tall-skinny FP8 tensor-core Phase~B that removes the
reconstruction deposits (though not the integer epilogue) from the
integer pipe are, to the author's knowledge, also new.

\subsection{What changed from the arXiv version}
\label{sec:changelog}

This manuscript revises the public arXiv:2606.23698v2 (26 June 2026;
v1 28 May 2026); the companion Part~1 is arXiv:2606.06510. Between v2
and this version the paper went through five independent technical
review rounds (two solicited reviews of the July 29--30 drafts, an
adversarial audit, and two correction rounds), a release-candidate
sign-off, and a cross-paper audit, and the changes are substantial
enough that a reader of v2 should treat the quantitative content as
superseded. The revision ledger at the end of this subsection dates
each state. In order of consequence:

\begin{enumerate}
\item \textbf{The modulus set was replaced (correctness).} The
Rev-33 set used through v2 has $\log_2 M = 111.84$ bits and fails
the CRT capacity budget \eqref{eq:capacity} by $2.16$ bits: a
constant (DC) input field already reconstructs wrong by exactly
$M$. Rev-34 (\S\ref{sec:tme}) replaces three primes, restores
$3.88$ bits of margin at unchanged $r=12$, and is now sized by an
explicit budget and regression-tested; every derived constant
($S=15$, slice tables, lift bounds) changed accordingly.
\item \textbf{Three further arithmetic repairs (correctness).} The
two-limb byte split now uses the shifted-column identity; the
signed Kulisch deposits use a biased carry-save accumulator; the
modulo-$M$ lift is two-sided. Each prior formulation fails on a
concrete counterexample now frozen in \texttt{verify.py}.
\item \textbf{Every hardware rate was re-audited (honesty of the
model).} v2 charged the INT32-SIMT epilogue at ``75~TOPS'', the
normalisation the companion papers adopt from the NVIDIA technical
memo (one issued \texttt{dp4a} per instruction at $75\cdot 10^{12}$/s);
this paper instead derives $41.7$~T~inst/s from the published FP32
TFLOP/s (which counts FMA${}=2$) under the assumption that \fp{32} and
integer issue share lanes, and every epilogue time doubled accordingly.
The two normalisations differ by $1.8\times$; which is physically
right is the integer-pipe census the companions defer to Part~3, and
the companions now carry the $41.7$ case as an explicit sensitivity
(their conversion-bound verdicts tighten under it; nothing reverses). v2 also mixed B300 SKUs (all columns
are now GB300-NVL72), costed an INT8 MMA at the FP8 peak, and
charged Phase~A at $100\%$ of an MMA peak its $k{=}12$ tile shape
cannot reach. The headline moved from v2's $\sim$20~ms class
claims, through $33$--$46$~ms, to the present $63$--$87$~ms; the
earlier ``$594\,\Bmem$ tensor floor,'' the Scenario-B
$17.5$--$24.3$~ms headline, and the $186$~ms Garner figure are
withdrawn, with retraction notes at the relevant sections.
\item \textbf{Phase~A moved to an FP16 tensor carrier}
(\S\ref{sec:classical-phasB}): B300/Rubin collapsed INT8-TC
$30$--$70\times$, killing v2's \bint{8} hosting; one canonical
residue per FP16 operand is exact by the $2^{24}$ column-sum bound.
\item \textbf{The software/hardware boundary is now stated per
part} (\S\ref{sec:swhw}): software alone gives B300
$63$--$87$~ms ($1.3$--$1.9\times$ native at peak issue, possibly no
gain at realised issue) and is an $8$--$11\times$ \emph{loss} on
Rubin; no software route reaches any roof.
\item \textbf{The co-design proposal was rebuilt around minimality}
(\S\ref{sec:inttc}): restore B200's INT8 tensor core with split
$6\times$INT64 positional accumulation ($4$ raw $+$ $2$ fractional
groups, H1), add a load-path
deconstruction datapath and multiword-carry ISA idioms (H2)---minor
hardware reaching $16.0$--$23.5$~ms against the $12.9$~ms
roof, conditional on a position-weighted accumulation
primitive---and narrow the interval with modular reduction at the
MMA output (H3). v2's bespoke output-stage ladder is retained as the
dominated alternative (\S\ref{sec:codesign}). New: the Rubin-class
$22$-TB/s extension (\S\ref{sec:rubin22}), the accumulator
width/split analysis (Table~\ref{tab:accwidth}), and the
\fp{32}-precision variant.
\item \textbf{Fourth and fifth (final-correction) review rounds}
tightened the model further: the $41.7$-T~inst/s integer rate is now labelled an
\emph{optimistic add-equivalent endpoint} with an instruction-mix
parameter $\varepsilon_{\text{mix}}\in[0.57,1]$; H1's accumulation
is specified as a \emph{position-weighted cross-column} primitive
\eqref{eq:h1-group} (new hardware, not a wider MMA register), and
the ladder is conditional on it; the positional-sum bounds are now
exact ($4360M{-}1$, $8724M{-}1$; a two's-complement register needs
132 bits, so INT128 is 4 bits short);
the pseudo-Mersenne claim is scoped to the 7 of 12 friendly moduli
and the H0 factor re-derived accordingly; H1's grouping is derived
from the \emph{two-limb} \bint{8} layout the ledger actually times
($6\times$INT64 groups, $12$ assembly inst/output---the one-limb
form's operand is not \bint{8}-representable), moving the up-to-H2
endpoint $13.0$--$19.5 \rightarrow 16.0$--$23.5$~ms and the H3
terminus to $12.9$--$15.0$~ms; $63$--$87$~ms is the
sole shipping-software baseline; Appendix~B's fractional CRT gained
its missing termwise reduction; and a manuscript-level gate
(\texttt{check\_manuscript.py}) now fails the build if any retired
constant reappears in the rendered PDF.
\item \textbf{A B200 cross-check was added} (\S\ref{sec:swhw}): the
design run on B200 as shipped with the \fp{64} pipe unused---Phase~A
exact on the native INT8-TC---lands at $70$--$96$~ms, $5.4$--$7.5\times$
its roof yet $1.2$--$1.7\times$ \emph{faster} than B300's native
\fp{64} path; B200 is the reference platform for measuring every
sensitivity interval in the paper.
\item \textbf{Revision 9 (August 19, 2026): a reconciliation, not a
correction.} A cross-paper audit of the three-paper bundle found that
this paper's $41.7$-T~inst/s integer-issue normalisation and the
companions' memo-sourced $75$ were stated as if one were simply wrong;
both are now framed as the two candidate normalisations of the same
pipe, differing by $1.8\times$, with the companions carrying the
$41.7$ case as an explicit sensitivity and the Part-3 pipe census
named as the arbiter (\S\ref{sec:tme}).  One intro clause now
qualifies the four amortising primitives ``at the engineered-$c_q$
bound'', matching Part~1's shipping-$c_q$ verdict for SpMV.  No
number in this paper changes.
\item \textbf{All constants are machine-generated}: the
\texttt{ledger.py}/\allowbreak\texttt{verify.py}/\allowbreak\texttt{check\_manuscript.py}
triple ships with the paper;
$34$ bit-exact unit tests include every review counterexample as a
regression guard, and the manuscript gate checks the rendered text
itself.
\item \textbf{Revision 10 (September 3, 2026): final pre-submission
pass.} A full read of the three-paper bundle by a fresh reviewer
model found no error that changes a headline, and the following
repairs: (i)~a \emph{layout caveat} is added to \S\ref{sec:tme}---the
$(3r{+}1)$ MMA count assumes a three-plane E4M3 layout that Ozaki 2.5
proves only up to modulus $577$, which the three $11$-bit Rev-34
primes exceed; the conservative $\alpha_{\mathbb{C}}=138$ alternative
($+2.2$~ms of tensor time, $1.19\times$ the roof) is now stated
wherever the tensor path is called ``at the roof'', and the
integer-epilogue headline is unaffected; (ii)~a stale naive-Ozaki
coefficient ($3.5$ for $3.75$) at three sites, two stale FP16-TC
floor cells in Table~\ref{tab:four-floors} (H100 $0.32$~PF, Rubin
$2.11$~PF), one stale $c_{\text{epi}}$ interval, the post-lift width
($<2^{117}$, not $112$ bits), and the fp32-variant slice count are
corrected; (iii)~the pure-SIMT ladder prose of
\S\ref{sec:codesign} is reconciled with the ledger's rung
assignment (the load-path converter reaches only the streamed
operand half; the residue-domain output stage carries deposits,
combines and canonicalisation); (iv)~the Kulisch-beneficiary
appendix now uses the table's $k=7$ throughout and gives the FFT's
epilogue-charged ratio; (v)~future-work items (1d) and (7) no
longer describe the demoted tall-skinny route as primary or
attribute strong integer tensor throughput to B300/Rubin; and
(vi)~cross-references to Part~1 name its current Revision~37 and
correct equation and section numbers. Revision numbers themselves are
new in this version: earlier states of this manuscript were labelled
only by review round or arXiv version, and the ledger below assigns
retroactive numbers from the review-response record.
\end{enumerate}

\paragraph{Revision ledger (reconstructed September 3, 2026).}
Revision numbers were assigned after the fact from the dated review
responses, the \texttt{ledger.py}/\texttt{verify.py} headers, and the
companion papers' synchronisation records; states before the July~29
draft other than the two public arXiv versions, and the exact date of
the fourth review-round state, are not individually recoverable and
are not numbered.
\begin{center}\small
\begin{tabular}{@{}lp{0.14\textwidth}p{0.66\textwidth}@{}}
\toprule
Revision & Date (2026) & State \\
\midrule
1 & May 28 & arXiv v1 (original snapshot; three-parameter TME model) \\
2 & June 26 & arXiv v2 \\
3 & July 29 & first post-v2 draft; first solicited technical review \\
4 & July 30 & response draft; second solicited review (\texttt{verify.py} T3/T4 counterexamples) \\
5 & July 31 & adversarial-audit round (\texttt{ledger.py} v3: Rev-34 moduli, $41.7$~T~inst/s, SKU consistency, tile padding, manuscript gate) \\
6 & early August & fourth review round (position-weighted H1 primitive; exact width bounds) \\
7 & August 3 & fifth (final-correction) round: two-limb H1 layout, $6\times$INT64, $16.0$--$23.5$ / $12.9$--$15.0$~ms \\
8 & August 4 & release-candidate sign-off; B200 cross-check; Part~1 Rev~34 synchronised \\
9 & August 19 & cross-paper reconciliation ($P_{\text{int}}$ normalisations; engineered-$c_q$ clause) \\
10 & September 3 & this version: final pre-submission pass \\
\bottomrule
\end{tabular}
\end{center}

\section{Background}

We recap the minimum machinery needed; the full TME exposition is
in \cite{matsuoka2026tme}.

\paragraph{Ozaki Scheme II.}
Given $A \in \mathbb{F}^{m\times k}_{64}$ and $B \in \mathbb{F}^{k\times n}_{64}$,
\OII\ \cite{ozaki2025ii} chooses scale factors $s_A, s_B$ such that
$\tilde A = \mathrm{round}(s_A \cdot A)$ and $\tilde B = \mathrm{round}(s_B \cdot B)$
are integer matrices, then computes
$\tilde C^{(i)} = \tilde A \bmod m_i \cdot \tilde B \bmod m_i \pmod{m_i}$
for $r$ \emph{pairwise-coprime} moduli $m_1,\dots,m_r$ (coprimality,
not primality, is what CRT requires; the shipping Rev-34 set
$\{2039, 2029, 2027, 1089, 1024, 961, 841, 625, 529, 511, 491, 487\}$
contains five perfect squares; seven of the twelve members are
reduction-friendly---$2039, 2029, 2027$ ($c = 2^{11}{-}m \le 21$),
$1024 = 2^{10}$, and $511, 491, 487$ ($c = 2^9{-}m \le 25$)---while
$1089, 961, 841, 625, 529$ have large complements and keep Barrett
reduction) on \fp{8}/\bint{8} tensor
cores. The product $C = AB$ is reconstructed by CRT:
\begin{equation}
C = \mathrm{CRT}(\tilde C^{(1)},\dots,\tilde C^{(r)};\,\{m_i\}) / (s_A s_B). \label{eq:ozaki}
\end{equation}
This paper works throughout with the actual Rev-34 modulus set, whose
product is $M = \prod_i m_i$ with $\log_2 M = 117.88$
(bit length 118). The set is \emph{sized by an explicit capacity
budget}, not chosen for convenience: exact CRT recovery of a
length-$K$ Bailey contraction of $k$-bit signed significands requires
\begin{equation}
\log_2 M \;>\; \underbrace{1}_{\text{sign}} + \log_2 K
   + 2(k+1) \;=\; 114 \ \text{bits}
   \qquad (K = 32,\ k = 53),
\label{eq:capacity}
\end{equation}
where the $+1$ per operand is the complex-Karatsuba third product
$(A_R{+}A_I)(B_R{+}B_I)$, which doubles each operand before
multiplication. \emph{The Rev-33 set used in every draft of this
paper before the present revision had $\log_2 M = 111.84$ and failed
\eqref{eq:capacity} by $2.16$ bits}: a constant (DC) input field
already reconstructs wrong by exactly $M$, because the $k=0$ DFT row
is all ones and the contraction attains $K\cdot 2^{2k} = 2^{111} >
M/2$. Rev-34 replaces the three primes $509, 503, 499$ by
$2039, 2029, 2027$, which restores $3.88$ bits of margin while
keeping $r = 12$---so $\alpha_{\mathbb{C}}$, and with it the entire
\fp{8} Bailey budget, is unchanged \emph{at the $(3r{+}1)$ count}. The new moduli remain below
$2048$, preserving the \fp{16}-carrier exactness argument of
\S\ref{sec:classical-phasB}, and are pseudo-Mersenne, which is also
the reduction-friendliness criterion rung H0 of the co-design
ladder (\S\ref{sec:inttc}) needs.

\emph{Layout caveat (Revision 10).} The $(3r{+}1)$ MMA count presumes
that each residue is carried in the three E4M3 planes of the
Ozaki-2/FP8 layout, and Ozaki 2.5~\cite{matsuoka2026ozaki25} proves
that layout exact only inside an envelope---nonsquare moduli up to
$321$ in canonical form and $577$ in redundant form, perfect squares
up to $1089$. Nine of the twelve Rev-34 moduli sit inside it; the
three $11$-bit primes $2039, 2029, 2027$ do not, and no exact
three-plane layout is proven for them. This paper therefore states
its \fp{8} Bailey cost as a pair: the $\alpha_{\mathbb{C}} = 111$
figure used throughout \emph{assumes} a three-plane layout for the
three large moduli (a layout obligation this paper does not
discharge), and the conservative alternative---one additional plane
product for each of the three, $\alpha = 46$,
$\alpha_{\mathbb{C}} = 138$---raises the Bailey time from $9.2$ to
$11.4$~ms on B300 (FP8-TC floor $552\,\Bmem$ in place of $444$) and
the summed tensor time from $13.1$ to ${\approx}15.3$~ms, i.e.\
$1.19\times$ the $12.9$~ms roof rather than at it. Every statement
below that the tensor path ``sits at the roof'' (or, after rung H1,
``under'' it at $12.3$~ms) should be read with this $+2.2$~ms
conditional overshoot in mind; the integer-epilogue wall of
\S\ref{sec:deconstruction} ($63$--$87$~ms) is unaffected, since it
is set by per-output scalar work and not by $\alpha$. Replacing the
three primes by envelope-conforming moduli (which forces $r = 13$
and $\alpha_{\mathbb{C}} = 120$) lands between the two figures and is
left to the implementation. Every derived constant below (slice counts,
accumulator widths, lift bounds) is generated from this set by the
supplied \texttt{ledger.py} script, not from the earlier
seven-bit-per-modulus approximation $\log_2 M\approx 7r$, which is
obsolete for this set. On Blackwell the FP8
substrate~\cite{uchino2026fp8} requires
$r\in[11,14]$ for fp64-equivalent precision in Ozaki-II (recommended
$r=12$); the INT8 substrate~\cite{ozaki2025ii} requires a larger
byte-modulus count $r_{\textsc{i}}\geq 15$ (an INT8 operand cannot hold
moduli $>256$). For FP8, the cost of one emulated \emph{real} GEMM is
$\alpha=(3r+1)$ MMAs of the same shape, because Karatsuba is used to
emulate signed int9 internally; for
INT8 it is $\alpha_{\textsc{int8}}=(r_{\textsc{i}}+1)=16$
MMAs~\cite{matsuoka2026ozaki25}, so the FP8/INT8 cost ratio is
$37/16\approx 2.31\times$ (corrected in Part~1's July 2026 revision cycle from the earlier
$37/13$; the INT8 substrate is cheaper per modulus but needs more
moduli). The substrate choice is, however, no longer free:
\emph{Blackwell Ultra and Rubin have collapsed INT8 \emph{tensor}
throughput exactly as B300 collapsed \fp{64}}---B300 lists 167~TOPS
dense INT8-TC against 5~PF FP8~\cite{uchino2025complex} ($30\times$),
and Rubin 250~TOPS against 17.5~PF~\cite{nvidia2026rubin}
($70\times$)---so every exact-arithmetic component in this paper must
ride the FP8/FP16 tensor path or the SIMT pipe.
\emph{This paper uses $r=12$ throughout for Ozaki-2/FP8.}
The complex-CRT extension~\cite{uchino2025complex} executes a
\emph{complex} GEMM as three real GEMMs (a Karatsuba split into
$D=A_RB_R$, $E=A_IB_I$, $F=(A_R{+}A_I)(B_R{+}B_I)$); this outer
factor of three is algebraically independent of the inner
signed-int9 Karatsuba inside $\alpha$, and the two must not be
conflated. The per-\emph{complex}-GEMM cost is therefore
$\alpha_{\mathbb{C}} = 3(3r+1) = 111$ MMAs at $r=12$
($\approx 109$ if the max-magnitude pass is shared across the three
real products); an earlier draft of this paper charged $3r+1=37$ for
the complex GEMM, an error identified by independent review and
corrected throughout this revision.

\paragraph{The Bailey six-step FFT.}
Given an array $X$ of length $N = pq$, index the input as
$j = j_1 + p\,j_2$ ($j_1\in[0,p)$, $j_2\in[0,q)$) and the output as
$k = k_2 + q\,k_1$ ($k_1\in[0,p)$, $k_2\in[0,q)$)---these index
conventions, made explicit at review request, are what the fusion
identity of \S\ref{sec:obfft-cost} relies on. Compute
$Y_k = \sum_{j=0}^{N-1} X_j \omega_N^{jk}$ via:
\begin{enumerate}[leftmargin=2em, topsep=2pt, itemsep=0pt]
\item Transpose to $q\times p$.
\item Length-$q$ DFT down each column:
$Y'_{j_1, k_2} = \sum_{j_2}\omega_q^{j_2k_2} X_{j_1,j_2}$
(matrix $F_q$).
\item Twiddle Hadamard: $Y''_{j_1,k_2} = \omega_N^{j_1 k_2} Y'_{j_1,k_2}$.
\item Transpose back to $p\times q$.
\item Length-$p$ DFT down each column:
$Z_{k_1,k_2} = \sum_{j_1}\omega_p^{j_1k_1} Y''_{j_1,k_2}$
(matrix $F_p$; fused form $F_p^{(k_2)}$ in \S\ref{sec:obfft-cost}).
\item Final transpose / permutation.
\end{enumerate}
(An earlier draft's stage list swapped the $p$/$q$ roles relative to
Algorithm~\ref{alg:obfft}---numerically hidden by $p=q=32$, flagged
by review, fixed here.)
Steps~2 and~5 are dense GEMMs with the DFT matrix when $p$ and $q$
are small (typically zero-padded to powers of two); the
GEMMs are tensor-core-amenable. For $N=1024$ the natural choice
is $p=q=32$, and the DFT$_{32}$ matrix fits in shared memory.

\paragraph{Related work: Ozaki-style FFT via Bluestein + NTT
(Kawakami--Takahashi).}
\label{para:kawakami}
Concurrently with the present work, Kawakami and
Takahashi~\cite{kawakami2026bluestein} independently propose an
Ozaki-scheme-based FFT method, taking a fundamentally different
algorithmic path: they reduce the DFT to a cyclic convolution via
Bluestein's algorithm~\cite{bluestein1970}, apply Ozaki splitting
to the convolution inputs, and compute the split component
convolutions \emph{exactly} using the number-theoretic transform
(NTT)~\cite{pollard1971}---an FFT over a finite field
$\mathbb{Z}/p\mathbb{Z}$---combined with CRT reconstruction. By
replacing the inner floating-point FFT with an NTT, they eliminate
rounding error in the convolution stage entirely. Their
implementation uses 32-bit NTTs and reports relative errors lower
than FFTW's double-precision FFT, requiring at most 96 NTT calls
(or 64 with NTT-domain accumulation, conceptually analogous to our
Phase~A/Phase~B split discussed in \S\ref{sec:tcgarner}). On Intel
Xeon Platinum 8468 for lengths $n=2^{10}\text{--}2^{18}$, their
method's execution time is $107$--$1315\times$ \emph{longer} than
FFTW's---the price of exactness on a CPU---with NTTs accounting for
$\sim 80\%$ of total time.

The two works are complementary rather than competing:
\begin{itemize}[topsep=2pt, leftmargin=2em]
\item \emph{Kawakami--Takahashi (Bluestein--Ozaki--NTT)} targets
CPUs (Intel Xeon) where integer and modular arithmetic is
bandwidth- and cache-bound but FP64 silicon is natively available,
and the goal is to \emph{reuse existing optimised
lower-precision FFT infrastructure} (NTTs) to compute
target-precision FFTs.
\item \emph{This paper (Ozaki-Bailey FFT with Kulisch Phase B)}
targets modern AI-optimised GPUs (B300, Rubin) where FP8 tensor
cores deliver PFLOPS-scale throughput but FP64 vector silicon has
been heavily reduced, and the goal is to \emph{exploit FP8 tensor
cores plus surviving integer-vector (INT32, or any width
$\geq 8$~bits; \S\ref{sec:noint32}) SIMT throughput} to recover
memory-roof parity for full \fp{64} FFT.
\end{itemize}
\begin{sloppypar}
Both routes apply the Ozaki splitting idea to the inner stage of
an FFT factorisation; the key algorithmic distinction is that
Kawakami--Takahashi factorise via Bluestein's cyclic convolution
(amenable to exact NTT arithmetic), while we factorise via Bailey's
six-step decomposition (amenable to dense-GEMM tensor-core
arithmetic). The architectural conclusions
of~\S\ref{sec:parity}---the bandwidth-parity floors and the
four-floor codesign rule---are agnostic to which Ozaki-FFT
formulation is used, and could in principle be derived for the
Kawakami--Takahashi route on architectures with high integer
tensor-core throughput (a natural extension noted
in~\S\ref{sec:future}).
\end{sloppypar}

\section{The TME Model: Recap}
\label{sec:tme}

We recap the Tensor--Memory Equilibrium (TME) model of
\cite{matsuoka2026tme}, paying particular attention to the
parameter $\gamma$, which becomes the binding cost in the FFT case
treated by this paper.

\paragraph{Operation-counting conventions.} Three different kinds of
``operation'' appear in this paper, and every division by a vendor
peak states which one it uses; an earlier draft mixed them (charging
some multiply-accumulates as one operation against peaks that count
them as two), an inconsistency identified by independent review and
repaired throughout this revision.
\begin{center}
\small
\begin{tabular}{@{}lll@{}}
\toprule
unit & definition & matching peak \\
\midrule
FLOP & one FP add \emph{or} multiply; FMA $=2$ & \fp{64}/\fp{32}
TFLOP/s (FMA $=2$) \\
tensor op & one low-precision mul \emph{or} add; MMA MAC $=2$ &
the \emph{class-matched} tensor peak \\
INT32 inst & one issued SIMT integer instruction & $41.7$-T-inst/s-class
issue rates \\
\bottomrule
\end{tabular}
\end{center}
Crucially---and this repairs a defect of a prior draft identified
by independent review---tensor work is divided by the peak of the
\emph{instruction class that executes it}, which on Blackwell
Ultra/Rubin differ by up to $30$--$70\times$:
\begin{center}
\small
\setlength{\tabcolsep}{4pt}
\begin{tabular}{@{}lcccccc@{}}
\toprule
GPU (dense, per GPU) & $\Bmem$ & FP8-TC & FP16-TC & INT8-TC &
INT32-SIMT & \fp{64} / \fp{32} vec \\
\midrule
B200~\cite{uchino2025complex} & 8 & 4.5~PF & 2.25~PF & 4500~T &
37.5~T & 37 / 75~TF \\
B300~\cite{nvidia2026gb300} & 8 & 5~PF & 2.5~PF & \textbf{167~T} &
\textbf{41.7~T} & 1.39 / 83.3~TF \\
Rubin~\cite{nvidia2026rubin} & 22 & 17.5~PF & 4~PF & \textbf{250~T} &
65~T$^\dagger$ & 33 / 130~TF \\
\bottomrule
\end{tabular}
\end{center}
{\footnotesize $^\dagger$No vendor publishes a SIMT INT32 rate. Under
this paper's convention (one \emph{issued} instruction, no factor 2)
it is $P_{\text{FP32}}/2$, since an FP32 TFLOP/s figure counts FMA as
two flops on the same lanes; that is the derived column above, and it
holds whether Blackwell's 128 cores/SM are unified FP32/INT32 or
separate co-issuing pipes. Drafts before this revision carried
``75~TOPS'' for B300---the NVIDIA memo's normalisation, which the
companion papers~\cite{matsuoka2026tme,matsuoka2026ozaki25} retain as
their reference constant and reconcile against this paper's $41.7$ in
their provenance appendices; under this paper's FMA${}=2$ convention
the memo's figure reads as an FP32 \emph{TFLOP/s} number reused as an
instruction rate, which halved every epilogue time in earlier drafts
of this paper. The Part-3 pipe census decides between them. All B300
columns are the GB300-NVL72
per-GPU dense figures; the B300-SXM variant of
\cite{uchino2025complex} Table~I (4.5~PF, 2.25~PF, 150~TOPS, 37.5~T,
1.2~TF) scales the epilogue by $1.11\times$ and raises the native bar
to $\approx 134$~ms.\par}
All work terms below are quoted in these units, and the supplied
\texttt{ledger.py} generates every count, time, and floor from this
table; \texttt{verify.py} unit-tests the arithmetic identities.

\subsection{Native Roofline}

Following the classical Roofline model~\cite{williams2009roofline},
for a kernel performing $W$ \fp{64} flops on $Q$ bytes of memory
traffic, with operational intensity $\mathrm{OI} = W/Q$, the
native (un-emulated) wall time is bounded below by
\begin{equation}
T_{\text{nat}} = \max\!\left(\frac{W}{\Pnat},\
                              \frac{Q}{\Bmem}\right)
                 + L_{\text{mem}}, \label{eq:roofline}
\end{equation}
where $\Pnat$ is the peak native \fp{64} throughput and
$L_{\text{mem}}$ is fixed kernel-launch and latency overhead. The
ridge point of the Roofline lies at
$\mathrm{OI} = \Pnat / \Bmem$; below it the kernel is
memory-bound, above it compute-bound.

\subsection{Emulated Roofline: definitions of $\alpha$, $\beta$,
$\gamma$}

Under \OII\ emulation the same fp64-equivalent work is performed,
but on \fp{8}/\bint{8} tensor cores at peak throughput
$\Plow \gg \Pnat$, plus a residue \emph{deconstruction} step on input
and a Garner \emph{reconstruction} step on output. Following Part~1
(Revision~32 onward)~\cite{matsuoka2026tme} and the Ozaki 2.5
companion~\cite{matsuoka2026ozaki25}, the TME model parameterises the
overhead by \emph{four} multipliers, and the emulated wall time is a
\emph{three-way} service maximum plus the reconstruction tail:
\begin{equation}
T_{\text{emu}} = \max\!\left(\frac{\alpha W_{\text{mma}}}{\Plow},\
                              \frac{\beta Q}{\Bmem},\
                              \frac{c_q\, r\, \varphi\, n_{\text{in}}}{P_{\text{int}}}\right)
                 + \gamma\, n_{\text{out}}. \label{eq:emuroofline}
\end{equation}
The third service term---the input-side deconstruction cost, absent from
the two-term form of the original snapshot---was identified as a
first-order omission by the NVIDIA technical review of Bayraktar,
Gunnels and Caday~\cite{bayraktar2026tme}. Here $W_{\text{mma}}=
\max(1,8/B)\,W$ charges the FP8 MMA's minimum 8-column accumulator (a
tile-granularity penalty that bites small batch $B<8$; see below),
$\varphi\in[0,1]$ is the fraction of streamed inputs actually converted,
$P_{\text{int}}$ is the integer-issue rate, and the reported \emph{rate}
uses the fuller service-min $P_{\text{svc}}$ of Part~1 (its Eq.~(14) in Revision~37), of
which \eqref{eq:emuroofline} is the serial endpoint. Concretely:
\begin{itemize}[topsep=2pt, leftmargin=2em]
\item \textbf{$\alpha$ (compute multiplier)}: the number of
\fp{8}/\bint{8} tensor-core MMAs required to emulate one fp64
\emph{real}-GEMM MMA. For Ozaki-2/FP8 with $r$ moduli, $\alpha = (3r+1)$
because each fp64-equivalent real inner product expands into $(3r+1)$
\fp{8} matmul invocations: the factor $r$ comes from the residue
planes, and the extra $(2r+1)$ comes from the Karatsuba structure
used internally to emulate signed int9 on the \fp{8} substrate.
For a \emph{complex} GEMM, the outer 3-real-GEMM Karatsuba split of
\cite{uchino2025complex} multiplies this again:
$\alpha_{\mathbb{C}} = 3(3r+1) = 111$ at $r=12$ ($\approx 109$ with a
shared max pass). The two Karatsuba layers are algebraically
independent; the FFT accounting below uses $\alpha_{\mathbb{C}}$.
For Ozaki-2/INT8 the cost is $\alpha_{\textsc{int8}} =
(r_{\textsc{i}}+1) = 16$ at $r_{\textsc{i}}\geq 15$, so the FP8/INT8
ratio is $37/16 \approx 2.31\times$ (corrected in Part~1's July 2026 revision cycle from the
earlier $37/13$; the INT8 substrate is cheaper per modulus but needs
more of them).\footnote{This $(3r+1)/(r_{\textsc{i}}+1)$ distinction
between the FP8 and INT8 substrates was clarified by Imamura
(private communication) and the modulus-count correction by the
Ozaki 2.5 companion~\cite{matsuoka2026ozaki25}; see Acknowledgments.}
\item \textbf{$\beta$ (bandwidth multiplier)}: the byte-ledger ratio
$\beta = Q_{\text{emu}}/Q_0$ of emulated to native HBM traffic. Part~1
(Revision~32 onward) fixes its definition and range: with the register-level fusion
discipline of \cite{matsuoka2026tme} the residue planes never reach
global memory and $\beta = 1$; in the unfused-materialised regime
$\beta \approx (8+2p)/8$ per streamed operand, where $p$ is the stored
plane-bytes per scalar ($p = 30$--$44$ for the candidate modulus sets),
so $\beta$ ranges over $[1,(8+2p)/8]\approx[1,12]$. \emph{The earlier
statement ``$\beta$ can grow to $r$'' conflated the byte multiplier with
the modulus count; $r$ is kept strictly as the modulus count, and
$\beta$'s ceiling is set by $p$, not $r$.} We flag at the outset that
$\beta = 1$ is an \emph{idealisation} whose realisability must be
established by measurement, not assumed. Driving $\beta\to 1$
requires that two distinct costs be hidden behind the tensor-core
MMA stream: (i)~the \emph{pre-computation}---scaling,
rounding-to-integer, and per-modulus reduction of the operands
into $r$ residue planes---and (ii)~the \emph{post-computation}
storage of the $r$ residue accumulators that feed Garner
reconstruction. Neither is free. The pre-computation is INT32-pipe
arithmetic that competes with the SIMT resources the reconstruction
also needs, and the $r$ residue accumulators consume register and
shared-memory capacity that, at production GEMM tile sizes, may
exceed what the SM can hold without spilling---in which case the
fused planes spill to HBM and $\beta>1$. A throughput model that
neglects this storage implicitly assumes unbounded registers; the
honest reading is that $\beta$ lies somewhere in $[1, (8+2p)/8]$, set by
the tile shape, the per-thread register budget, and the achievable
overlap, and is a \emph{per-kernel measured quantity}. The
Bailey-FFT case is more favourable here than general Ozaki-II
GEMM (\S\ref{sec:obfft}): its Phase~A is a fixed small
DFT$_{32}$ matrix resident in shared memory and reused across all
columns, so only the data operands---not a large streaming GEMM
tile---are decomposed in registers. Even so, the realised $\beta$
(and the realised $\alpha$, once pre-computation is charged in
full) is an empirical target of the validation programme of
\S\ref{sec:future}, not a settled constant.
\item \textbf{$\gamma$ (reconstruction cost per output)}:
the cost, per output element of the emulated GEMM, of
reconstructing the integer-valued product from its $r$ residues.
Operationally this is a fixed amount of small-modulus integer
arithmetic that runs on the INT32 SIMT pipe rather than tensor
cores. Its value depends on the reconstruction formulation used:
\begin{itemize}[topsep=1pt, leftmargin=2em]
\item Recursive Garner~\cite{ozaki2025ii}: $\gamma \sim
2.5\,r^2/P_{\text{INT32}}$ seconds, since the mixed-radix
recursion does $O(r^2)$ small-modulus integer operations per
output (derivation in \S\ref{sec:obfft-cost}).
\item Tensor-core direct CRT (this paper, \S\ref{sec:tcgarner};
\emph{not} Garner's mixed-radix recursion---the reformulation uses
precomputed CRT idempotents, and we name it accordingly):
$\gamma$ splits into a tensor-core Phase~A of $\sim r\cdot S /
\Plow$ seconds plus a Phase~B reduction whose cost depends on the
reduction scheme chosen, plus a per-output modulo-$M$ lift
(\S\ref{sec:classical-phasB}).
\end{itemize}
In the reported-rate ($P_{\text{svc}}$) form, reconstruction enters not
as a flat serial $\gamma\, n_{\text{out}}$ but as an overlapped
host tax $N_{\text{rec}}(r)/2k$ ($N_{\text{rec}} = r(r{-}1)/2 + 2r = 90$
at $r=12$) folded into whichever host pipe carries it; we retain the
serial $\gamma\, n_{\text{out}}$ as the endpoint of \eqref{eq:emuroofline}.
\item \textbf{$c_q$ (deconstruction cost per input, per modulus)}: the
SIMT integer-issue cost of converting one streamed \fp{64} operand
element into one residue plane---scale, round-to-integer, and reduce
modulo $m_i$. Total deconstruction work is $c_q\, r$ integer
instructions per converted element, the input-side dual of $\gamma$
(reconstruction is charged per output, deconstruction per input;
\cite{bayraktar2026tme,matsuoka2026ozaki25}). The Ozaki 2.5 anatomy
decomposes the per-element conversion into six stages---scale \&
truncate, byte-plane marshal, the reduction $\sum_k b_k(2^{8k}\bmod m)$,
final mod, Karatsuba split, and an ESC/max pass---of which the
\emph{reduction} stage (the bulk) is linear and \emph{migrates} onto
\bint{8} limb-GEMMs or packed dp4a, leaving a SIMT residual
$c_q^{\text{res}}\!\approx\!6$ per modulus (lazy variant $\approx\!4.3$,
codesigned $\approx\!5.3$) once migration is applied; the un-migrated
shipping value is $c_q = 16$. Unlike $\gamma$, whose SIMT keep-up budget
per byte of bandwidth \emph{shrinks} with each GPU generation (Part~1:
$8P_{\text{int}}/\Bmem$ falls as bandwidth outgrows integer rate),
$c_q$ is the term that makes the integer pipe a first-order constraint
rather than a rounding error.
\end{itemize}

\paragraph{The conversion-bound regime $\mathrm{OI}^{*}$.}
The deconstruction term makes a third binding regime possible,
alongside compute- and memory-bound: \emph{conversion-bound}. Part~1
defines the threshold operational intensity
\begin{equation}
\mathrm{OI}^{*} = \frac{c_q\, r\, \Pnat}{8\, P_{\text{int}}},
\label{eq:oistar}
\end{equation}
below which (at $\mathrm{OI} < \varphi\,\mathrm{OI}^{*}$) a kernel cannot
pay for emulation even against a collapsed native \fp{64} pipe. At the
B300 working point ($r=12$, $c_q=16$, $\Pnat = 1.4$~TF,
$P_{\text{int}} = 41.7$~T~inst/s), $\mathrm{OI}^{*} \approx 0.806$~flops/B.
The FFT's $\mathrm{OI}_{\text{FFT}} = 1.56$~flops/B sits above this, so deconstruction does not by itself unseat the memory
roof for FFT---but, as \S\ref{sec:deconstruction} shows, it still
roughly doubles the integer-pipe sub-floor once charged, which is what
moves the memory-roof path from the integer pipe to the \fp{8} tensor
cores.

\subsection{The $\gamma n_{\text{out}}$ regime}

The emulated wall time is bounded \emph{below} by the three-way
maximum of \eqref{eq:emuroofline} plus $\gamma n_{\text{out}}$: the
maximum is a service bound under ideal overlap, so real kernels run
at or above it, never below. The
TME model of \cite{matsuoka2026tme} treats both conversion terms---the
serial $\gamma n_{\text{out}}$ \emph{and} the deconstruction service
$c_q r\varphi n_{\text{in}}/P_{\text{int}}$---as small corrections, valid
when the per-GEMM operational intensity $k = W/n_{\text{out}}$ (flops
per output) is large enough that the $\alpha W_{\text{mma}}/\Plow$ or
$\beta Q/\Bmem$ term dominates.
For dense $M\times N\times K$ GEMM, $k=K \gtrsim 100$ and this
holds comfortably; for batched-small GEMV ($k \sim 10$) it is
marginal but still satisfied on Blackwell-class hardware.

\emph{The Bailey-FFT case breaks this assumption}: each Bailey
GEMM has inner dimension $k = q \approx \sqrt N$, which for
$N=1024$ gives $k = 32$. At this $k$, the per-output Garner
cost $\gamma$ becomes comparable to or larger than the per-output
compute and memory cost, and the third term in
\eqref{eq:emuroofline} becomes binding. Section~\ref{sec:obfft}
develops this analysis.

\section{The Ozaki--Bailey FFT Kernel: Costs and the Conversion Obstruction}
\label{sec:obfft}

\subsection{Algorithm and operation count}
\label{sec:obfft-cost}

Algorithm~\ref{alg:obfft} states the \OBFFT\ kernel. Each 1-D
FFT of length $N$ is realised as a Bailey six-step decomposition
in which each DFT$_p$ and DFT$_q$ is a complex GEMM emulated by
\OII\ via the 3-real-GEMM Karatsuba split of
\cite{uchino2025complex}.

\paragraph{Twiddle fusion (no \fp{64} Hadamard).}
An earlier draft kept the inter-stage twiddle Hadamard
$Y''_{j_1,k_2}=\omega_N^{j_1k_2}Y'_{j_1,k_2}$ in \fp{64}. Independent
review pointed out that this step was absent from the cost model,
and charging it is fatal on B300: $3N^3$ complex multiplies
$=18N^3\approx 1.9\times 10^{10}$~FLOPs, i.e.\ $\approx 13.8$~ms at
$1.4$~TFLOP/s---\emph{by itself exceeding the $12.9$~ms memory roof}.
We therefore \emph{eliminate} the step instead: the twiddle factor
depends on $(j_1,k_2)$ and the second-stage DFT contracts over $j_1$,
so the twiddle folds into the second-stage operand,
\begin{equation}
Z_{k_1,k_2} = \sum_{j_1} \omega_p^{j_1k_1}\,\omega_N^{j_1k_2}\,
Y'_{j_1,k_2}
= \sum_{j_1} \bigl[F_p^{(k_2)}\bigr]_{k_1,j_1} Y'_{j_1,k_2},
\qquad
\bigl[F_p^{(k_2)}\bigr]_{k_1,j_1} := \omega_p^{j_1k_1}\omega_N^{j_1k_2},
\label{eq:twiddle-fusion}
\end{equation}
turning the single shared DFT$_p$ matrix into $q$ per-column matrices
$F_p^{(k_2)}$: $q=32$ matrices of $32\times 32$ complex entries
($32^3 = 32{,}768$ values; $\approx 64$~KiB in raw complex E4M3, but
$\approx 0.75$~MiB as $12$ real/imag residue byte-planes and
$\approx 2.25$~MiB in the full plane representation, plus scale
metadata---L2-resident, \emph{not} shared-memory-resident, with
traffic amortised over the $N^2$ batched columns that reuse each
matrix; a prior ``$\approx 100$~KB'' stated no
representation and is corrected at review request). The MMA count is
unchanged---the second-stage GEMM becomes a batched GEMM with one
matrix per $k_2$ column group---and one \fp{64} rounding per element
per stage is \emph{removed}: the integer product is exact \emph{for
the quantised fused coefficient}; the coefficient quantisation error
itself remains, at the same bound as quantising the DFT matrix
(all twiddle entries are unit-modulus), and is part of the end-to-end
accuracy budget of \S\ref{sec:future}. The same fusion argument
disposes of the final scaling: with $s_A s_B$ restricted to powers of
two (as the Rev-34 configuration provides), the scale is an exponent
adjustment inside the final conversion, not an \fp{64} multiply. On a
part with a collapsed \fp{64} pipe, \emph{any} per-element \fp{64}
work is prohibitive ($\approx 0.7$~ms per surviving FLOP per element
at $N=1024$)---this, not elegance, is why the design eliminates the
category entirely. The fused batched-GEMM schedule is an
implementation obligation of the measurement programme
(\S\ref{sec:future}).

\begin{algorithm}[t]
\caption{Ozaki-Bailey 1-D FFT (length $N=pq$, $p\approx q\approx
\sqrt N$), twiddle-fused, no \fp{64} arithmetic}
\label{alg:obfft}
\begin{algorithmic}[1]
\Require Input $X \in \mathbb{C}^N$; DFT matrix $F_q$ and twiddle-fused
matrices $\{F_p^{(k_2)}\}_{k_2=1}^{q}$ (Eq.~\ref{eq:twiddle-fusion}),
all precomputed with residue planes $\{\cdot \bmod m_i\}_{i=1}^r$.
\emph{Scaling invariant (per review): the shared row/column
power-of-two scale vectors of the complex \OII\
scheme~\cite{uchino2025complex} are used, so that for every output
the three real products $D,E,F$ carry a \emph{common} scale---the
residue-domain combines below are valid only under this invariant;
the per-output exponent is carried as metadata to the final pack.}
\Ensure Output $Y = \mathrm{DFT}_N(X)$
\State Load $X$ tiled as $p\times q$, real/imag separated
\State Decompose $\mathrm{Re}(X)$, $\mathrm{Im}(X)$ into $r$ residue planes (fused in registers)
\For{$i = 1, \ldots, r$}
  \State $(\tilde D^{(i)},\tilde E^{(i)},\tilde F^{(i)}) \gets \text{cplxKaratsuba}(F_q\bmod m_i,\ X\bmod m_i)$
         \Comment{$3\times(3$ \fp{8} MMAs$)$}
  \State $\tilde R^{(i)} \gets \tilde D^{(i)}{-}\tilde E^{(i)}$;\quad
         $\tilde I^{(i)} \gets \tilde F^{(i)}{-}\tilde D^{(i)}{-}\tilde E^{(i)}$
         \Comment{residue combine (valid under the shared-scale
         invariant), then canonicalise to $[0,m_i)$ for the FP16
         carrier}
\EndFor
\State $Y' \gets \mathrm{tcCRT}(\tilde R^{(1..r)},\tilde I^{(1..r)})$
       \Comment{Phase A tc-GEMM $+$ exact Kulisch $+$ mod-$M$ lift (\S\ref{sec:tcgarner}, \S\ref{sec:classical-phasB})}
\State Decompose $\mathrm{Re}(Y')$, $\mathrm{Im}(Y')$ into $r$ residue planes
\For{$i = 1, \ldots, r$}
  \State $(\tilde D^{(i)},\tilde E^{(i)},\tilde F^{(i)}) \gets \text{cplxKaratsuba}(Y'\bmod m_i,\ \{F_p^{(k_2)}\}\bmod m_i)$
  \State combine as above
\EndFor
\State $Z \gets \mathrm{tcCRT}(\cdots)$
\State \Return permuted $Z$ \Comment{scale $=$ power-of-two exponent adjustment in the final conversion}
\end{algorithmic}
\end{algorithm}

For 3-D $N^3$ FFT with $N=1024$, $p=q=32$, we now derive each cost
component in turn. The TME model
parameters~\eqref{eq:emuroofline} for this kernel are
$\alpha_{\mathbb{C}} = 3(3r+1) = 111$ (Ozaki-2/FP8 with $r=12$,
complex; \S\ref{sec:tme}), $\beta = 1$ (registered-fused
residue planes), and $\gamma$ to be computed from the recursive
Garner cost below.

\paragraph{Memory traffic $Q$.}
The 3-D FFT visits each of the $N^3$ complex-fp64 elements once
per axis: read on entry, write back permuted on exit. Each complex
element is 16 bytes (two fp64 words). Across all three axes,
\begin{equation}
Q = 3 \cdot N^3 \cdot 16\,\text{B} \cdot 2 = 96 N^3\,\text{B}
   \approx 103\,\text{GB} \text{ for } N=1024. \label{eq:Q-fft}
\end{equation}
At $\Bmem = 8$~TB/s, the memory roof is
$Q/\Bmem \approx 12.9$~ms.

\paragraph{Bailey MMA count: the issued-MMA ledger.}
Each 1-D FFT in the Bailey six-step decomposition requires 2
complex GEMMs (the length-$q$ and length-$p$ DFTs). An earlier
draft charged each complex GEMM at $\alpha = 3r+1 = 37$, asserting
that the formula ``already absorbs'' the complex 3-real-GEMM split;
independent review showed this conflates two algebraically distinct
Karatsuba layers, and we correct it here with an explicit ledger.
Per complex Bailey GEMM of shape $(p\times p)\cdot(p\times pN^2)$:
the outer complex Karatsuba forms \emph{three} real GEMMs
($D,E,F$); each real GEMM expands into $3r$ \fp{8} plane-product
MMAs plus a max-magnitude pass ($3r{+}1$ per Part~1); nothing in
either layer cancels the other, so the issued count is
$\alpha_{\mathbb{C}} = 3(3r+1) = 111$ MMAs per complex GEMM
($\approx 109$ if the max pass is shared across $D,E,F$). Each
complex GEMM contributes $p^2\cdot pN^2 = 32N^3$ complex MACs, so
across the 3-D FFT (6 stage-GEMMs):
\begin{equation}
W_{\text{MMA}} = \underbrace{6}_{\text{stage GEMMs}}
        \cdot\, \alpha_{\mathbb{C}} \cdot 32N^3\ \text{MACs}
        \;=\; 2\cdot 6\cdot 111\cdot 32N^3
        \;\approx\; 4.6\times 10^{13}\ \text{tensor ops}
\quad(\text{MAC}=2\ \text{ops}),
\label{eq:wmma}
\end{equation}
i.e.\ $\approx 9.2$~ms at $\Plow = 5$~PF tensor ops/s---three times
the prior $1.5\times 10^{13}$ / $3$~ms, and no longer
negligible against the $12.9$~ms memory roof. The sensitivity is
central, not cosmetic: if a fused complex/residue identity
eliminating the outer factor of three exists, it must be derived
and proved; we know of none, and all figures in this paper use
$\alpha_{\mathbb{C}}=111$.

\paragraph{Recursive Garner count.}
Garner reconstruction~\cite{garner1959} converts the
$r$-tuple of residues $\{v'_k = v \bmod m_k\}$ back to the integer
$v$. The mixed-radix recursion expresses $v$ in a positional
notation with mixed bases $m_1, m_2, \ldots$:
\begin{equation}
v = v_1 + m_1(v_2 + m_2(v_3 + \cdots + m_{r-1} v_r)),
\quad v_k \in [0, m_k).
\end{equation}
The digits $v_k$ are computed in sequence by
\begin{equation}
v_k = \left(v'_k - \sum_{j<k} v_j \prod_{i<j} m_i \right)
      \cdot \left(\prod_{i<k} m_i\right)^{-1}
      \pmod{m_k}, \label{eq:recursive-garner}
\end{equation}
so each digit $v_k$ requires $k-1$ multiply-add-mod operations
against the previously-computed lower digits. The total per-output
work is $\sum_{k=1}^{r} (k-1) = r(r-1)/2$ such triples. Each
triple involves one INT32 multiply, one Barrett-reduced modulus
(itself $\sim 3$ INT32 ops: multiply, shift, conditional
subtract), and one INT32 add---roughly $5$~ops per triple.
Including the inverse-mod multiplication step at the end of each
iteration, the per-output count is
\begin{equation}
\text{ops/output} \approx 2.5 \, r^2 \quad \text{INT32 ops, at } r=12: \approx 360.
\end{equation}

\paragraph{How many scalar reconstructions? ($n_{\text{out}}=12N^3$.)}
A prior formulation reconstructed the three Karatsuba products $D,E,F$
independently ($18N^3$ scalar reconstructions) while its algorithm
combined them before reconstruction; the dataflow is fixed explicitly. The Karatsuba
combines are \emph{linear}, so they are performed in the residue
domain (Algorithm~\ref{alg:obfft}, line~5):
$\tilde R^{(i)} = \tilde D^{(i)}{-}\tilde E^{(i)}$ and
$\tilde I^{(i)} = \tilde F^{(i)}{-}\tilde D^{(i)}{-}\tilde E^{(i)}$,
computed as \emph{signed} INT32 values, with canonicalisation
applied \emph{once}, after the combines and before the adopted
FP16/FP32 Phase-A carrier (which requires canonical nonnegative
residues; the epilogue ledger charges exactly this single pass).
The congruence $C\equiv\sum_k v_ku_k \pmod M$ itself holds for
signed residues, and the $z$-range budget in \texttt{ledger.py}
covers the intermediate $|v_k|<2m_k$ signed bound. Only \emph{two} scalar components (Re, Im) per
complex output are then reconstructed:
\begin{equation}
n_{\text{out}} = 3\ \text{axes}\cdot 2\ \text{stage GEMMs}\cdot
2\ \text{components}\cdot N^3 = 12N^3 \approx 1.29\times 10^{10},
\end{equation}
not $18N^3$. The combine cost ($\approx 3$ INT32 ops per modulus per
complex output, $\approx 2.3\times 10^{11}$ instructions total) is
charged to the integer epilogue in \S\ref{sec:deconstruction}.
The recursive-Garner rung then costs
\begin{equation}
W_{\text{Garner}} = 360\cdot 12N^3 \approx 4.6 \times 10^{12} \text{ INT32 ops}.
\end{equation}

\paragraph{Deconstruction count (input side).}
Dual to reconstruction, the kernel must convert each operand into its
$r$ residue planes before every Bailey GEMM. Algorithm~\ref{alg:obfft}
decomposes twice per 1-D FFT---the input $X$ and the reconstructed
intermediate $Y'$ (the twiddle now lives inside the second-stage
operand matrices, Eq.~\ref{eq:twiddle-fusion})---each over the complex array (real and imaginary
parts separately, the $\times 2$ that the complex-Karatsuba split later
reuses). The number of scalar decompositions is therefore
\begin{equation}
n_{\text{dec}} = 3\text{ axes}\cdot N^2\text{ 1-D FFTs}\cdot 2\text{ decompose steps}\cdot 2\ (\mathrm{re},\mathrm{im})\cdot N
              = 12 N^3,
\end{equation}
equal to the $12 N^3$ reconstruction outputs. After migrating the
linear reduction stage onto the \bint{8}/dp4a pipe (Ozaki 2.5), the
SIMT residual is $c_q^{\text{res}}\!\approx\!6.3$ integer ops per
modulus in the general case. The FFT admits a plausibly cheaper
special case: its operands have bounded, statically-known dynamic
range (unit-modulus twiddles; bounded per-stage growth), so the
per-element exponent scan and scale logic can be replaced by a
static scale, leaving an estimated
$c_q^{\text{res,static}}\!\approx\!3$ (final conditional subtract
$+$ shift; \emph{our estimate, unmeasured}---the general $6.3$ is
Ozaki 2.5's). We carry both endpoints throughout:
\begin{equation}
W_{\text{dec}} = n_{\text{dec}}\cdot c_q^{\text{res}}\cdot r
              \approx 4.6\text{--}9.7\times 10^{11}\ \text{INT32 ops}
              \quad (c_q^{\text{res}}\!\in\![3.0,6.3],\ r\!=\!12),
\end{equation}
i.e.\ $6.2$--$13.0$~ms alone on B300's $41.7$-T-inst/s pipe against the
$12.9$~ms roof (\S\ref{sec:deconstruction}). Ozaki 2.5's
lazy-reduction variant would lower the residual to $\approx\!4.3$,
but at the disclosed cost of roughly doubling the tensor-side plane
count ($\alpha'\approx 6r{+}1$)---a trade its own paper excludes,
and which the FFT, already tensor-tight
(\S\ref{sec:parity}), cannot afford either; we do not use it.
The migratable bulk ($\sim r\, n_{\text{dec}}$ byte MACs) was
destined by Ozaki 2.5 for ``INT8 tensor or dp4a''; with B300's
INT8-TC collapsed to 167~TOPS (\S\ref{sec:tme}), it lands on
\emph{packed dp4a on the SIMT pipe}, adding a small
($\approx 2$--$4$ inst/element, $\approx 0.3$--$0.7$~ms) charge that
the epilogue ledger of \S\ref{sec:deconstruction} carries
explicitly.

\paragraph{Tile-granularity note.}
The Bailey-MMA count above assumes the FP8 MMA's 8-column accumulator is
filled. A length-$32$ butterfly processed with batch $B<8$ pays the
$W_{\text{mma}}=\max(1,8/B)\,W$ penalty of \eqref{eq:emuroofline}; we
assume the implementation batches $\geq 8$ columns (FFTs) per MMA tile
to stay in the $B\!\geq\!8$ regime, and flag the padded-lane cost as a
measured quantity of \S\ref{sec:future} where it cannot.

\paragraph{One rate convention, applied once.}
The $\approx 360$-instruction count above already prices Barrett
reduction at $\approx 3$ issued instructions per modular operation;
it must therefore be divided by the \emph{nominal} $41.7$-T-inst/s issue
rate. (Earlier drafts also divided by an ``effective 25~TOPS''
obtained from the same factor of three---charging Barrett twice, an
error identified by independent review; the recursive-Garner rung
improves from the previously printed 186~ms to 111~ms, which changes
no conclusion but is now stated correctly.)

\paragraph{Wall-time summary.}
At B300's 8~TB/s HBM, 5~PF FP8-TC, and $41.7$-T-inst/s INT32 issue:
\begin{center}\small
\begin{tabular}{lrl}
\toprule
Phase & Time (ms) & Derivation\\
\midrule
HBM traffic & 12.9 & $Q/\Bmem = 103\,\text{GB}/8\,\text{TB/s}$ \\
Bailey MMAs & 9.2 & $W_{\text{MMA}}/\Plow = 4.6\!\times\!10^{13}/5\,\text{PF}$\\
\textbf{Recursive Garner (SIMT)} & \textbf{$\sim 111$} &
$W_{\text{Garner}}/P_{\text{INT32}} = 4.6\!\times\!10^{12}/41.7\,\text{T inst/s}$\\
\bottomrule
\end{tabular}
\end{center}
The $\gamma$ term dominates the kernel by a factor of $\sim 8.6$
over the memory roof ($111/12.9$). The throughput-amortised per-output Garner
cost is
\begin{equation}
\gamma \approx \frac{2.5 r^2}{P_{\text{INT32}}}
       = \frac{360 \text{ inst}}{41.7 \times 10^{12}\,\text{inst/s}}
       \approx 8.6\,\text{ps/output},
\end{equation}
so $\gamma n_{\text{out}} \approx 8.6\,\text{ps} \cdot
1.29\!\times\!10^{10} \approx 111$~ms. \emph{The TME assumption
$\gamma n_\text{out} \ll W/\Plow$ fails decisively at $k \approx
32$}: the right-hand side is $W/\Plow = $ Bailey MMAs $= 9.2$~ms,
while $\gamma n_{\text{out}} = 111$~ms is roughly $12\times$ larger.

\subsection{Tensor-core direct CRT: Phase A / Phase B split}
\label{sec:tcgarner}

To attack the $\gamma$-bottleneck, we replace Garner's mixed-radix
recursion by \emph{direct CRT with precomputed idempotents} (an
earlier draft called this reformulation ``tensor-core Garner''; the
formula below is not Garner's, and we rename it tc-CRT):
\begin{equation}
C \equiv E := \sum_{k=1}^r v_k \cdot u_k \pmod M,
\qquad u_k = \frac{M}{m_k}\cdot \left(\frac{M}{m_k}\right)^{-1}_{\!m_k},
\quad M = \prod_i m_i. \label{eq:fcrt}
\end{equation}
Note carefully that \eqref{eq:fcrt} is a \emph{congruence}, not an
equality: the raw sum $E$ equals $C + zM$ for some integer $z$
(bounded by $|z| \le r\max_k|v_k| < 2^{15}$ for the signed
post-combine residues, per \texttt{ledger.py}), and an explicit
modulo-$M$ lift is required before conversion---a step an earlier
draft omitted entirely, identified as a correctness blocker by
independent review and specified in full in
\S\ref{sec:classical-phasB}. The $\{u_k\}$ are large integers---for
the actual Rev-34 set, $\log_2 u_k \le \log_2 M = 117.88$
(\emph{not} the obsolete $7r=84$-bit approximation)---but are
\emph{constant} in any given Ozaki-II configuration. Slicing each
$u_k$ into
\begin{equation}
S = \lceil 118/8\rceil = 15
\end{equation}
unsigned \bint{8} chunks $u_k^{(s)}\in [0, 256)$ and substituting
into~\eqref{eq:fcrt}:
\begin{equation}
\sum_{k} v_k \cdot u_k = \sum_{s=0}^{S-1} 256^s \cdot
\underbrace{\sum_{k=1}^r v_k\cdot u_k^{(s)}}_{=:P_s} . \label{eq:slicing}
\end{equation}

\paragraph{Phase A: inner products on the FP16 tensor path.}
Two defects of a prior draft, both identified by independent
review, are repaired here together. \emph{First (the rate defect)}:
that draft specified Phase~A as an s8$\times$u8 IMMA but divided its
count by the FP8 peak; on B300 the INT8-TC rate is 167~TOPS
(\S\ref{sec:tme}), so the INT8-tensor hosting actually costs
$\approx 85$~ms---dead on arrival, on B300 and Rubin alike. \emph{The
repair uses the tensor cores' FP16 path}: after canonicalising each
residue to $v_k\in[0,m_k)$ ($\le 2038 < 2048$, the FP16
consecutive-exact-integer bound), one residue fits \emph{one} FP16
operand with no limb split at all; the $u$-slices $[0,255]$ are
likewise FP16-exact; every product is $\le 2038\cdot 255 = 519{,}690$
and every $r$-term column sum $\le 12\cdot 519{,}690 = 6{,}236{,}280 < 2^{23}$, far below
the FP32 accumulator's $2^{24}$ exact-integer bound
(\texttt{verify.py} T6)---so the contraction is exact with no
banding and no digitisation. The canonicalisation
($\approx 2$--$3$ inst per residue per output) is charged to the
integer epilogue (\S\ref{sec:deconstruction}). \emph{Second (the
radix defect)}: a two-limb formulation must shift its high-limb
column by one radix position ($v_1$ multiplies $u^{(s-1)}$, not
$u^{(s)}$; \texttt{verify.py} T2)---on the FP16 carrier no limbs
exist and the issue vanishes for the raw columns, but we retain the
shifted-column convention ($S{+}1$, $S_{\text{frac}}{+}1$) in the
byte-limb fallback ledger. The Phase~A count, including the
$S_{\text{frac}}=6$ fractional-lift columns of
\S\ref{sec:classical-phasB}, is
\begin{equation}
\begin{aligned}
W_{\text{A}} &= 2\cdot r\,(S{+}S_{\text{frac}})\, n_{\text{out}}
 = 6.49\times 10^{12}\ \text{tensor ops (unpadded; MAC}=2),\\
W_{\text{A}}^{\text{pad}} &= 2\cdot 16\cdot 24\cdot n_{\text{out}} = 9.90\times 10^{12}\ \text{(padded)},
\end{aligned} \label{eq:phaseAcost}
\end{equation}
i.e.\ $2.60$~ms unpadded / $3.96$~ms padded at B300's 2.5-PF dense FP16-TC rate (1.55~ms unpadded
on Rubin's 4~PF). An E4M3-pure alternative---Phase~A as an
Ozaki-plane GEMM reusing Part~1's carry-corrected E4M3 layouts---costs
$\approx 2\times$ more ($\approx 4.1$~ms at 5~PF) and is retained in
Appendix~\ref{app:garner-detail} as the fallback should the FP16
path's exact-integer accumulation contract fail measurement. The
Phase-A GEMM's output width ($21$ unpadded, $24$ padded) sits at the MMA tile granularity, so
a tile-granularity efficiency factor applies; we flag it as a
measured quantity.

\paragraph{Phase B: per-output reduction on FP vector pipe.}
Phase~B naively computes $\sum_s P_s\cdot c_s$ where
$c_s = 256^s / (s_A s_B)$ is a precomputed fp64 constant. The
operation count is $2\,S\cdot N_{\text{out}}$ FLOPs (FMA $=2$; an
earlier draft divided MAC counts by an FMA$=2$ peak without the
factor, understating this baseline by half). On B300's
1.4~TFLOPS FP64 vector pipe, Phase~B for $1024^3$ at $r=12$, $S=15$
takes
\begin{equation}
T_{\text{Phase B}} = \frac{2\cdot 12N^3\cdot S}{\eta_{\text{fp64-vec}}}
 = \frac{3.87\times 10^{11}\ \text{FLOPs}}{1.39\ \text{TFLOP/s}}
                   \approx 278~\text{ms}. \label{eq:phaseBcost}
\end{equation}
(This naive form also skips the modulo-$M$ lift and rounds at every
step; it is the accuracy-and-speed baseline, not a candidate.)
The Bernstein fractional formulation (Appendix~\ref{app:bernstein})
reduces $S$ to $r$ but gives the same scaling, so the conclusion is
robust: \emph{naively reduced, Phase~B is bounded by the FP64 vector
pipe, not the INT32 modular pipe}---and on B300 that pipe is collapsed.
The kernel's costs are now fully on the table: Bailey MMAs
($9.2$~ms, FP8-TC), memory traffic ($12.9$~ms roof), deconstruction
($6.2$--$13.0$~ms of integer residue), Phase~A reconstruction
($4.0$~ms, FP16-TC), and a Phase~B reduction with no acceptable
naive home. Which substrate carries Phase~B---and how the input side
constrains the choice---is the design question the next section
answers directly.

\section{The Direct Design: Full-FP64 FFT on FP8 Silicon}
\label{sec:design}

The obstruction of \S\ref{sec:obfft} and the four-parameter cost model
of \S\ref{sec:tme} fix the design almost uniquely among the substrate
assignments we evaluate, and we now derive it directly rather than by
refinement. The model dictates where each cost must run:
\begin{itemize}[topsep=2pt, leftmargin=2em, itemsep=1pt]
\item \emph{The Bailey GEMMs} run on the \fp{8} tensor
cores---this is the premise of \OII\ emulation and needs no argument.
At the corrected $\alpha_{\mathbb{C}}=111$ they cost $9.2$~ms against
the $12.9$~ms roof, so the tensor pipe is \emph{tight}, not abundant:
nothing else of first order can be added to it.
\item \emph{Every per-element \fp{64} step is eliminated by
construction}---twiddles fused into the second-stage operands
(\S\ref{sec:obfft-cost}), combines in the residue domain, scaling by
exponent adjustment---because on B300 each surviving \fp{64} FLOP per
element costs $\approx 0.7$~ms per $N^3$ and the twiddle Hadamard
alone would exceed the roof.
\item \emph{CRT reconstruction} cannot run serially on the integer
SIMT pipe (the $\gamma$-roof, $\sim 111$~ms) nor per-output on the
collapsed \fp{64} vector pipe ($\sim 278$~ms), nor on the collapsed
INT8 tensor path ($\sim 85$~ms). Its bulk contraction
(Phase~A) rides the FP16 tensor path; its per-output remainder---the
signed Kulisch deposits, the two-sided modulo-$M$
lift, and the final conversion---is scalar work with no tensor
formulation known to us, and lands on the integer SIMT pipe.
\item \emph{Residue deconstruction} is, in its bulk, linear, and
migrates onto the \bint{8}/dp4a path per Ozaki
2.5~\cite{matsuoka2026ozaki25}; a nonlinear residual
($c_q^{\text{res}}\!\in\![3.0,6.3]$/modulus, \S\ref{sec:obfft-cost})
remains on the integer SIMT pipe. Ozaki 2.5 states explicitly that
this residual is elementwise-nonlinear and \emph{not} further
migratable to tensor cores.
\end{itemize}
The result is a machine that touches \emph{no \fp{64} silicon at
all}, whose tensor cores run near their own floor, and whose binding
resource is the \emph{integer-SIMT epilogue}: deconstruction residual
$+$ canonicalisation $+$ combines $+$ deposits $+$ modulo-$M$ lift
$+$ conversion, totalling
$\approx 63$--$87$~ms on B300's $41.7$-T-inst/s pipe against the $12.9$~ms
memory roof ($4.9$--$6.7\times$ off; \emph{at most}
$1.3$--$1.9\times$ the $116$~ms native path if attained). This section
derives each component in turn---the exact Phase~B with its
two-sided modulo-$M$ lift (\S\ref{sec:classical-phasB}), the
integer-epilogue charge (\S\ref{sec:deconstruction}), the
tall-skinny tensor deposit variant, retained qualitatively
(\S\ref{sec:fp8-tallskinny})---and
then assembles the codesign floors (\S\ref{sec:parity}); the
hardware co-design that removes the epilogue wall is
\S\ref{sec:inttc}. The
alternatives this design supersedes are instructive, and they are what
shows the substrate assignment to be \emph{forced within the
alternatives evaluated} rather than chosen; they are treated as
ablations in \S\ref{sec:noint32}.

\subsection{Exact Phase~B: the Kulisch fixed-point reduction}
\label{sec:classical-phasB}

The Phase~B reduction is, per output, a sum of $S=15$ signed
INT32 slice sums each scaled by a fixed positional weight, followed
by a modulo-$M$ lift:
$\text{result} = (s_A s_B)^{-1}\bigl[\bigl(\sum_{s=0}^{S-1} P_s\cdot
256^s\bigr) - zM\bigr]$,
with $\log_2 |P_s| \le 23$~bits. This is the canonical
multiply-accumulate-with-positional-weights structure studied
extensively in the floating-point summation literature. We surveyed
the classical options, focusing on their fit to B300 where the FP64
vector pipe is the binding bottleneck. The full survey is in
Appendix~\ref{app:garner-detail}; the qualitative finding is direct:

\paragraph{All classical methods other than Kulisch trade speed
against precision in the wrong direction.} Kahan, Neumaier, and
Ogita-Rump-Oishi Sum2/SumK compensation all run on the same FP64
vector pipe with $4$--$6\times$ overhead and so are strictly slower
than naive fp64 sum on B300 (without gaining precision-vs-cost on a
collapsed pipe). Double-double on fp32 substrate yields $\sim 48$
bits at $\sim 30$~ms---fast but not full fp64. Quad-double on fp32
yields $\sim 72$ bits at higher cost than full fp64 needs.
Reproducible BLAS (ReproBLAS) uses a bin-based signed-integer scheme
but still bottlenecks on the fp64 vector pipe for its bin floats.
Tall-skinny structured DGEMV on cuBLAS does not invoke tensor cores
for fp64 reductions on Blackwell/Rubin. None of these routes Phase~B
around the FP64 vector collapse.

\begin{figure}[h]
\centering
\includegraphics[width=\textwidth]{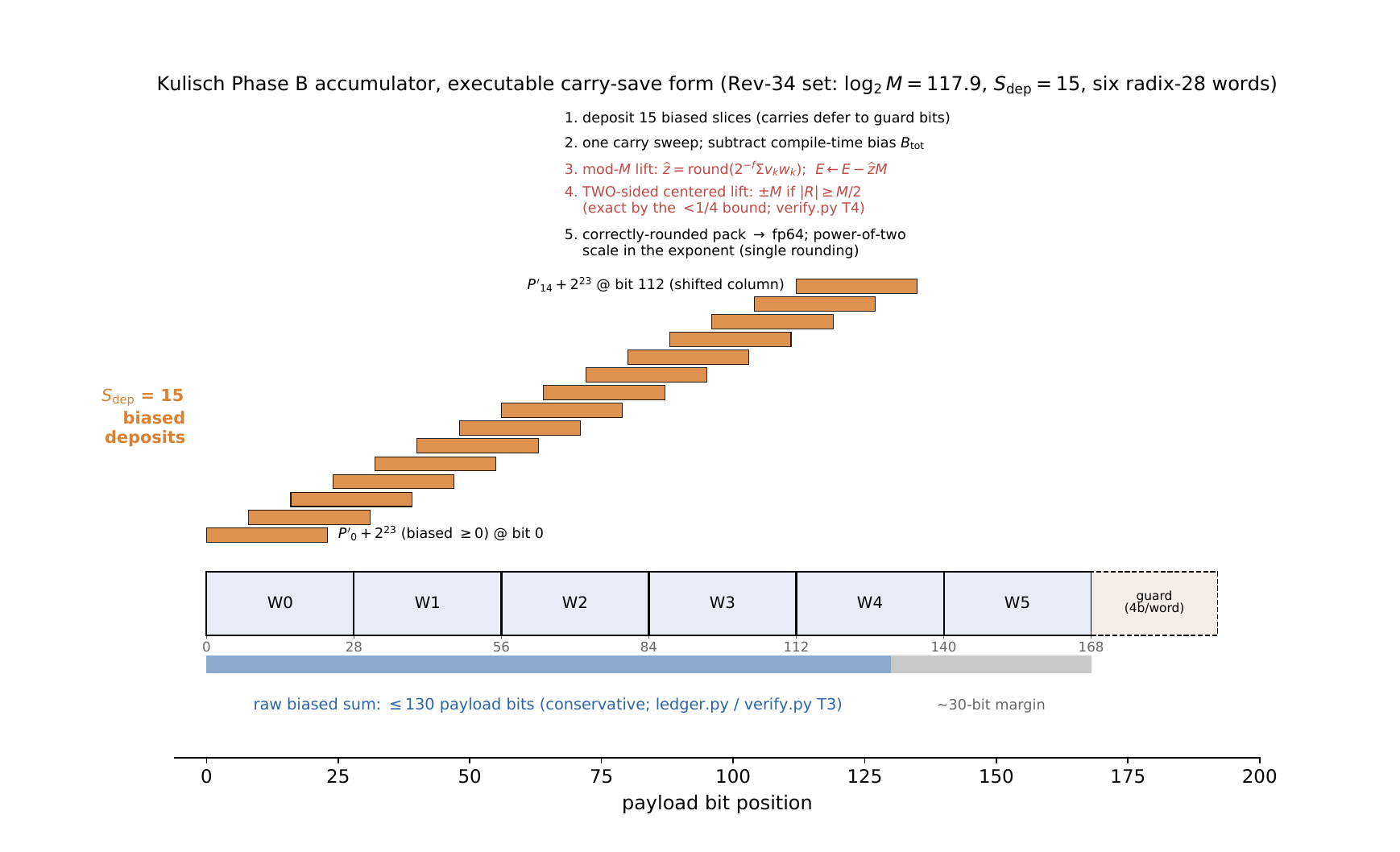}
\caption{Kulisch fixed-point Phase~B accumulator (Rev-34 modulus
set; executable carry-save form). The $S_{\text{dep}}=15$
shifted-column slice sums $P_s$ produced by Phase~A are signed INT32
values ($\le 23$ bits); each is \emph{biased} non-negative at
deposit and lands at bit position $8s$ across six radix-28
carry-save words (W0--W5, 4 guard bits each; logical width 160+
bits). The exact signed raw-sum magnitude of $131$~bits ($133$ in the
loose envelope) leaves ${\approx}27$--$29$~bits of margin. After the deposits: one guard-bit
carry sweep, the compile-time bias and $\widehat z M$ subtractions,
the \emph{two-sided} centered correction, and a single
correctly-rounded pack into \fp{64} with the power-of-two scale
folded into the exponent.}
\label{fig:kulisch}
\end{figure}

\paragraph{The Kulisch route.}
Kulisch's complete-arithmetic
proposal~\cite{kulisch1976,kulisch1986} maintains a wide fixed-point
register that accumulates products of floating-point values
\emph{exactly}, with a single conversion to \fp{64} at the final
readout. For our Phase~B specifically, the structure is unusually
clean. Recall the per-output reduction:
\[
y = (s_A s_B)^{-1}\,\bigl[E - zM\bigr],\qquad
E := \sum_{s=0}^{S-1} P_s\cdot 256^s,
\qquad P_s\in\mathbb{Z},\ \log_2 |P_s| \le 23\text{ bits},
\quad S = 15.
\]
Each term $P_s\cdot 256^s$ is an INT32 value shifted to a known bit
position---specifically bit $8s$, since $256^s = 2^{8s}$---so the
raw sum $E$ is bounded in width by
$\log_2(S\cdot \max |P_s|\cdot 256^{S-1}) \approx 133$~bits
(loose shifted-sum envelope; the \emph{exact} signed magnitude is
$131$~bits $= 8724M{-}1$, \texttt{ledger.py}). A logical 160-bit
accumulator suffices with $\approx 27$~bits of margin; its
\emph{executable} form is six
radix-28 carry-save words with 4 guard bits each and a biased
deposit convention, resolved by one carry sweep before the lift
(Figure~\ref{fig:kulisch}, Algorithm~\ref{alg:kulisch}). An earlier
draft claimed $\approx 104$~bits and $56$~bits of margin from the
obsolete $7r$-bit modulus model, and deposited signed values into
only two words---which \texttt{verify.py} T3 shows corrupts the
accumulator for $P_0=-1$.

\paragraph{The two-sided modulo-$M$ centered lift.}
Exact accumulation of $E$ is \emph{not} the reconstructed value:
Eq.~\eqref{eq:fcrt} is a congruence, and $E = C + zM$ with
$0 \le z \le r\max_k v_k < 2^{14}$ for canonical residues.
Sizing the accumulator prevents
\emph{overflow}; it does not select the congruence representative.
The lift is computed by the classical fractional-CRT
device~\cite{bernstein2008fast} (cf.\ Appendix~\ref{app:bernstein}):
precompute $w_k := \operatorname{round}(2^{f}\,u_k/M)$ at $f=48$
fractional bits and take
$\widehat z = \operatorname{round}\bigl(2^{-f}\sum_k v_k
w_k\bigr)$. The accumulated fixed-point error is below
$r\max v_k\,2^{-f} < \tfrac14$, which guarantees only that
$R = E - \widehat z M$ is within \emph{one} $M$ of the centered
representative---\emph{not} that $\widehat z$ is exact. The July-30
review exhibited the concrete failure: for $C = M/2-1$ under the
Rev-34 set, the fixed-point sum lands 10 units above a rounding
boundary, $\widehat z$ is one too large, and a one-sided correction
returns a value wrong by $M$ (reproduced bit-exactly in
\texttt{verify.py} T4). The repair, due to the review, is the
symmetric \emph{two-sided} correction, which the $<\tfrac14$ bound
renders exact with no guard band and no data-dependent fallback:
\[
R \leftarrow E-\widehat zM;\qquad
\text{if } R \ge M/2:\ R \leftarrow R-M;\qquad
\text{if } R < -M/2:\ R \leftarrow R+M.
\]
The contraction $\sum_k v_k w_k$ has \emph{exactly the Phase-A
shape}, so its $S_{\text{frac}} = 6$ columns
are appended to the Phase-A GEMM (already counted in
Eq.~\ref{eq:phaseAcost}); the surviving scalar work is forming
$\widehat z$ from the tensor partials, the $\widehat z\times M$
limb multiplies, the multiword subtract, and the two-sided
correction---estimated at $25$--$35$ issued INT32 instructions per
output, \emph{carried as a sensitivity interval, not a settled
count} (per review; SASS-level counts are a measurement deliverable,
\S\ref{sec:future}). The final conversion is a correctly-rounded
normalise-and-pack of the signed post-lift value ($|R| < M/2 < 2^{117}$) into \fp{64}
(sign handling, cross-word CLZ, guard/round/sticky formation,
round-to-nearest-even, exponent add folding the power-of-two
$s_As_B$ scale)---estimated at $20$--$30$ instructions, likewise a
sensitivity interval. \emph{No
floating-point arithmetic appears anywhere;} the accumulation is
exact, the lift is exact by the two-sided argument, and the only
rounding is the final conversion. We present the accumulator in its
integer-word form because that is where the exactness argument is
most transparent; \S\ref{sec:fp8-tallskinny} discusses a tensor-core
deposit variant, retained qualitatively.

\paragraph{The per-output kernel.}
Algorithm~\ref{alg:kulisch} gives pseudocode for the per-output
reduction. The two non-trivial steps are (i)~aligning each $P_s$ to
its bit position $8s$, which is a fixed-amount shift known at compile
time and so collapses to a constant-offset deposit into one or two
INT32 words, and (ii)~propagating the carry from the lower word to
the higher word, which uses a single \texttt{addcarry} idiom (one
\texttt{add} plus one conditional \texttt{add} for the carry, two
\texttt{add}s into adjacent words).

\begin{algorithm}[h]
\caption{Kulisch Phase~B for one Bailey-FFT output, \emph{including
the modulo-$M$ lift}. Constants
$\mathit{word}[s]$, $\mathit{shift}[s]$, the 4-word $M$, and the
$f{=}48$-bit fractions $w_k$ are compile-time;
$P_s$ are the $S_{\text{dep}}{=}15$ signed INT32 slice sums (the
shifted-column convention of \S\ref{sec:tcgarner}) and
$z_{\text{acc}}$
the fractional-lift partials, both produced by the same Phase-A
GEMM; $B_{\text{tot}} = \sum_s 2^{23}\,256^s$ is the compile-time
deposit bias; $s_A s_B$ is a power of two; output is full \fp{64}.
The biased representation and two-sided correction repair the
signed-deposit and one-sided-centering defects exhibited by the
July-30 review (\texttt{verify.py} T3/T4).}
\label{alg:kulisch}
\small
\begin{algorithmic}[1]
\State $\mathit{W}[0..5]\gets 0$ \Comment{six radix-$28$ carry-save words ($\geq$160 usable bits, 4 guard bits each)}
\For{$s = 0,1,\ldots,S_{\text{dep}}{-}1$}
  \State deposit $(P_s + 2^{23}) \ll 8s$ into $\mathit{W}$
        \Comment{biased $\Rightarrow$ non-negative; carries defer into guard bits}
\EndFor
\State normalise $\mathit{W}$ (one guard-bit carry sweep);\quad
       $\mathit{W} \gets \mathit{W} - B_{\text{tot}}$
       \Comment{compile-time bias off; merged with the $zM$ subtract below}
\State $\widehat z \gets \operatorname{round}(2^{-f} z_{\text{acc}})$
       \Comment{fractional-CRT lift from tensor partials}
\State $\mathit{W} \gets \mathit{W} - \widehat z\cdot M$
       \Comment{4 limb multiplies $+$ multiword subtract}
\State \textbf{if} $\mathit{W} \ge M/2$ \textbf{ then } $\mathit{W}\gets \mathit{W}-M$;\quad
       \textbf{if} $\mathit{W} < -M/2$ \textbf{ then } $\mathit{W}\gets \mathit{W}+M$
       \Comment{TWO-sided centered lift (exact by the $<\tfrac14$ bound)}
\State \textbf{return} $\mathrm{fp64}_{\mathrm{pack}}\!\left(\mathit{W};\ e_{s_As_B}\right)$
       \Comment{correctly-rounded pack; power-of-two scale in the exponent; the single rounding}
\end{algorithmic}
\end{algorithm}

Per deposit the work remains $\sim 4$ INT32 instructions (the bias
is folded into the deposit constant); $S_{\text{dep}}=15$ deposits
plus the guard-bit normalisation give $\approx 66$--$70$
instructions, the two-sided lift $25$--$35$, and the
correctly-rounded pack $20$--$30$---an integer-hosted Phase~B of
$\approx 111$--$135$ issued instructions per output, \emph{every
component carried as a sensitivity interval} pending SASS-level
counts (\S\ref{sec:future}). All work executes on the
INT32 SIMT pipe; \emph{no FP64 vector instruction appears anywhere},
which is the critical structural property. (There is no native
wide-integer$\to$\fp{64} conversion instruction; the pack sequence
\emph{is} the conversion, and its cost is charged---a fact the
co-design ladder of \S\ref{sec:inttc} turns into a hardware
ask.)

\paragraph{Why exact, and the accuracy claim's scope.}
\emph{Exactness}: each shifted slice
$P_s\cdot 256^s$ is exactly representable in the
accumulator; the additions and the $zM$ subtraction are integer
operations with no rounding; the lift is exact by the $2^{-f}$
error bound; the only rounding is the final conversion, bounded by
$2^{-53}\cdot |y|$ ($\leq 1$ ulp; the power-of-two scale is
error-free). The naive fp64 sum, by contrast, has a worst-case
bound that grows with its $S$ unmasked additions. The claim is a
\emph{worst-case-bound} comparison for the reduction step---a naive
sum can be exact on favourable inputs---and covers the CRT
reduction only, not end-to-end transform accuracy
(\S\ref{sec:future}).
\emph{Speed}: B300's INT32 SIMT issue rate is $41.7$~T~inst/s
($\sim 30\times$ its 1.39-TFLOPS FP64 vector throughput). For
$1024^3$ Bailey FFT, the integer-hosted Phase~B has
$12 N^3 = 1.29\times 10^{10}$ outputs at $111$--$135$ instructions
each:
\begin{equation}
T_{\text{Kulisch}}^{\text{B300}} \approx \frac{12 N^3 \cdot (111\text{--}135)}{\eta_{\text{INT32}}}
\approx 34\text{--}42\text{~ms at peak}.
\label{eq:kulisch-time}
\end{equation}
\emph{This omits the deconstruction residual, canonicalisation, and
residue combines} (\S\ref{sec:deconstruction}), which share the same
integer pipe: with them the full epilogue reaches
$\approx 2.6$--$3.6\times 10^{12}$
instructions ($\approx 63$--$87$~ms at peak). This epilogue---not
the choice of reconstruction host, not the tensor cores, and not
memory---is what separates the kernel
from the $12.9$~ms roof; \S\ref{sec:inttc} shows what removes
it.

\paragraph{Implementation considerations.}
Three practical notes about CUDA realisation:
\begin{itemize}[topsep=2pt, leftmargin=2em]
\item \emph{Register pressure.} The 6 INT32 accumulator registers are
held thread-local. Combined with the inputs $P_s$ (which can stream
in from shared memory, not be register-resident), this fits within
the per-thread register budget on Blackwell SMs without spilling.
\item \emph{Warp-level reduction.} If multiple threads in a warp
compute outputs sharing the same accumulator (e.g., when several
outputs are reduced across a Bailey GEMM column), the per-thread
160-bit accumulators must be combined via a 6-word \texttt{warp.reduce.add}, costing $\sim 25$~ops per warp once per
output. This overhead is one of the contributors to the realised
integer-issue efficiency whose derating is discussed in
\S\ref{sec:walltime} (a $50$--$70\%$ realised efficiency stretches
the epilogue bounds accordingly); it is not included in the
peak-issue instruction counts.
\item \emph{Double-buffering for overlap.} Because the integer-SIMT
work (both reconstruction and deconstruction) and HBM traffic use
disjoint pipes, one tile's integer work can be scheduled concurrently
with the next tile's HBM load. We stress what this does and does not
buy. The bandwidth-parity floors of this paper are already
\emph{ideal-overlap} (service-maximum) quantities---each is derived by
setting a pipe's compute time equal to the memory-roof time---so
double-buffering is the \emph{mechanism} that realises the model's
assumption, not a further relaxation of its floors. In particular, no
overlap schedule lowers the integer-epilogue demand of
\S\ref{sec:deconstruction}: even under perfect
overlap the integer-hosted path walls at
$\max(T_{\text{int}},T_{\text{mem}})\approx 63$--$87$~ms on B300
(Appendix~\ref{app:amdahl}).
\end{itemize}

\paragraph{Bridge to the design.}
The integer-word Kulisch kernel is exact and, in the
precision-runtime plane
(Figure~\ref{fig:precision}, \S\ref{sec:precision}), the only
full-precision candidate off the collapsed FP64 pipe. But
reconstruction does not run in isolation: \S\ref{sec:deconstruction}
next charges the same integer pipe for residue deconstruction,
canonicalisation, the
combines, the lift, and the conversion---an epilogue that exceeds
what B300's pipe can deliver at the roof. The integer-word form
presented here is the
\emph{specified}-construction realisation (algebraically complete
and unit-tested; not implemented); the tensor-hosted variant of
\S\ref{sec:fp8-tallskinny} is retained qualitatively only. The wall
times are projected, not measured, and no
production library yet incorporates a Kulisch Phase~B.

\subsection{The deconstruction tax: charging the input side}
\label{sec:deconstruction}

The Kulisch analysis of \S\ref{sec:classical-phasB} charges only the
\emph{output} side---the per-output Garner reconstruction. But the same
integer-SIMT pipe must also perform the per-\emph{input} residue
deconstruction (\S\ref{sec:tme}, the $c_q$ term): before any Bailey MMA
issues, every operand element is scaled, rounded, and reduced modulo
each of the $r$ moduli. In the original snapshot this cost was folded
implicitly into $\beta$ and $\alpha$; following the NVIDIA technical
review~\cite{bayraktar2026tme} and Ozaki 2.5~\cite{matsuoka2026ozaki25}
we charge it explicitly, and for the Bailey-FFT kernel it is not a
correction but a term of the same order as the reconstruction itself.

\paragraph{The deconstruction sub-floor.}
From the input count $n_{\text{dec}} = 12 N^3$ (\S\ref{sec:obfft-cost})
and the migrated SIMT residual $c_q^{\text{res}}\!\in\![3.0,6.3]$ per
modulus, the integer-pipe deconstruction work is
$W_{\text{dec}} = 12 N^3\, c_q^{\text{res}}\, r$. Setting it equal to
the memory-roof time $Q/\Bmem = 96N^3/\Bmem$ gives
\begin{equation}
\boxed{\;\etaopt^{\text{dec}}
     = \frac{12\, c_q^{\text{res}}\, r}{96}\,\Bmem
     = \frac{c_q^{\text{res}}\, r}{8}\,\Bmem
     \approx 4.5\text{--}9.5\,\Bmem \quad (c_q^{\text{res}}\!\in\![3.0,6.3],\ r\!=\!12),\;}
\label{eq:decon-floor}
\end{equation}
i.e.\ $\sim 36$--$76$~TOPS at 8~TB/s. Deconstruction alone demands up
to as much integer throughput as B300 has.

\paragraph{The full integer epilogue.}
Deconstruction is only one of five integer-SIMT charges; the others,
derived in \S\ref{sec:obfft-cost} and \S\ref{sec:classical-phasB} and
generated by \texttt{ledger.py} and unit-tested by
\texttt{verify.py}; every count is a stated sensitivity interval,
not a settled constant:
\begin{center}\small
\begin{tabular}{lrrr}
\toprule
integer-epilogue term & inst/unit [lo,hi] & floor ($\Bmem$) & B300 ms @$41.7$T\\
\midrule
deconstruction residual ($c_q^{\text{res}}\!\in\![3.0,6.3]$) & $[36,76]$/elem & 4.50--9.45 & 11.1--23.4\\
dp4a-hosted decon bulk & $[2,4]$/elem & 0.25--0.50 & 0.6--1.2\\
canonicalise residues to $[0,m_k)$ & $[24,36]$/out & 3.00--4.50 & 7.4--11.1\\
residue combines: out $36$ $+$ in $24$ & $60$/cplx out ($=30$/out-eq) & 3.75 & 9.3\\
biased carry-save deposits $+$ normalise & $[66,70]$/out & 8.25--8.75 & 20.4--21.6\\
two-sided modulo-$M$ lift & $[25,35]$/out & 3.12--4.38 & 7.7--10.8\\
correctly-rounded pack ($+$exp scale) & $[20,30]$/out & 2.50--3.75 & 6.2--9.3\\
\midrule
\textbf{total (single quantitative scenario)} & $[203,281]$/out-eq
 & \textbf{25.4--35.1} & \textbf{62.8--86.8}\\
\bottomrule
\end{tabular}
\end{center}
\begin{equation}
\boxed{\;\etaopt^{\text{epi}}
     = \frac{c_{\text{epi}}}{8}\,\Bmem
     \approx 25\text{--}35\,\Bmem
     \quad (c_{\text{epi}}\in[203,281]\ \text{inst/output}),\;}
\label{eq:combined-floor}
\end{equation}
i.e.\ $\approx 203$--$281$~T~inst/s at 8~TB/s against B300's $41.7$.
This is the central quantitative finding, and its closed form is
also its constructive content: like Ozaki 2.5's
deconstruction-$\lambda$ floor, \eqref{eq:combined-floor} is a
per-output instruction coefficient times bandwidth, so \emph{each
term of $c_{\text{epi}}$ names the hardware mechanism that would
remove it}---the co-design ladder of \S\ref{sec:inttc}. The
epilogue terms survive every \emph{software} substrate reassignment
we evaluated, because they are per-output scalar work with no known
tensor formulation. No overlap schedule softens any of it: like
every floor in this paper, \eqref{eq:combined-floor} is derived by
setting the pipe's compute time equal to the memory-roof time---i.e.\
it is already an \emph{ideal-overlap} (service-maximum) quantity, and
double-buffering is merely the mechanism that attains it
(Appendix~\ref{app:amdahl}).

\paragraph{Consequence for B300: the integer-epilogue wall.}
B300's $41.7$~T~inst/s ($=5.2\,\Bmem$) integer pipe misses the
epilogue floor by $4.9$--$6.7\times$ under perfect overlap: the
integer pipe carries $2.6$--$3.6\times 10^{12}$ instructions and the
transform walls at $\approx 63$--$87$~ms at peak issue, against a
$12.9$~ms roof and a $13.1$~ms tensor total. (The earlier
tensor-hosted-deposit variant, whose numbers previous drafts
carried, is demoted entirely in \S\ref{sec:fp8-tallskinny}: its
digitisation cost was uncosted, and it is not yet quantitative.) The honest statement of the thesis: eliminating
\fp{64} is \emph{achieved}---no branch of the design touches
it, and the collapsing INT8 tensor path is routed around too---but
the memory roof is \emph{not} reached on B300 by software alone.
What removes the wall is hardware: each term of
\eqref{eq:combined-floor} names its own escape, and the recommended
ladder of \S\ref{sec:inttc} reaches the roof with its up-to-H2
minor-hardware package plus one moderate ask. The
floor table of \S\ref{sec:parity} and the ablation ladder of
Figure~\ref{fig:walltime} (\S\ref{sec:walltime}) reflect this
accounting.

\subsection{Carrying the deposits on the \fp{8} tensor cores: the
tall-skinny formulation (design hypothesis)}
\label{sec:fp8-tallskinny}

The epilogue accounting of \S\ref{sec:deconstruction} shows that
moving the Kulisch \emph{deposit loop} off the integer pipe is worth
$\approx 5$--$6$~ms on B300. This subsection specifies, at first
order, how the \fp{8} tensor cores can carry it exactly. We label
the construction a \emph{design hypothesis} throughout: its operand
digitisation and band-stitching are specified below but neither
implemented nor measured, and (per independent review) an earlier
draft's version of this section rested on a false representability
claim, corrected here. One
false start must be dispatched first, because the mechanism that makes
Kulisch exact---accumulation with no rounding until readout---is
incompatible with FP8 \emph{floating-point} arithmetic.

\paragraph{FP8 vector arithmetic is ruled out.}
E4M3 carries a $3{+}1$-bit significand. Every FP8 add rounds to four
significant bits, so a literal fixed-point accumulation in FP8 is a
contradiction in terms. Even error-compensated summation
(Kahan/2Sum/Dekker) recovers at best of order twice the significand,
$\sim\!8$ bits, against the $53$ required. No amount of compensation
closes a $45$-bit gap; FP8 vector arithmetic cannot carry Phase~B.

\paragraph{FP8 as an exact integer carrier: the tall-skinny tensor
route.}
There is, however, a second way to use FP8 that sidesteps the FP8
adder entirely---provided the operands are digitised correctly. An
earlier draft asserted that ``E4M3 represents integers up to
$\pm 448$ exactly''; that statement is \emph{false} (independent
review): E4M3's four-bit significand represents \emph{consecutive}
integers only up to $\pm 16$---selected larger integers, 448
included, are representable only at increasingly coarse
spacing---and the $\approx 23$-bit slice sums $P_s$ are far outside
any exact E4M3 range. A correct construction must digitise first:
each $P_s$ decomposed into signed nibble digits (every digit and
every within-band power-of-two weight E4M3-exact), the contraction
banded so each FP32 accumulator stays below its $2^{24}$
exact-integer limit, and the band outputs stitched back into the
wide accumulator.

\paragraph{Why this variant is now qualitative only.}
The July-30 review identified two defects that together remove this
route from the quantitative comparison. \emph{First}, the
digitisation is not free and was uncosted: extracting, biasing, and
packing $\sim 120$ digit positions per output (at the full
$S{+}S_{\text{frac}}$ column count) is integer-SIMT work of
plausibly $\sim 200{+}$ instructions per output---more than the
deposit loop it was meant to replace---unless it can be folded into
the Phase-A constant structure, a restructuring for which no ledger
yet exists. \emph{Second}, the exactness bound was computed at
contraction length $84$ ($S\cdot T$) while the operation count used
$120$ ($(S{+}S_{\text{frac}})\cdot T$); at $120$ the band-span bound
still clears the FP32 limit but with only $\approx 6\%$ margin, and
the raw and fractional columns need distinct band layouts. Both
reviews therefore concur: the tensor-hosted deposit route is a
\emph{research direction}, not a quantified scenario, and every
wall-time or floor number previously attached to it (including the
former ``Scenario~B'' $17.5$--$24.3$~ms headline and its
$594\,\Bmem$-based $1.05\times$ tensor-margin claim) is withdrawn
from the headline accounting. The quantitative route of this paper
is the integer-hosted realisation of \S\ref{sec:classical-phasB};
the corrected per-class tensor floors are in \S\ref{sec:parity}.
What survives of the idea is its co-design shadow: a
\emph{hardware} wide-accumulate epilogue at the tensor-core output
achieves what this software variant attempted, exactly and with no
digitisation, and is rung S2 of the alternative ladder in \S\ref{sec:codesign} (and, in tensor-core form, part of rung H1 of \S\ref{sec:inttc}).

\subsection{Bandwidth-parity floors}
\label{sec:parity}

With every component of the design in hand, its hardware demands
reduce to a clean architectural rule. The design---and each of the
lineage variants it supersedes---activates some subset of the
machine's compute pipes, and---each
instruction class is now floored against \emph{its own} peak: the
FP8-TC pipe (Bailey MMAs), the FP16-TC pipe (Phase~A), the \fp{64}
vector pipe (naive Phase~B, if used), and the INT32-SIMT pipe (the
epilogue). Each carries its own
bandwidth-parity floor below which it becomes the binding bottleneck.

For memory-roof FFT parity at full \fp{64}, a GPU must satisfy
\emph{either}: (i) the native FP64 floor alone (no emulation needed);
\emph{or} (ii) the direct design's \emph{route-complete} condition,
which is \emph{additive in time across the tensor classes and
conjunctive with the integer floor}:
\begin{equation}
\boxed{\;\frac{W_{\text{MMA}}}{\Pfpeight}
      + \frac{W_{A}}{\Pfpsixteen} \;\le\; \frac{Q}{\Bmem}
      \quad\textbf{and}\quad
      \eta_{\text{INT32}} \;\ge\; \etaopt^{\text{epi}}.\;}
\label{eq:route-complete}
\end{equation}
\emph{This correction matters.} Prior formulations stated (ii) as the
conjunction of the per-class floors $\etaopt^{\text{FP8-TC}}$ and
$\etaopt^{\text{FP16-TC}}$ of \eqref{eq:parity-fp8}. That form is
\emph{unsound}: each per-class floor is defined by setting \emph{that
class alone} equal to the memory-roof time, so a part meeting both at
equality spends $2\times$ the roof in the tensor cores. The per-class
coefficients are individually necessary and jointly insufficient; only
\eqref{eq:route-complete} is the parity condition. We derive each
floor in turn; Table~\ref{tab:four-floors} evaluates them.

\paragraph{Floor 1: Native FFT parity (FP64 vector).}
Setting the native FFT compute-bound time $T_{\text{nat}} = W/\eta$
equal to the memory-roof time $T_{\text{mem}} = Q/\Bmem$ and solving
for $\eta$:
\begin{equation}
\boxed{\;\etaopt^{\mathrm{native}}
     = \mathrm{OI}_{\mathrm{FFT}}\cdot \Bmem
     \approx 1.56\,\Bmem.\;}\label{eq:parity-binding}
\end{equation}

\paragraph{Floor 2: Naive Ozaki-Bailey Phase~B parity
(FP64 vector, informational).}
For the \OBFFT\ kernel with the naive fp64 Phase~B reduction
($2Sn_{\text{out}}$ FLOPs, FMA $=2$, $n_{\text{out}}=12N^3$):
\begin{equation}
\etaopt^{\mathrm{naive\text{-}OBFFT}}
     = \frac{2\cdot 12\,S}{96}\,\Bmem = \frac{S}{4}\,\Bmem
     = 3.75\,\Bmem \quad (r=12,\ S = 15).
\label{eq:parity-naive-ozaki}
\end{equation}
This is strictly stronger than~\eqref{eq:parity-binding}: native FFT
demands less FP64-vector work per byte than \OBFFT\ Phase~B does. At
any $\eta_{\text{FP64}} \geq 1.56\,\Bmem$ the native FFT path is
already memory-bound, the emulated path is not required for FFT, and
its strictly higher Phase~B demand is moot.
Equation~\eqref{eq:parity-naive-ozaki} therefore identifies the
threshold at which the naive Ozaki path \emph{also} reaches the memory
roof---useful for sanity-checking architectures but not the binding
floor.

\paragraph{Floor 3: the integer-epilogue floor (INT32 SIMT pipe).}
For the full epilogue of \S\ref{sec:deconstruction}
(canonicalisation, combines, biased deposits, two-sided lift, pack,
deconstruction residual, dp4a bulk), with
$c_{\text{epi}}\in[203,281]$ instructions per output:
\begin{equation}
\boxed{\;\etaopt^{\text{epi}}
      = \frac{12\,c_{\text{epi}}}{96}\,\Bmem
      = \frac{c_{\text{epi}}}{8}\,\Bmem
      \approx 25.4\text{--}35.1\,\Bmem.\;}
\label{eq:parity-kulisch}
\end{equation}
The full derivation is in Appendix~\ref{app:kulisch-floor}; the
reconstruction-only share (deposits$+$lift$+$pack,
$c\in[111,135]$) is $14$--$17\,\Bmem$. The coefficient is an order
of magnitude above the native-FFT coefficient because the epilogue
does far more per-output work---but on a pipe $30\times$ faster than
B300's \fp{64} vector unit.

\paragraph{Floor 4: tensor floors, per instruction class.}
Per the July-30 review, tensor work is floored against the
instruction class that executes it (all in tensor ops, MAC $=2$):
\begin{itemize}[topsep=2pt, leftmargin=2em]
\item FP8-TC (Bailey MMAs): $W_{\text{MMA}} = 2\cdot 6\cdot
\alpha_{\mathbb{C}}\cdot 32N^3 \approx 4.6\times 10^{13}$ at
$\alpha_{\mathbb{C}}=111$ (Eq.~\ref{eq:wmma}) $\Rightarrow$
$\etaopt^{\text{FP8-TC}} = 444\,\Bmem$ ($3.55$~PF at 8~TB/s;
B300's 5~PF clears by $1.41\times$, Rubin's 17.5~PF by
$1.79\times$).
\item FP16-TC (Phase~A, incl.\ lift columns): $W_{\text{A}} \approx
9.90\times 10^{12}$ padded (Eq.~\ref{eq:phaseAcost}) $\Rightarrow$
$\etaopt^{\text{FP16-TC}} = 96\,\Bmem$ ($0.77$~PF at 8~TB/s;
B300's 2.5~PF clears this individual floor by $3.26\times$).
\item INT8-TC: \emph{not used}---at 167/250~TOPS the class cannot
host any first-order term (\S\ref{sec:tcgarner}).
\end{itemize}
\begin{equation}
\boxed{\;\etaopt^{\text{FP8-TC}} = 444\,\Bmem,\qquad
\etaopt^{\text{FP16-TC}} = 96\,\Bmem\quad (1024^3,\ r=12),\;}
\label{eq:parity-fp8}
\end{equation}
summing (as time, per \eqref{eq:route-complete}, since the classes
share the physical tensor cores) to $11.8$~ms on B300 \emph{if Phase~A
is charged at $100\%$ of peak}. It cannot be. Phase~A is an
$(n_{\text{out}}\times r)\cdot(r\times 21)$ contraction: its $k$ is
$r = 12$ against an \fp{16} MMA $k$-granularity of $16$, and its $n$
is $21$ against an $n$-granularity of $8$. Charging that padding
($16/12 \cdot 24/21 = 1.52\times$) raises Phase~A from $2.6$ to
$4.0$~ms and the tensor total to $\mathbf{13.1}$~ms---\emph{marginally
above} the $12.9$~ms roof. The honest statement is therefore not
``the tensor path fits with margin'' (as prior formulations claimed, on an unpadded $11.9$~ms) but \emph{the tensor path sits at
the roof with no margin}, and any aggregate tensor efficiency below
$\approx 98\%$ makes it binding---and, under the layout caveat of
\S\ref{sec:tme}, the conservative $\alpha_{\mathbb{C}}=138$ count
puts it at ${\approx}15.3$~ms, $1.19\times$ the roof, outright. Two further defects of the July-29
draft are also repaired here: its single ``$594\,\Bmem$ tensor
floor'' merged the classes, and it costed an INT8 MMA at the FP8
peak.

\paragraph{Which floor binds.}
The floors evaluated at the relevant HBM tiers, against current
GPU specifications, are in Table~\ref{tab:four-floors}. On B300 the
\emph{integer-epilogue floor} binds ($4.9$--$6.7\times$ short), with
the summed tensor demand \emph{at} the roof (no margin once tile
padding is charged); on H100/B200 native FP64 wins outright; on Rubin
native FP64 is essentially at parity (within 4\%), making emulation
unnecessary---fortunately, since Rubin's integer-epilogue requirement
($558$--$772$~T~inst/s at 22~TB/s) is far beyond any plausible SIMT
provisioning.

\begin{table}[h]
\centering
\caption{Bandwidth-parity floors for $1024^3$ FP64 FFT, evaluated at
the HBM tiers of current architectures, \emph{per instruction class}
(dated sources in \S\ref{sec:tme}). ``$\checkmark$'' satisfied with
margin; ``$\sim$'' near parity; ``$\times$'' missed. The integer
column is the epilogue floor
$\etaopt^{\text{epi}}=25$--$35\,\Bmem$
\eqref{eq:combined-floor}; tensor columns are FP8-TC $444\,\Bmem$
and FP16-TC $96\,\Bmem$ \eqref{eq:parity-fp8}. B300 \fp{64} is the
frozen 1.4-TF value (\cite{uchino2025complex} lists 1.2). Entries
marked $^{\rm d}$ are \emph{derived proxies}
($P_{\text{FP32}}/2$, \S\ref{sec:tme}), not vendor specifications;
no vendor publishes a SIMT INT32 issue rate. All
specifications are dense, per-GPU values.
}\label{tab:four-floors}
\footnotesize
\setlength{\tabcolsep}{3pt}
\resizebox{\textwidth}{!}{%
\begin{tabular}{lccccccccc}
\toprule
GPU & $\Bmem$ & F1 (FP64) & F3 (int epilogue) & F4a (FP8-TC) & F4b (FP16-TC)
    & FP64-vec & INT32-SIMT & Memory-roof path \\
    & (TB/s)  & $1.56\Bmem$ & $25$--$35\Bmem$ & $444\Bmem$ & $96\Bmem$
    & spec    & spec       & \\
\midrule
H100   & 3.35 & 5.2~TF  & $85$--$117$~T~inst/s  & 1.5~PF & 0.32~PF & 34~TF
       & $33.5$~T$^{\rm d}$  & native ($\checkmark$ F1) \\
B200   & 8    & 12.5~TF & 203--281~T~inst/s & 3.55~PF & 0.77~PF & 37~TF
       & $37.5$~T$^{\rm d}$  & native ($\checkmark$ F1) \\
B300   & 8    & 12.5~TF & 203--281~T~inst/s & 3.55~PF & 0.77~PF & \textbf{1.4~TF}
       & \textbf{$41.7$~T$^{\rm d}$} & none at roof; emulated walls $4.9$--$6.7\times$ \\
Rubin  & 22   & 34.3~TF & $558$--$772$~T~inst/s & 9.77~PF & 2.11~PF & 33~TF
       & $65$~T$^{\rm d}$ & native ($\sim$ F1, 4\%) \\
\bottomrule
\end{tabular}}
\end{table}

The reading:
\begin{itemize}[leftmargin=2em, topsep=2pt]
\item \emph{H100 and B200} satisfy Floor~1 comfortably---native FFT
is the right path, no emulation needed. (Their integer pipes would
also sit far below the epilogue band, but the
emulated path is moot when native FFT is already memory-bound.)
\item \emph{B300} fails Floor~1 by $\sim 9\times$ and Floor~2 by
$\sim 22\times$ (both FP64 vector). Its $41.7$~T~inst/s integer pipe misses
the epilogue floor by $4.9$--$6.7\times$; its summed tensor demand
sits at the roof ($13.1$~ms, no margin once tile padding is charged). \emph{No
evaluated route reaches the roof on B300}: the projection is
$63$--$87$~ms against the $12.9$~ms roof---\emph{at most}
$1.3$--$1.9\times$ the $116$~ms native path if attained, and
with no \fp{64} arithmetic anywhere.
\item \emph{Rubin} sits within 4\% of Floor~1 at 22~TB/s (native FFT
essentially at parity), which is fortunate: the emulated
integer-epilogue requirement at its bandwidth
($558$--$772$~T~inst/s) is far beyond any plausible SIMT provisioning.
\end{itemize}

The binding constraints in practice are therefore Floor~1 (for
native-capable GPUs) and, for FP64-collapsed GPUs, the
integer-epilogue floor, which B300 misses. The quiet asymmetry the
earlier draft leaned on---abundant FP8 silicon---is real but not
sufficient: the FP8-family tensor classes fit under the roof, yet
the scalar epilogue does not, and \emph{that} floor no software
reassignment moves. What removes it is hardware, and the epilogue
floor's closed form tells us exactly which hardware: the co-design
ladder of \S\ref{sec:inttc}.

\begin{figure}[t]
\centering
\includegraphics[width=0.85\textwidth]{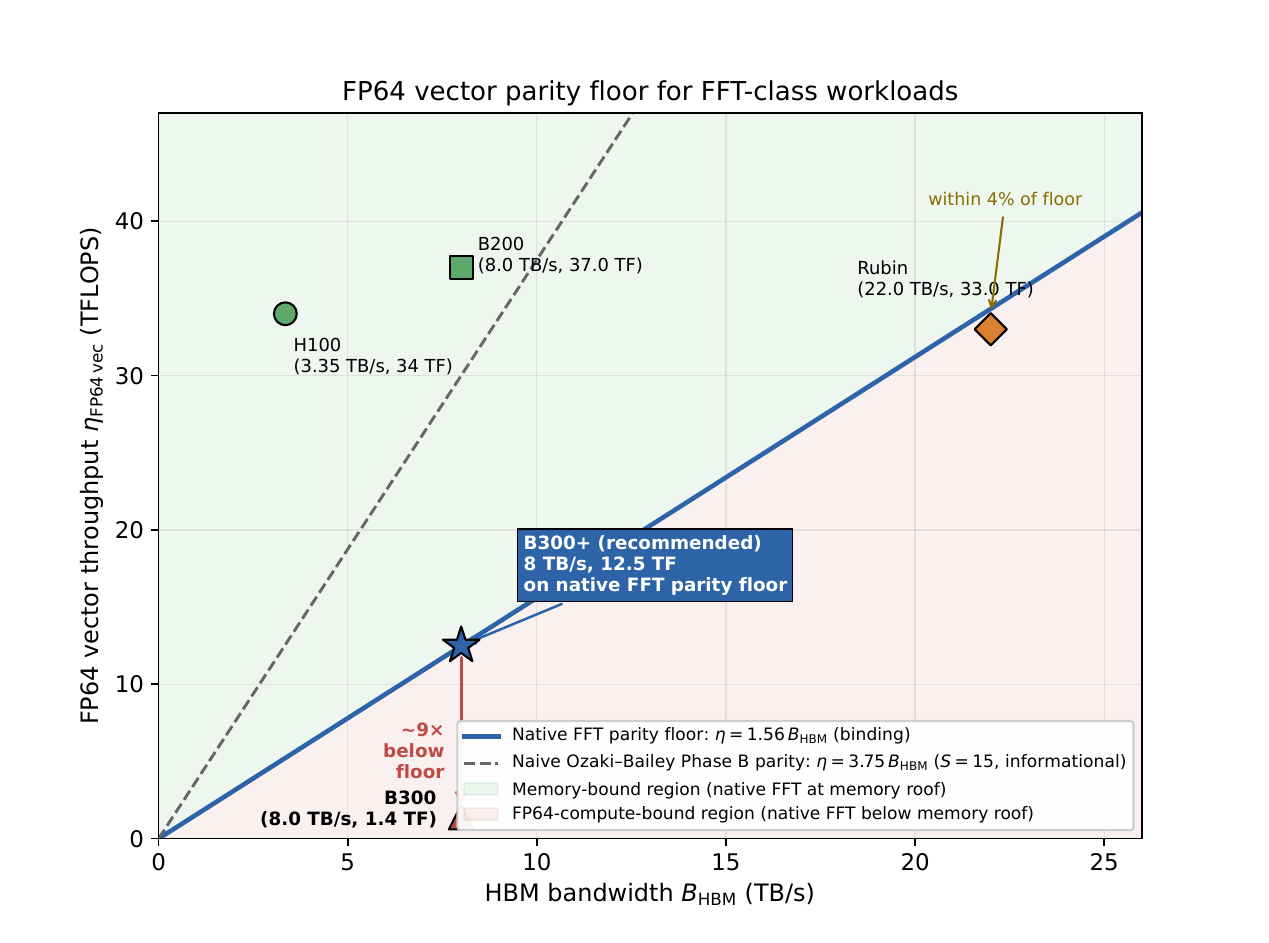}
\caption{FP64 vector parity floor for FFT-class workloads. The solid
line is the binding floor $\eta = 1.56\,\Bmem$ (native FFT parity); the
dashed line is the naive-\OBFFT\ Phase~B parity $\eta = 3.75\,\Bmem$
($S=15$, FMA $=2$), shown
for reference but strictly stronger than the binding floor and therefore
not operationally relevant. Current architectures: H100 (3.35~TB/s,
34~TF) and B200 (8~TB/s, 37~TF) above the floor; \textbf{B300
(8~TB/s, 1.4~TF) $\sim 9\times$ below}; Rubin (22~TB/s, 33~TF) 4\%
below the floor (essentially at parity). The blue star at
(8~TB/s, 12.5~TF) marks the minimum recommended \fp{64} vector spec for
a hypothetical B300+ at 8~TB/s HBM: this brings native FFT to memory-roof
parity without any reliance on emulation.}
\label{fig:parity}
\end{figure}

Figure~\ref{fig:parity} plots~\eqref{eq:parity-binding} and the
informational~\eqref{eq:parity-naive-ozaki} as lines on the
$(\Bmem,\eta_{\text{FP64}})$ plane, with current architectures marked.
The epilogue floor~\eqref{eq:parity-kulisch} and the per-class
tensor floors~\eqref{eq:parity-fp8} live on the
$(\Bmem,\eta_{\text{INT32}})$ and $(\Bmem,\eta_{\text{TC}})$ planes
respectively, and are captured by Table~\ref{tab:four-floors} rather
than by additional figures.

\subsection{Architectural implications}

Three implications follow:
\begin{enumerate}[leftmargin=2em, topsep=2pt]
\item \emph{B300 is the architectural outlier, with a partial
software escape route.} The 1.4~TFLOPS FP64 vector spec is
$\sim 9\times$ below the native-FFT parity floor of 12.5~TF at
8~TB/s; among naive reductions, no kernel engineering recovers the
memory roof at full \fp{64}. The Ozaki-Bailey-Kulisch design
eliminates \fp{64} entirely and cuts the projection from $116$~ms
to $63$--$87$~ms (at peak integer issue)---at most a
$1.3$--$1.9\times$ recovery---but the integer epilogue walls it
$4.9$--$6.7\times$ above the $12.9$~ms roof
(\S\ref{sec:deconstruction}). B300 need not be a dead end for
FFT-heavy codes, but on the evaluated routes it does not reach the
roof either; closing the remainder is the hardware co-design of
\S\ref{sec:inttc}, not a kernel-engineering item.
\item \emph{Rubin is at native FFT parity by design or by
coincidence.} The 33~TFLOPS FP64 vector spec at 22~TB/s sits 4\%
below the 34.3~TF native floor---essentially at parity. For
spectral codes, Rubin is balanced; \OII\ provides no further
speedup for FFT (and indeed $\etaopt^{\text{naive-\OBFFT}} =
3.75\times 22 = 82.5$~TF far exceeds
Rubin's 33~TF, while the exact route's integer-epilogue floor,
$558$--$772$~T~inst/s at 22~TB/s, is beyond plausible SIMT
provisioning---unless the co-design rungs of \S\ref{sec:inttc}
discharge it).
\item \emph{The FP64 floor for future GPUs is set by FFT, not by
HPL.} Designers tempted to cut FP64 vector silicon further on
post-Rubin generations should treat $\eta \geq 1.56\,\Bmem$ as a
hard floor---because the software escape route is no longer free:
after correct accounting it is integer-epilogue-bound at
$4.9$--$6.7\times$ the roof on B300-class parts. The exactness of
the fallback does not depend on INT32 specifically
(\S\ref{sec:noint32}: any integer width $\geq 8$~bits carries the
reduction \emph{correctly}), but its \emph{throughput} does depend
on integer provisioning. The safe procurement target is
$\eta \geq 1.56\,\Bmem$ \emph{plus}, if emulated FFT is to reach
the roof, either an integer-SIMT budget of $\gtrsim 25$--$35\,\Bmem$
or the epilogue-discharging hardware of \S\ref{sec:inttc}.
\end{enumerate}

\section{Lineage, Ablations, and Fallback Substrates}
\label{sec:noint32}

The direct design of \S\ref{sec:design} did not appear from nowhere: it
is the endpoint of an algorithmic lineage---recursive Garner, direct
CRT with naive \fp{64} summation, integer-hosted Kulisch---each stage
of which lands further from the memory roof for an identifiable reason. This section
retains that lineage, for two purposes. First, read as an
\emph{ablation study}, it is the evidence that the substrate
assignment is forced within the alternatives evaluated: each variant
below undoes one ingredient of the direct design, and each lands
further from the roof
(\S\ref{sec:walltime}--\ref{sec:precision}). Second, it supplies the
design's \emph{fallback}: the integer-hosted deposits generalise to
any integer
vector pipe of width $\geq 8$~bits,
exactly (\S\ref{sec:width-subfloor}--\ref{sec:noint32-perf}),
resting on classical reduced-radix arithmetic.

One candid observation frames the fallback discussion. B300 retains a
wide INT32 SIMT pipe not because anyone provisioned it for
double-precision science, but because INT32 remains useful for address
arithmetic, indexing, and integer-heavy AI preprocessing: its
throughput relative to FP64 is an \emph{accident} of the AI design
pivot, and the next turn of the same pivot could narrow it as readily
as it narrowed FP64. This is precisely why the direct design leans on
the \fp{8} tensor cores---the one substrate the AI ratchet reliably
grows---and treats the integer pipe as a fallback, sized only for the
deconstruction residue, rather than as a load-bearing component.

\subsection{The ablation ladder: wall time across the lineage}
\label{sec:walltime}

\begin{figure}[t]
\centering
\includegraphics[width=\textwidth]{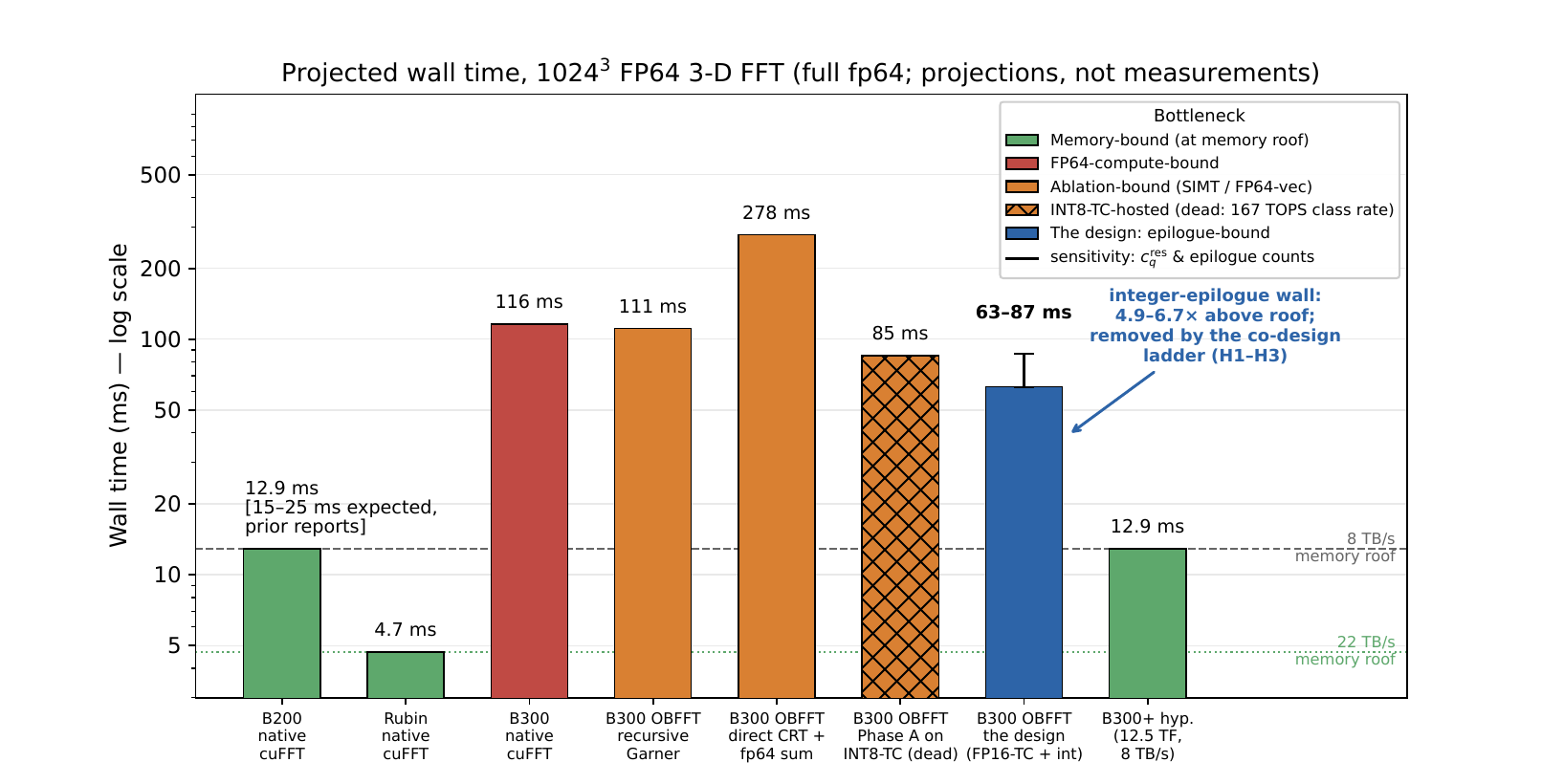}
\caption{The ablation ladder as service-maximum lower bounds on wall
time for $1024^3$
FP64 3D FFT, all at full \fp{64} precision. All bars are
Roofline-model projections from peak rates; no hardware measurement of the proposed kernel exists. B200 (37~TF, 8~TB/s) and Rubin (33~TF, 22~TB/s) are
memory-bound near the 12.9~ms and 4.7~ms roofs of their HBM tiers
(the bracketed annotation gives expected cuFFT performance from prior
reports). B300 native is compute-bound at $\sim 115$~ms (collapsed
\fp{64} vector pipe). The orange bars are the ablations:
recursive Garner at $\sim 111$~ms (Barrett double-count corrected
this revision); direct CRT $+$ naive fp64
sum on the collapsed \fp{64} vector pipe at
$\sim 278$~ms; INT8-tensor-hosted Phase~A at $\sim 85$~ms (dead:
167-TOPS INT8-TC). The blue bar is the design (integer-hosted
deposits, FP16-carrier Phase~A), walling on the integer epilogue at
$63$--$87$~ms at full \fp{64}---\emph{above the $12.9$~ms roof by
$4.9$--$6.7\times$; the whisker spans the
$c_q^{\text{res}}\in[3.0,6.3]$ and epilogue-count sensitivity
ranges}. The recommended co-design ladder of \S\ref{sec:inttc}
(Figure~\ref{fig:codesign}) takes it to $16.0$--$23.5$~ms with
minor hardware (up to H2) and to $12.9$--$15.0$~ms with H3---the
roof at the optimistic end of the epilogue interval, $1.17\times$ at
the conservative end.
The hypothetical
B300+ at 12.5~TF native FFT would reach the memory roof through the
native path.}
\label{fig:walltime}
\end{figure}

Figure~\ref{fig:walltime} lays the lineage out as wall times: the
theoretical decomposition for $1024^3$ FP64 3D FFT across every path,
all derived from the same Roofline framework with no measured numbers
mixed in, and \emph{all at full \fp{64} precision}. Each bar of the
ladder is the direct design of \S\ref{sec:design} with one ingredient
removed---read it as the ablation chart. The 8~TB/s memory roof of
$12.9$~ms is shown as a dashed reference; the 22~TB/s memory roof of
$4.7$~ms applies to the Rubin column. The B200 and Rubin theoretical
numbers reflect the bandwidth-bound regime (compute is comfortably above
the memory roof). The B300 native bar at $\sim 115$~ms is compute-bound
on the collapsed 1.4~TFLOPS \fp{64} vector pipe.

The reading is direct. Among the \emph{naive} reductions on B300, the
paths (116~ms native, 111~ms recursive Garner, 278~ms direct
CRT + fp64 sum, 85~ms INT8-TC Phase~A) all sit far above the 12.9~ms
memory roof. The B200,
Rubin, and hypothetical B300+ paths all hit their respective memory
roofs natively.
The design bar is the crux: with every removable cost migrated to
the FP8/FP16 tensor path (which is itself at the roof, $13.1$~ms),
the integer epilogue walls the transform at $63$--$87$~ms.
Each rung of the ladder differs from the design by one substrate
assignment---which is the sense in which the design is forced within
the alternatives evaluated---and Figure~\ref{fig:codesign} previews
how far the co-design ladder of \S\ref{sec:inttc} narrows it.
This is
the drill-down of \S\ref{sec:deconstruction} and
\S\ref{sec:classical-phasB}.

\paragraph{The integer rate is an optimistic add-equivalent
endpoint.} One further derating deserves its own name. The
$41.7$~T~inst/s figure is \emph{derived} ($P_{\text{FP32}}/2$) and is
at best an add-equivalent optimum: the CUDA programming guide's
instruction-throughput table lists lower per-SM result rates for
several classes the epilogue actually uses---integer multiply-add,
funnel shifts, extended-precision carry chains, and type
conversions---and Algorithm~\ref{alg:kulisch} and the pack lean on
the slower ones. Splitting $c_{\text{epi}}$ by class (add/carry
$\approx 45\%$ at full rate; mul/mad $\approx 30\%$ and shift/logic
$\approx 15\%$ at an assumed $0.5$--$1\times$; cvt/clz $\approx 10\%$
at $0.25$--$0.5\times$; \texttt{ledger.py}, pending a SASS-level mix)
gives a mix-efficiency parameter
$\varepsilon_{\text{mix}} \in [0.57, 1.0]$ multiplying the endpoint,
i.e.\ a mix-degraded projection of $63$--$152$~ms \emph{before} the
occupancy/issue derating above. For this reason every peak-issue
figure in this paper---$63$--$87$~ms included---should be read as an
\emph{optimistic service floor}, not an expected wall time; the class
split also states exactly which slow instruction classes each rung of
the co-design ladder removes.

\paragraph{From theoretical to measured.}
Real implementations achieve $\sim 50\text{--}80\%$ of the theoretical
roofs. Production cuFFT on B200 is expected to reach $\sim$15--25~ms for
$1024^3$ FP64 FFT against the 12.9~ms theoretical roof, based on prior
reports (the bracketed annotation in Figure~\ref{fig:walltime}). B300 native
FFT, being compute-bound, is expected to run closer to its 116~ms
theoretical roof (compute-bound workloads more readily saturate). The
\OBFFT\ ablation rungs on B300 will likely measure above their
service-maximum bounds due to kernel-launch overhead and the
$\beta\to 1$ register-fusion discipline that real implementations will
have to maintain. The design bound of $63$--$87$~ms assumes
\emph{peak} integer issue on the epilogue; at $50$--$70\%$ realised
integer efficiency it stretches to $\approx 90$--$174$~ms.
Empirical measurement is the natural next step.

\subsection{Trading precision for speed on B300}
\label{sec:precision}

\begin{figure}[t]
\centering
\includegraphics[width=\textwidth]{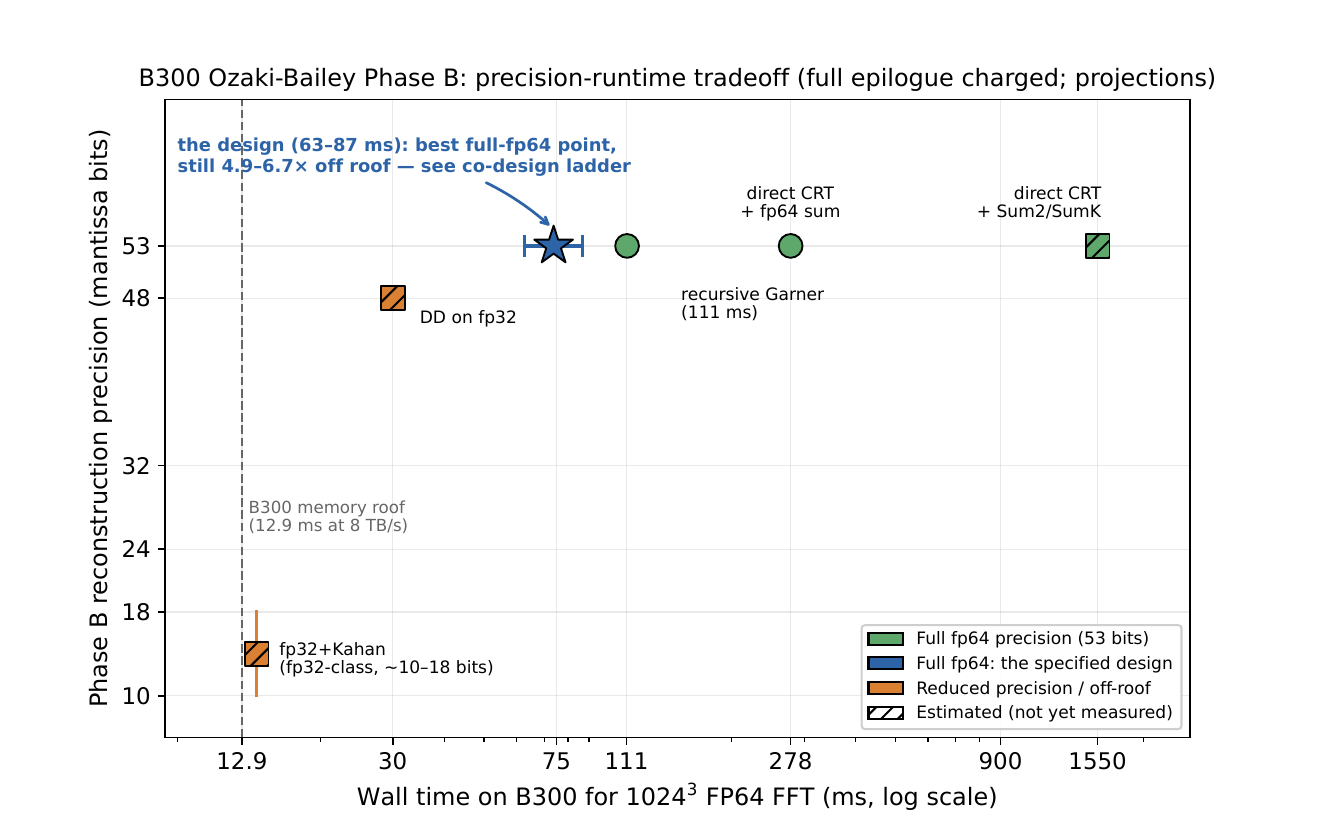}
\caption{B300 Ozaki-Bailey Phase~B precision-runtime tradeoff, with
the full integer epilogue charged. Full-\fp{64} (53-bit) schemes on
the collapsed
FP64 vector pipe sit far right: direct CRT + \fp{64} sum ($278$~ms),
recursive Garner ($111$~ms, Barrett double-count corrected), Sum2/SumK
($\sim 1550$~ms, estimated).
Reduced-precision options trade bits for speed: DD on fp32
($\sim 30$~ms, $\sim 48$~bits), fp32+Kahan (projected $\sim 14$~ms
near the roof but fp32-class, $\sim 10$--$18$-bit; the whisker marks
the measured-to-analytic range). The star is the full-\fp{64}
integer-hosted design at $63$--$87$~ms, the Pareto point among
full-precision candidates---\emph{still $4.9$--$6.7\times$ above the
$12.9$~ms roof} (\S\ref{sec:deconstruction}); the range bar spans
the $c_q^{\text{res}}$ and epilogue-count sensitivity intervals.
Hatched markers are
estimates. Detailed per-scheme analysis is in
Appendix~\ref{app:garner-detail}.}
\label{fig:precision}
\end{figure}

A separate question is what happens if \emph{reduced} precision is
acceptable. Phase~B can be carried out in \fp{32} with Kahan
compensation (mapping onto B300's $83.3$-TF FP32 vector pipe) or in
double-fp32 (DD) arithmetic~\cite{dekker1971} for $\sim 48$~bits of
precision. These reduce Phase~B time substantially but at the cost of
bits of precision. Figure~\ref{fig:precision} summarises the
precision-runtime tradeoff for the available Phase~B reduction
schemes. Full per-method numerics, prototype-measured reconstruction
errors, and a per-scheme cost analysis are in
Appendix~\ref{app:garner-detail}.

The qualitative picture: the classical schemes
(Kahan/Neumaier/Sum2/SumK compensation, DD on fp32, structured DGEMV,
ReproBLAS) all trade speed against precision in one direction or the
other---reaching memory-roof speed only by sacrificing bits, or
preserving \fp{64} precision only by remaining on the FP64-vector
pipe. The Kulisch fixed-point accumulator
of~\S\ref{sec:classical-phasB} is the one plausible escape: it
produces a full 53-bit \fp{64}-mantissa result with exact accumulation
(stronger worst-case reduction bound than naive \fp{64} sum---a single
final rounding) while running on the integer SIMT pipe. Figure
\ref{fig:precision} shows the full-\fp{64} Kulisch point at 53
bits: the integer-hosted design at $63$--$87$~ms---the Pareto point
among full-precision candidates, though still $4.9$--$6.7\times$
above the memory roof on B300.

Read as an ablation, this tradeoff chart is the second half of the
argument for the direct design: within the classical
exact-accumulation literature, Kulisch is the single scheme that
routes Phase~B around the collapsed FP64 vector pipe at full
precision (\S\ref{sec:classical-phasB}). No point in the plane is
simultaneously full-\fp{64} and at the roof on B300: the epilogue
wall of \S\ref{sec:deconstruction} keeps even the best
full-precision candidate $4.9$--$6.7\times$ above it, and closing
that gap is the hardware co-design of \S\ref{sec:inttc}, not a
scheme-selection question.

\subsection{The accumulator is substrate-agnostic: a width-parameterised
sub-floor}
\label{sec:width-subfloor}

The Kulisch accumulator is a fixed-point bin; nothing in its definition
requires the bin to be tiled by 32-bit words. The running sum of
\S\ref{sec:classical-phasB} occupies $\approx 130$~bits and is currently held in $6\times$INT32 words. It can equally be held in $w$-bit words for
any $w$, provided (i)~enough words are used to span the $130$-bit range
and (ii)~carry propagation between words is handled. This is the
classical \emph{reduced-radix} (redundant) representation long used in
multiprecision and modular arithmetic on SIMD units: a wide integer is
stored as an array of words each carrying fewer than $w$ ``real'' bits,
so carries accumulate in the spare guard bits and explicit
carry-propagation is deferred to the end~\cite{gueron2021avx512}. The
Kulisch accumulator in a high-radix carry-save form has been realised on
GPUs before, for reproducible
summation~\cite{collange2015repro,uguen2017kulisch}; what is new here is
the deliberate use of word width as a \emph{lane-count lever} on a pipe
whose width is at a premium.

The lever is the following identity. A vector datapath exposing lanes
of $w$-bit elements multiplies its lane count by $32/w$ as $w$
narrows, but the $130$-bit
accumulator then needs more words, and each $\le 23$-bit slice payload
$P_s$ spans more words on deposit. Writing $g$ for the guard bits
reserved per word and $\rho = w-g$ for the usable radix, the
word-level operation count per output scales as
\begin{equation}
\;n_{\text{word}}(w)
 \approx S\,\lceil b_P/\rho\rceil + \lceil W_{\text{acc}}/\rho\rceil,
\label{eq:kulisch-width}
\end{equation}
where $b_P \le 23$ is the slice payload width, $W_{\text{acc}}\approx
130$ the accumulator width, and $S=15$ the slice count (deposit $+$
final carry-resolve). \emph{Scope note (this revision).} An earlier
draft converted \eqref{eq:kulisch-width} into per-width throughput
ratios via a fixed-bit-width vector-lane model ($V=512$); independent
review correctly observed that NVIDIA SIMT issue is not a fixed-width
vector datapath, that narrow scalar types do not automatically yield
independent carry-chain lanes, and that the stated ratios did not
derive cleanly from \eqref{eq:kulisch-width}. We therefore retain
\eqref{eq:kulisch-width} only as an \emph{instruction-count} scaling
and defer all cross-width \emph{performance} claims to the
ISA-specific ledger and measurements of \S\ref{sec:future}.

\subsection{INT16 and INT8: narrowing the integer word}
\label{sec:int16-int8}

Two strategies present themselves, and they differ sharply.

\paragraph{Strategy A: keep the moduli, split slices across narrow words.}
Retain the existing $9$--$11$-bit Rev-34 moduli and therefore the
existing $S=15$, $\le 23$-bit slices, and simply deposit each slice
across the $\lceil b_P/\rho\rceil$ narrow words it now spans. The lane
count rises by $32/w$ while the per-slice deposit work rises by
$\lceil b_P/\rho\rceil$; the two partially cancel.

\paragraph{Strategy B: co-design the moduli to fit one narrow word.}
Shrink the moduli so each slice payload fits a single $w$-bit word's
radix. This restores one-word-per-slice deposits, but smaller moduli
require \emph{more} of them to cover the same $\log_2 M = 117.88$
bits of CRT range ($r$ grows from $12$ toward $r_{\textsc{i}}\geq 15$
at byte moduli, per Part~1), which both lengthens the slice loop and
inflates the Phase~A tensor cost. Strategy~B is therefore
counterproductive: it
trades cheap SIMT relief for expensive tensor work and lengthens Phase~B
as well. We reject it and adopt Strategy~A.

\begin{table}[h]
\centering
\caption{Width-parameterised Kulisch Phase~B under Strategy~A (Rev-34
moduli, split deposit): word-level instruction counts per output from
\eqref{eq:kulisch-width}. \emph{Per-width throughput and sub-floor
ratios are deliberately not stated} (this revision): they require an
ISA-specific issue model that no public NVIDIA documentation
supplies, and are deferred to measurement (\S\ref{sec:future}). All
three substrates reach full \fp{64} accuracy---the accumulation is
exact in every case.}\label{tab:substrate}
\small
\setlength{\tabcolsep}{5pt}
\begin{tabular}{lcccc}
\toprule
Substrate & word $w$ & guard/radix $\rho$ & words for $130$b
 & exact? \\
\midrule
INT32 (baseline) & 32 & $\sim$28 & 5 ($6$ as implemented$^{\rm a}$) & yes \\
INT16            & 16 & $\sim$12 & 11 & yes \\
INT8             & 8  & $\sim$6  & 22 & yes \\
\bottomrule
\end{tabular}\\[2pt]
{\scriptsize $^{\rm a}$~$\lceil 130/28\rceil = 5$ words suffice for the range; the executable form of Algorithm~\ref{alg:kulisch} uses six radix-28 words so that the $\ge 160$-bit logical accumulator keeps $\approx 27$--$30$ bits of margin above the exact $131$-bit signed magnitude.}
\end{table}

The table's one load-bearing column is the last: \emph{all rows are
exact}. Narrowing the word changes only
how the $130$-bit accumulator is tiled and how often carries resolve; it
never rounds before the final \fp{64} readout. The full-\fp{64}
guarantee of \S\ref{sec:classical-phasB} is preserved verbatim. That
correctness statement---any integer width $\geq 8$~bits carries the
exact reduction---is what this section establishes; which width is
\emph{fastest} on a given part is an open measurement question.

\subsection{Performance: where each substrate lands}
\label{sec:noint32-perf}

On the $1024^3$ B300 footing of \S\ref{sec:walltime} and
Figure~\ref{fig:walltime}, the INT32-hosted Phase-B component
(deposits $+$ lift $+$ conversion) is $\approx 34$--$42$~ms at peak (Eq.~\ref{eq:kulisch-time}); the full Scenario-A epilogue is
$\approx 63$--$87$~ms. For INT16 and INT8 hostings we state only the
qualitative expectation from \eqref{eq:kulisch-width}---the word-count
growth and lane-count growth partially cancel, so narrow widths are
plausibly within a small factor of INT32---and defer the numbers to
measurement; the prior specific $1.2\times$/$1.1\times$
ratios rested on a vector-lane model inapplicable to NVIDIA SIMT and
are withdrawn, as are the tall-skinny tensor hosting's numbers
(\S\ref{sec:fp8-tallskinny}: digitisation uncosted).
Figure~\ref{fig:walltime}'s integer-hosted bar should be
read as representative of the whole integer-substrate family, not of
INT32 alone; the binding question for every member is the epilogue
floor of \eqref{eq:combined-floor}, not the word width.

\subsection{The two-tier architectural recommendation}
\label{sec:noint32-recommendation}

The design's components sit at stated levels of confidence, and the
distinction matters for how the title should be read. The
\emph{specified} construction (integer-hosted deposits, FP16-carrier
Phase~A) is complete at the algebraic level and unit-tested in
exact arithmetic (\texttt{verify.py}); it is not implemented on
hardware, so ``specified,'' not ``demonstrated.'' The tensor-hosted
deposit variant is qualitative only (\S\ref{sec:fp8-tallskinny}).

The integer-width fallback of this section is
\emph{assertive about correctness}: it rests on classical
reduced-radix carry-save
arithmetic~\cite{gueron2021avx512,collange2015repro,uguen2017kulisch},
established and widely deployed, and it changes only how an exact
$130$-bit accumulator is tiled across lanes. \emph{If the fallback is
invoked, the correctness requirement is
not for INT32 specifically, but for any integer-vector pipe of width
$\geq 8$~bits}---sized, per \S\ref{sec:deconstruction}, at the
epilogue floor ($25$--$35\,\Bmem$). That sizing is what B300
itself cannot meet, which is
why the software fallback alone is a target for
differently-provisioned parts, not
B300's own route to the roof---the co-design rungs of
\S\ref{sec:inttc} are.

The honest formulation of the thesis is therefore graded, with the
grading now anchored on the design rather than discovered at the end.
Native FP64 is \emph{not} required---established on every branch, by
construction, and the collapsing INT8 tensor path is routed around
too. The memory roof itself is \emph{not} reached on B300 by
any evaluated software route: the integer epilogue walls the design
at $4.9$--$6.7\times$ the roof.
What remains falsifiable-and-plausible is the useful weaker claim:
full-\fp{64} FFT at no more than $1.3$--$1.9\times$ the
native-B300 rate, on
FP8-generation tensor cores plus the existing integer pipe, with the
roof itself recoverable through the co-design ladder of
\S\ref{sec:inttc}. \fp{8}-generation tensor cores as the sole
\emph{floating-point}
primitive for double-precision FFT is established at the level of
construction; \fp{8}-plus-nothing-else is not, and the earlier
draft's contrary phrasing is withdrawn.

\subsection{Beyond FFT: the generalised Kulisch rescue}
\label{sec:kulisch-general}

The substrate argument above concerns where Phase~B \emph{runs}; a
separate generalisation concerns which \emph{kernels} it serves. The
Kulisch Phase~B is not specific to FFT: any kernel whose binding cost
is a many-term \fp{64} reduction over operands of bounded dynamic range
admits the same treatment, and the same substrate freedom. The
sub-floor generalises to
$\etaopt^{\text{Kulisch}} = c\cdot \mathrm{OI}_{\text{red}}\cdot\Bmem$
(derived in Appendix~\ref{app:beneficiary-analysis}), recovering the epilogue floor of~\eqref{eq:parity-kulisch}. The rescue helps
when a per-output \fp{64} reduction on the vector pipe is the binding
bottleneck and its Amdahl fraction $f$ is large: the strong cases
beyond \OBFFT\ are Ozaki-II SpMV ($f\!\sim\!0.8$, $2.5$--$4\times$) and a
ReproBLAS replacement ($f\!\sim\!0.7$, $2$--$3\times$), while dense
DGEMM ($f\!\approx\!0.08$) is correctly predicted not to benefit. Every
beneficiary inherits the substrate freedom of
\S\ref{sec:int16-int8}--\ref{sec:fp8-tallskinny}: the same INT16/INT8
width relaxation (exact at any width $\geq 8$~bits) and the same
tensor-hosted deposit hypothesis apply, as does the same
integer-epilogue charge, so the conclusions of this paper transfer to
the whole low-OI reduction-bound class. The full beneficiary
analysis, Amdahl treatment, and \texttt{libKulisch} library notes are
in Appendix~\ref{app:beneficiary-analysis}.

\section{Parallel FP32 Analysis}
\label{sec:fp32}

Many spectral scientific codes operate at FP32 precision, either
because the underlying physics tolerates it (turbulence
simulations using fp32 for the bulk and fp64 for sensitive
reductions) or because mixed-precision techniques have made
fp32-dominant pipelines viable~\cite{haidar2018harnessing}. We
apply the parity analysis to FP32 spectral codes.

\subsection{FP32 parity formula}

The FFT operation count is precision-independent
(it scales with $N^3 \log N$); the memory traffic is not: for FP32
each complex element is 8~B rather than 16~B, so the traffic
\emph{halves} and the operational intensity \emph{doubles}:
\begin{equation}
\mathrm{OI}_{\text{FP32}}
 = \frac{15N^3\log_2 N}{48N^3} = 3.125\ \text{flops/B},
\qquad
\etaopt^{\text{fp32, native}} = 3.125\,\Bmem,
\end{equation}
\emph{twice} the FP64-native $1.56\,\Bmem$. (An earlier draft
asserted the OI was ``the same as in fp64''---halving the
denominator while fixing the numerator cannot leave the ratio
unchanged; a factor-of-two error identified by independent review,
whose correction halves every FP32 headroom figure below.) The
architectural data is shown in Figure~\ref{fig:fp32}.

\begin{figure}[t]
\centering
\includegraphics[width=0.98\textwidth]{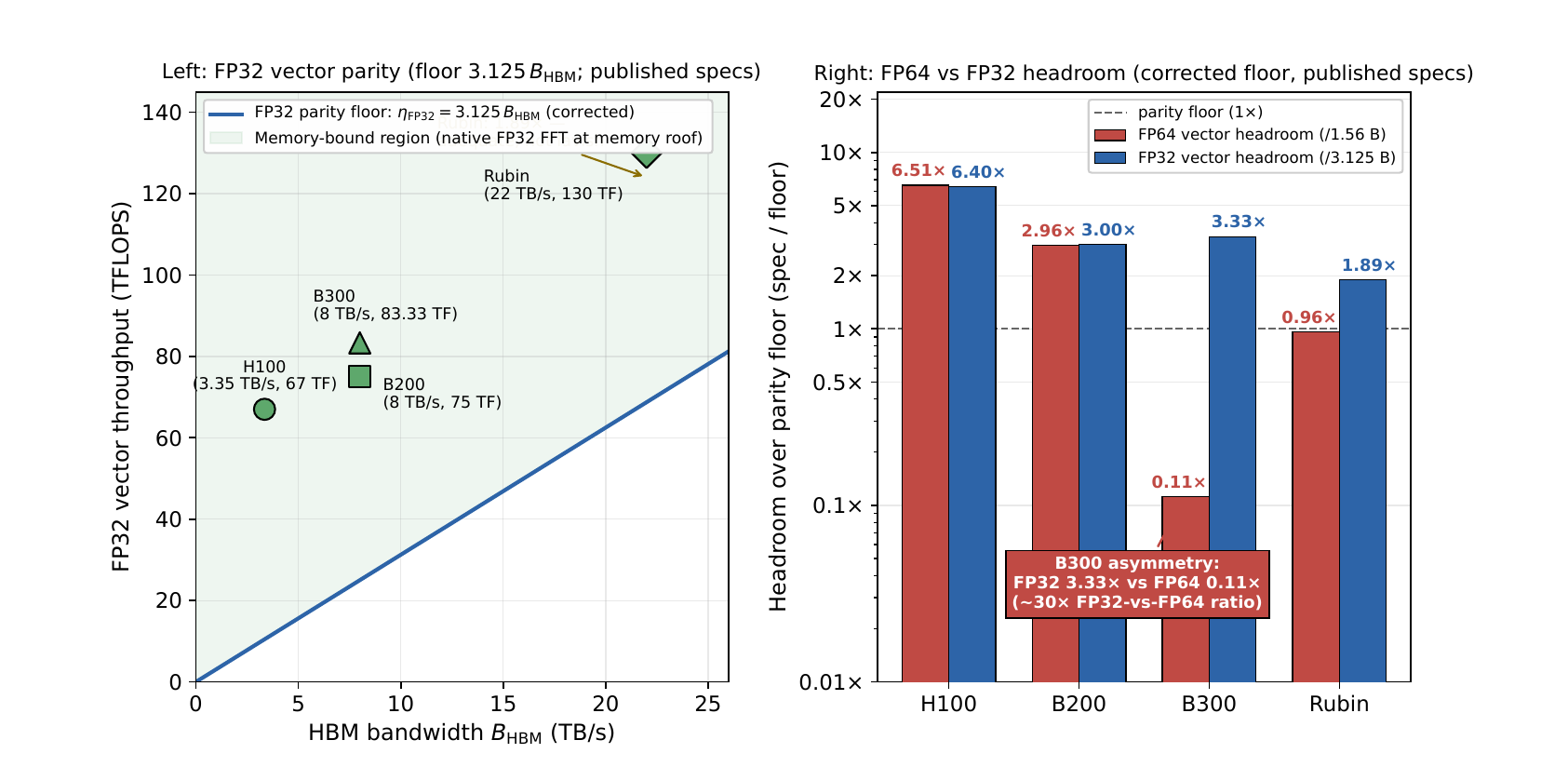}
\caption{FP32 parallel analysis (corrected: floor $=3.125\,\Bmem$).
Left: FP32 vector parity for FFT
workloads, with current architectures plotted. Right: headroom
ratio (spec / FFT parity) for FP64 (red) vs.\ FP32 (blue) on
each architecture. B300's FP64 headroom is $0.11\times$
(critically under-provisioned); its FP32 headroom is $3.33\times$.
Rubin's FP32 headroom, at its published 130~TF~\cite{nvidia2026rubin},
is $1.89\times$---adequate, not the $\sim 6\times$ comfort an
earlier draft claimed.}
\label{fig:fp32}
\end{figure}

The right panel makes the contrast vivid but graded: B300's FP32
vector is $3.33\times$ above the parity floor, while its FP64 vector
is $\sim 9\times$ below---and Rubin's FP32 headroom is
$1.89\times$ (130~TF against a 68.75-TF floor at 22~TB/s). The
procurement implication is two-sided: codes expressible in fp32 (or
fp32 + iterative refinement) are adequately served by
B300, but FP32 FFT is \emph{not} a free resource on
bandwidth-rich parts---Rubin-class bandwidth consumes half of
the FP32 provisioning---while codes requiring native fp64 throughout
need either the Ozaki-Bailey-Kulisch design of
\S\ref{sec:design} (with its integer-epilogue wall) or a future GPU
that restores the FP64 floor.

\subsection{Ozaki-II for FP32}

An \OII\ variant targeting fp32 precision requires roughly
$r\approx 6$ moduli (the capacity budget \eqref{eq:capacity} at
$k=24$ is $56$ bits; six moduli of the Rev-34 class give
$\log_2 M = 62.6$). The Phase~A contraction cost falls accordingly
($S = 8$ byte slices at $r=6$). The Phase~B reduction can be carried out in fp32 itself,
since the precision target is fp32 to begin with, so it runs on
the FP32 vector pipe at full throughput.

With the modular-combine dataflow ($n_{\text{out}}=12N^3$), FMA
counted as two FLOPs, and the halved FP32 traffic ($48N^3$ bytes),
the Ozaki-II-FP32 Phase-B parity formula becomes
\begin{equation}
\etaopt^{\text{Ozaki-II, fp32}}
 = \frac{2\cdot 12\,S}{48}\,\Bmem = \frac{S}{2}\,\Bmem
 = 4\,\Bmem
\quad (r=6,\ S=8),
\end{equation}
on the fp32 substrate with $\log_2 M = 62.6$ (a prior $2.25\,\Bmem$ mixed the $18N^3$ output count
with one-op-per-FMA counting, and a later one used the stale $S=6$
for $3\,\Bmem$), giving 32~TF at 8~TB/s and 88~TF at
22~TB/s---the latter within Rubin's published 130~TF but with only
$1.5\times$ margin. Combined
with the native floor above: \emph{on B300-class bandwidth, native
FP32 FFT is memory-bound with $\sim 3\times$ margin and Ozaki-II
is unnecessary; on Rubin-class bandwidth the margins thin to
$\approx 2\times$ for both native and emulated FP32 FFT.}

\subsection{The FP32 floor for future architectures}

The FP32 floor at 8~TB/s is 25~TF (native) / 32~TF (Ozaki, $S=8$). At
22~TB/s: 68.75~TF / 88~TF (Ozaki margins $\approx 2.6$ on B300 and
$1.5$ on Rubin). Current architectures sit at 67--130~TF
FP32 vector: B300-class parts have $\sim 3\times$ headroom, and
Rubin's published 130~TF leaves $1.89\times$. The
earlier draft's conclusion that ``an architect could plausibly trade
FP32 vector area for additional tensor area while keeping spectral
codes memory-bound'' is therefore withdrawn for high-bandwidth
parts: on post-Rubin bandwidth tiers, cutting FP32 vector silicon
would drop FP32 FFT below the memory roof exactly as B300's FP64
cut did for FP64 FFT. The natural
target for any post-Rubin GPU intended to serve mixed FP64/FP32
spectral codes is $\eta_{\text{fp32}} \geq 3.125\,\Bmem$ and
$\eta_{\text{fp64-vec}} \geq 1.56\,\Bmem$, with the route-complete
emulation floors of \S\ref{sec:parity} (FP8-TC $444\,\Bmem$,
FP16-TC $96\,\Bmem$, and
integer epilogue $25$--$35\,\Bmem$---or the co-design rungs of
\S\ref{sec:inttc}) as the fallback in case the FP64
vector is further reduced.

\section{Discussion and Implications}

\subsection{Updated TME picture}

Part 1~\cite{matsuoka2026tme} introduces $\gamma$ as a parameter
of the TME model from the outset (its emulated-execution
equation carries the term $\gamma\,n_{\text{out}}$ alongside
the $\alpha$ and $\beta$ contributions). It then argues---in its
\S5.1, and applied throughout its kernel analyses---that the
per-FMA Garner overhead $O(r^2/k)$ vanishes whenever the
inner-product reduction length $k$ satisfies $k \gg r^2 \sim 100$,
which Part 1 takes to hold for four of its five primitives:
dense GEMM ($k=K\gg 100$), batched GEMV ($k=N\gg 100$ for the
matrix dimension), structured stencils, and SpMV. Under that
amortisation, the kernel's wall time is bounded by
$\max(\alpha W/\Plow,\,\beta Q/\Bmem)$ and $\gamma$ drops out.

Two things change between the original snapshot of this paper and the
present analysis. First, the \emph{regime}: for Bailey-FFT-class kernels
the Bailey factor gives $k = q \approx \sqrt N$, which for $N=1024$ is
$k=32$---well below $r^2 = 144$ at the recommended $r=12$, so $\gamma$
no longer amortises and binds. Second---and this \emph{is} a change to
the model, not merely the regime---Part~1 (from Revision~32) and the Ozaki 2.5
companion \emph{extended} the TME model from three overhead parameters
to four, adding the input-side deconstruction cost $c_q$ (dual to
$\gamma$) after the NVIDIA technical review~\cite{bayraktar2026tme}; the
FFT is the primitive where both conversion terms bind at once. We drop
the earlier claim that ``the model is not extended''---it is, and the
extension is exactly what makes the integer pipe a first-order
constraint here. The updated picture (Table~\ref{tab:gamma-regimes}) is
therefore both a regime atlas and a record of the extended
four-parameter model:

\begin{table}[h]
\centering
\small
\caption{Regime atlas of the four-parameter TME model
$(\alpha,\beta,\gamma,c_q)$. Which term binds depends on the kernel's
inner-product length $k$ relative to $r^2\sim 100$ (for $\gamma$) and on
its operational intensity relative to $\mathrm{OI}^{*}$ (for $c_q$); the
FFT is the case where both conversion terms
bind.}\label{tab:gamma-regimes}
\begin{tabular}{p{0.22\textwidth}p{0.40\textwidth}p{0.28\textwidth}}
\toprule
Regime & Binding constraint & Where it applies\\
\midrule
$\alpha$-roof (compute) &
\fp{8} tensor throughput at $\alpha = 3r+1$ &
Dense GEMM in compute-bound regime ($k\gg r^2$)\\
$\beta$-roof (memory) &
HBM bandwidth $\Bmem/\beta$ &
Memory-bound kernels, $\beta=1$ (stencil, SpMV, batched GEMV; $\gamma$ amortised when $k\gg r^2$)\\
$c_q$-regime (conversion) &
integer issue at $c_q r/P_{\text{int}}$ per input; binds below $\mathrm{OI}^{*}$ &
Low-OI kernels with heavy input re-splitting\\
\textbf{$\gamma$-roof (reconstruction), with the $c_q$ tax} &
\textbf{Reconstruction $+$ the integer epilogue, on one of:}
FP64-vector (naive, binding on B300 at $278$~ms);
\emph{or} INT8-TC (dead at 167--250~TOPS: $\sim 85$~ms);
\emph{or} integer-SIMT at the epilogue floor
$25$--$35\,\Bmem$ (width $\geq 8$~bits) ---
which B300 \emph{misses} $4.9$--$6.7\times$; discharged only by
the co-design rungs of \S\ref{sec:inttc} &
\textbf{Bailey-FFT-class kernels with $k\!\sim\!\sqrt N\!\ll\!r^2$}\\
\bottomrule
\end{tabular}
\end{table}

The $\gamma$-roof is the third roofline of the TME model, made operative
by the small Bailey factor $k$, and the $c_q$ deconstruction tax rides
on the same integer pipe as the reconstruction epilogue. On B300 the
naive Phase~B binds on the collapsed FP64 vector pipe; the Kulisch
reformulation relocates the reduction onto the integer vector pipe
(\S\ref{sec:noint32}, INT32 today but any width $\geq 8$~bits,
exact)---but the epilogue
(deconstruction residual, canonicalisation, combines, deposits,
lift, conversion) survives every software
reassignment and exceeds B300's $5.2\,\Bmem$ by $4.9$--$6.7\times$.
The $\gamma$-roof thus admits a \emph{partial} software escape: the
FP64-vector constraint is eliminated entirely, the wall time drops
from $115$--$278$~ms to $63$--$87$~ms, and the rest is a
hardware question---which, because the floor is closed-form, is a
\emph{specific} hardware question. \S\ref{sec:inttc} answers it.

\subsection{What software alone achieves, and where silicon
becomes necessary}
\label{sec:swhw}

Before proposing hardware it is worth stating precisely what does
\emph{not} need any. Every construction in this paper---the twiddle
fusion of \eqref{eq:twiddle-fusion}, the residue-domain Karatsuba
combines, the \fp{16}-carrier Phase~A of
\S\ref{sec:classical-phasB}, the biased carry-save deposits and
two-sided centered lift of Algorithm~\ref{alg:kulisch}, the
\texttt{dp4a} migration of the deconstruction bulk, and the
Rev-34 modulus set with its reduction-friendly members---is ordinary CUDA on shipping
parts. Nothing before rung H1 of \S\ref{sec:inttc} asks for a
transistor that does not exist. Table~\ref{tab:swhw} states where
that gets us, per part, and it is the table a reader deciding
whether to \emph{implement} this should read first.

\begin{table}[h]
\centering
\caption{The software/hardware boundary. ``Software alone'' is
the shipping construction on today's silicon (the modelled H0
optimisation is quoted separately); ``up to H2'' is the minor-hardware package of \S\ref{sec:inttc}
(INT8-TC restoration $+$ split wide accumulation $+$ two small
datapath/ISA asks); ``full ladder'' adds the moderate H3. Ranges span
the epilogue-count sensitivity intervals; realised-issue figures apply
the $50$--$70\%$ SIMT issue efficiency of \S\ref{sec:walltime}.
Generated by \texttt{ledger.py}.}
\label{tab:swhw}
\small
\setlength{\tabcolsep}{4.5pt}
\begin{tabular}{lcc}
\toprule
 & \textbf{B300} & \textbf{Rubin/R200} \\
\midrule
memory roof & $12.9$~ms & $4.7$~ms \\
native \fp{64} & $116$~ms ($9.0\times$ its roof) & $4.9$~ms
   (\textbf{$1.04\times$: at parity}) \\
\midrule
\textbf{software alone}, peak issue & $\mathbf{63}$--$\mathbf{87}$~ms
   & $40$--$56$~ms \\
\quad vs.\ native \fp{64} & $\mathbf{1.3}$--$\mathbf{1.9\times}$
   \emph{faster} & $\mathbf{0.09}$--$\mathbf{0.12\times}$:
   \emph{$8$--$11\times$ slower} \\
\quad vs.\ memory roof & $4.9$--$6.7\times$ short & $8.6$--$11.9\times$
   short \\
\quad $+$H0 (modelled optimisation) & $58$--$78$~ms & $37$--$50$~ms \\
\quad at $50$--$70\%$ realised issue & $90$--$174$~ms
   ($0.7$--$1.3\times$ native) & $57$--$111$~ms \\
\midrule
\textbf{up to H2 (conditional on Eq.~\ref{eq:h1-group})} & $16.0$--$23.5$~ms & $10.3$--$15.0$~ms \\
\quad vs.\ memory roof & $\mathbf{1.2}$--$1.8\times$ &
   $2.2$--$3.2\times$ \\
\textbf{full ladder (${}+{}$H3)} & $\mathbf{12.9}$--$\mathbf{15.0}$~ms (roof--$1.17\times$) &
   $6.7$--$9.6$~ms \\
\quad vs.\ native \fp{64} & $7.7$--$9.0\times$ faster & $0.5$--$0.7\times$:
   \emph{still slower} \\
\midrule
\textbf{is emulation worth running?} & \textbf{yes, modestly, in
   software} & \textbf{no---use \fp{64}} \\
\bottomrule
\end{tabular}
\end{table}

Three consequences follow, and they are the practical content of
this paper.

\paragraph{On B300, software alone is a real but modest win, and it
is the only win available today.} The $1.3$--$1.9\times$ over native
at peak issue is worth having on a part whose \fp{64} pipe is
$9\times$ short of its own memory roof, and it requires nothing but
the kernel described here. But the honest range includes the
realised-issue band, where the same kernel lands at
$0.7$--$1.3\times$---a wash. An implementer should expect somewhere
between ``modestly better than native'' and ``no better than
native'', and should measure the integer-issue efficiency of the
epilogue first, because that single number decides which end of the
range applies. What software cannot do on B300, by any substrate
reassignment we evaluated, is reach the $12.9$~ms roof: the gap is
$4.9$--$6.7\times$ and every term of it is per-output scalar integer
work.

\paragraph{On Rubin/R200, do not emulate.} Rubin's $33$~TF \fp{64}
sits within $4\%$ of its own $4.7$~ms roof (Table~\ref{tab:four-floors}),
so the native path is already essentially optimal for this
primitive. The emulated route costs $40$--$56$~ms against it---an
$8$--$11\times$ \emph{loss}---because Rubin scaled bandwidth
($22$~TB/s) and tensor throughput far harder than it scaled SIMT
integer issue, which is what the epilogue actually consumes. Nor
does the hardware ladder rescue emulation on Rubin: the full-ladder
endpoint, $6.7$--$9.6$~ms (\S\ref{sec:rubin22}), is \emph{still}
slower than simply calling \fp{64}. This is a clean negative result and we state it as such:
\emph{nothing in this paper should be run on a part whose native
\fp{64} reaches its own memory roof}.

\paragraph{The B200 cross-check: the design with the \fp{64} pipe
switched off.} One shipping part permits a complete dress rehearsal.
B200 kept everything B300 cut---$4.5$~POPS of INT8-TC alongside
$4.5$~PF FP8-TC and $37$~TF \fp{64}---so the design can be run on it
\emph{as shipped, with the \fp{64} pipe deliberately unused}: Bailey
on FP8-TC ($10.2$~ms), Phase~A on the \emph{native} INT8 tensor core
(exactly---the two-limb column sums are $\le 416{,}160 < 2^{31}$, so
IMMA's INT32 accumulation is bit-exact; $3.2$~ms unpadded, $4.4$~ms
IMMA-padded), tensor total $13.3$--$14.6$~ms, and the epilogue on the
$37.5$-T-inst/s derived scalar pipe: $c_{\text{epi}}\in[203,281]
\Rightarrow$ \textbf{$70$--$96$~ms at peak issue}
($5.4$--$7.5\times$ the shared $12.9$~ms roof; $100$--$193$~ms at
$50$--$70\%$ realised issue; $1.7$--$2.3$~TF effective
\fp{64}-equivalent throughput). Three facts make this configuration
worth stating. \emph{First}, it is the purest expression of the
paper's thesis that exists on shipping silicon: full-\fp{64} output,
zero \fp{64} arithmetic, Phase~A on the integer tensor core the
co-design ladder must otherwise ask to have restored. \emph{Second},
it is quotable across parts: \emph{a B200 running this route with
its \fp{64} pipe idle is $1.2$--$1.7\times$ faster than a B300
running the native \fp{64} FFT it kept} ($70$--$96$ vs.\ $116$~ms).
Against B200's own memory-bound native path ($12.9$~ms) it loses
$5.4$--$7.5\times$, so it is a rehearsal, not a route. \emph{Third},
there is no FP32 side door: hosting Phase~B in double-double on the
$75$-TF FP32 pipe would cost $\approx 33$~ms of FP32 work at only
$\sim 48$-bit precision, and on Blackwell's unified cores that work
\emph{adds} to the surviving integer work rather than overlapping
it---the SIMT issue slots are the same. B200 is therefore the
reference platform on which every sensitivity interval this paper
carries---$c_{\text{epi}}$, $\varepsilon_{\text{mix}}$,
$c_q^{\text{res}}$, the IMMA tile efficiency, the FP16-vs-INT8
carrier delta---can be measured before any B300 silicon argument is
had. All values from \texttt{ledger.py} (\texttt{b200\_no\_fp64()}).

\paragraph{The construction's relevance is therefore conditional,
and the condition is a hardware trend, not a hardware ask.} What
makes the FFT emulation interesting is not B300 or Rubin
individually but the possibility of a part that inherits
Rubin-class bandwidth while inheriting B300's \fp{64} collapse. On
such a part the native path would be $\sim 9\times$ off a $4.7$~ms
roof and emulation would be the only route---and it is exactly there
that the software/hardware boundary bites, because software alone
would still leave an $8$--$11\times$ gap. That hypothetical part, not
B300 and not Rubin, is the audience for
\S\ref{sec:inttc} and \S\ref{sec:rubin22}.

\subsection{The recommended co-design ladder: restore the integer
tensor core, then two small asks}
\label{sec:inttc}

This is the paper's primary hardware proposal, and its organising
principle is \emph{minimality}: every rung is either a restoration of
something B200 already shipped, a modest extension of a structure
that already exists, or software. The bespoke SIMT-side alternative
(\S\ref{sec:codesign}) is retained for comparison, but it is
dominated by the route developed here.

The starting observation: B200 shipped 4500~TOPS of INT8-TC; B300 cut
it to 167 and Rubin to 250. Undoing that cut is a far smaller
institutional ask than any new epilogue structure. The analysis
separates into two capabilities that buy different things, and they
become rung H1 of the ladder below.

\paragraph{Ask (I): an integer tensor core with ordinary INT32
accumulation. This fixes the \emph{tensor} wall---and the hardware
already shipped.} Phase A needs $1.42\times 10^{13}$ INT8 tensor ops
in the two-limb form. To fit inside the Phase-A budget---the memory
roof less the \fp{8} Bailey time, $12.9 - 9.2 = 3.7$~ms---an INT8-TC
must run at
\begin{equation}
P_{\text{INT8-TC}} \;\geq\; \frac{W_A^{\text{int8}}}{Q/\Bmem - W_{\text{MMA}}/\Pfpeight}
   \;=\; 3.8\ \text{POPS}
   \qquad(\text{INT16-TC equivalent: } 2.7\ \text{POPS}),
\label{eq:inttc-breakeven}
\end{equation}
i.e.\ $22.8\times$ B300's 167~TOPS. \emph{But B200's 4.5-POPS INT8-TC
already exceeds it by $1.18\times$.} The hardware that removes this
paper's tensor overshoot is not speculative: it is one generation
old, and was deleted by the same AI-ratchet decision that deleted
\fp{64}. On Rubin the bar is higher ($6.9$~POPS, because its $4.7$~ms
roof leaves only $2.1$~ms after a $2.6$~ms Bailey pass) and B200's
part would \emph{not} suffice at $0.65\times$.

This is worth having---it retires the tile-padding overshoot of
\S\ref{sec:parity} and puts the tensor path back under the roof with
$0.6$~ms to spare---but it does not touch the binding term.
\emph{Restoring INT8-TC to B200 levels changes the projected wall
time by under $2\%$}, because the epilogue is unchanged: it is
positional carry work, and an INT32 accumulator digitises exactly the
sum an integer tensor core would otherwise carry. That is the same
defect that demoted the software tall-skinny variant of
\S\ref{sec:fp8-tallskinny}, reappearing in hardware.

\paragraph{Ask (II): give that integer tensor core wide accumulation.
This is what shifts the break-even.} With a $\geq 128$-bit
fixed-point accumulator the contraction itself carries the positional
sum, so the tensor core absorbs the deposits ($66$--$70$ inst/output),
the linear residue-domain combines ($30$/output-equivalent), and the
$r$-term inner product of the lift ($\approx 10$)---
$106$--$110$ of $c_{\text{epi}}$'s $203$--$281$. This is rung~H2
restated as a \emph{tensor-core capability} rather than a bespoke
output stage, and stated that way it is a considerably more
plausible product decision. What it leaves behind is
$c_{\text{epi}}^{\text{res}} = 97$--$171$: canonicalisation, the
scalar half of the lift, the pack, and the deconstruction
residual---all elementwise or carry-serial.

\begin{figure}[t]
\centering
\includegraphics[width=0.92\linewidth]{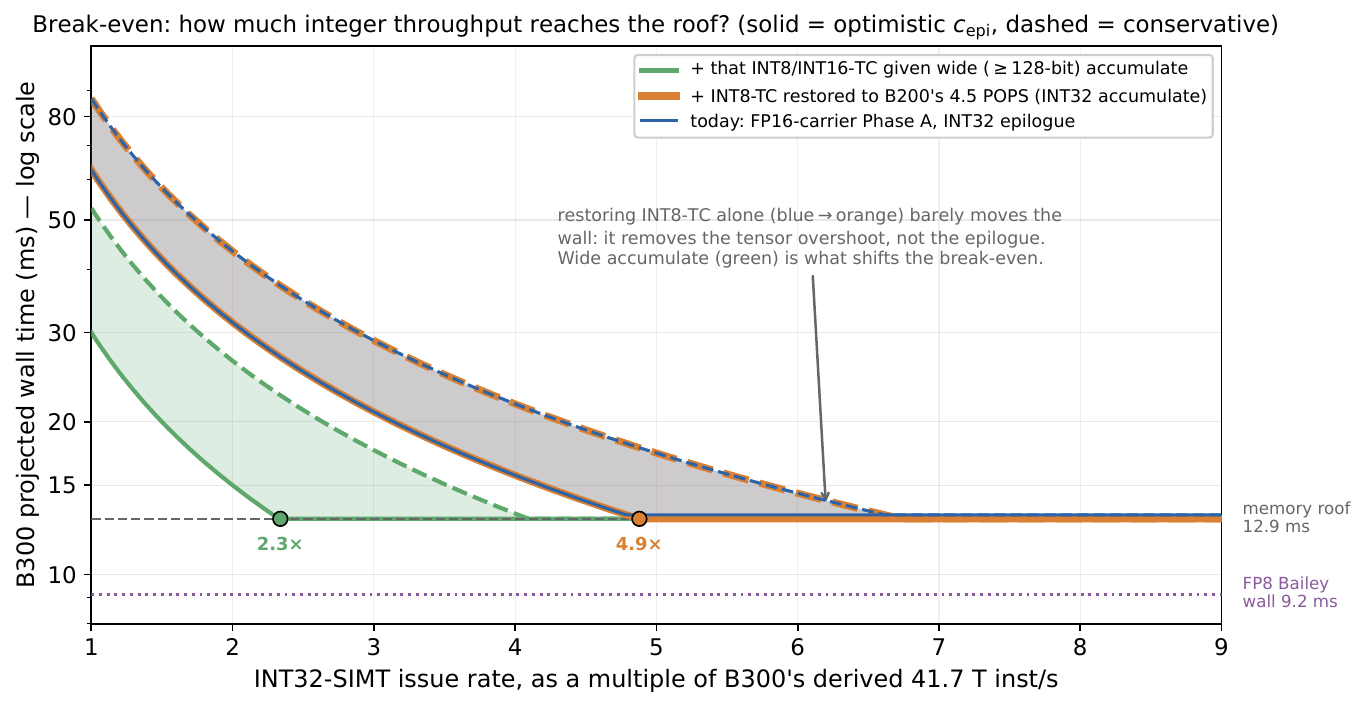}
\caption{\textbf{Break-even sensitivity to integer throughput
(B300).} Projected wall time against INT32-SIMT issue rate, in
multiples of B300's derived $41.7$~T~inst/s. Solid lines are the
optimistic $c_{\text{epi}}$ endpoint, dashed the conservative one;
bands span the interval. Restoring INT8-TC to B200's 4.5~POPS
(orange) all but coincides with today (blue)---it removes the tensor
overshoot, so the plateau drops from $13.1$ to $12.9$~ms, but the
epilogue is untouched. Giving that integer tensor core wide
accumulation (green) moves the break-even from $4.9$--$6.7\times$ to
$2.3$--$4.1\times$. Neither reaches the \fp{8} Bailey wall at any
plausible multiple.}
\label{fig:inttc}
\end{figure}

\paragraph{Break-even.} Table~\ref{tab:inttc} and
Figure~\ref{fig:inttc} give the sensitivity: the INT32-SIMT issue
rate still required, as a multiple of B300's derived
$41.7$~T~inst/s, to bring the transform to the memory roof and to the
stricter \fp{8}/Bailey wall.

\begin{table}[h]
\centering
\caption{Break-even integer throughput. Column~1 is the INT32-SIMT
multiple needed with today's tensor cores; column~2 with a
wide-accumulate integer tensor core absorbing
$106$--$110$ inst/output. Ranges span the $c_{\text{epi}}$
sensitivity interval. Generated by \texttt{ledger.py}.}
\label{tab:inttc}
\small
\setlength{\tabcolsep}{5pt}
\begin{tabular}{llcc}
\toprule
part & target & INT32-SIMT multiple & \emph{with} wide-accum.\ int-TC \\
\midrule
\multirow{2}{*}{B300}
 & memory roof ($12.9$~ms) & $4.9$--$6.7\times$ & $\mathbf{2.3}$--$\mathbf{4.1\times}$ \\
 & \fp{8}/Bailey wall ($9.2$~ms) & $6.9$--$9.5\times$ & $3.3$--$5.8\times$ \\
\midrule
\multirow{2}{*}{Rubin}
 & memory roof ($4.7$~ms) & $8.6$--$11.9\times$ & $4.1$--$7.2\times$ \\
 & \fp{8}/Bailey wall ($2.6$~ms) & $15.4$--$21.3\times$ & $7.4$--$12.9\times$ \\
\bottomrule
\end{tabular}
\end{table}

\paragraph{How wide must ``wide'' actually be? Not 128 bits---and not
one register.} Two different accumulators are involved and only one
of them is wide. The tensor core already forms the \emph{column} sum
$\sum_k v_k u_k[s]$, bounded by $r\cdot 2\max_k m_k\cdot 255 =
1.25\times 10^7$, i.e.\ \textbf{25 bits}: INT32 (equivalently FP32)
accumulation is \emph{already exact}, which is precisely why the
\fp{16} carrier of \S\ref{sec:classical-phasB} works on shipping
parts. What is wide is the \emph{positional} sum
$E = \sum_s c_s 256^s$ over the $S=15$ byte columns, and
\begin{equation}
\log_2 |E|_{\max} \;\le\; \log_2\!\big(r\cdot 2\max_k m_k\cdot M\big)
    = 133.5\ \text{bits (loose envelope)},
\label{eq:accwidth}
\end{equation}
while the \emph{exact} bounds from the actual idempotents are
$\sum_k (m_k{-}1)u_k = 4360M - 1$ ($130.0$ bits, canonical) and
$\sum_k (2m_k{-}1)u_k = 8724M - 1$ ($131.0$ bits, signed
$|v_k| < 2m_k$)---both machine-verified in \texttt{verify.py} (T9).
The exact signed \emph{magnitude} is therefore 131 bits; a
conventional two's-complement register representing both $+B$ and
$-B$ needs \textbf{132 total bits} (a signed $w$-bit register tops
out at $2^{w-1}{-}1$), so \textbf{INT128 is short by four total
bits}---three magnitude bits plus the sign. (A
magnitude-plus-separate-sign design needs 131${}+{}$1.) A detail
worth stating, because 128 is the width a reader would assume. The group-width table below deliberately uses
the safe per-group envelope $25 + 8(g{-}1)$; the three grades
(exact, safe, loose) are kept distinct.

But $E$ never needs to exist as one register. The positional sum
splits exactly at byte boundaries: a group of $g$ consecutive columns
is bounded by $25 + 8(g{-}1)$ bits, and $G=\lceil S/g\rceil$ such
groups are reassembled by $\approx 3(G{-}1)$ scalar shift/add-with-carry
instructions. That gives a clean width-versus-instruction trade
(Table~\ref{tab:accwidth}), and it answers the question directly:
\emph{yes, much smaller}. Table~\ref{tab:accwidth} gives the
\emph{conceptual one-limb} trade (full-width residues, $S=15$
columns)---three INT64 accumulators, or two INT88s, in place of one
INT144. The \emph{implemented} H1 layout is the two-limb \bint{8}
form of Eq.~\eqref{eq:h1-group}, whose $16{+}7$ shifted columns give
$4{+}2 = 6$ INT64 groups of $52$-bit width at $12$ assembly
instructions per output (last row). Either way, a handful of INT64
accumulators are an unremarkable addition to a datapath that already
carries INT32 accumulators; a 144-bit register is not.

\begin{table}[h]
\centering
\caption{Accumulator width versus scalar assembly cost. The upper
rows are the \emph{conceptual one-limb} trade ($S=15$ columns,
full-width residues---not an \bint{8} operand and not what H1
times); the bold final row is the \emph{implemented} two-limb H1
layout of Eq.~\ref{eq:h1-group}. The $g{=}15$ row is the monolithic
case a prior formulation's ``$160$--$192$-bit epilogue'' assumed.
Generated by \texttt{ledger.py}.}
\label{tab:accwidth}
\small
\begin{tabular}{ccccc}
\toprule
$g$ (cols/group) & groups $G$ & width $25{+}8(g{-}1)$ & natural register &
   assembly (inst/output) \\
\midrule
15 & 1 & 137 bits & INT144 & 0 \\
 8 & 2 &  81 bits & INT88  & 3 \\
 5 & 3 &  57 bits & INT64  & 6 \\
 4 & 4 &  49 bits & INT56  & 9 \\
 3 & 5 &  41 bits & INT48  & 12 \\
\midrule
\multicolumn{5}{l}{\emph{implemented H1 (two-limb \bint{8},
Eq.~\ref{eq:h1-group}): $16{+}7$ cols}} \\
\textbf{5} & \textbf{4${}+{}$2} & \textbf{52 bits} &
   \textbf{6$\times$INT64} & \textbf{12} \\
\bottomrule
\end{tabular}
\end{table}

\begin{figure}[t]
\centering
\includegraphics[width=0.88\linewidth]{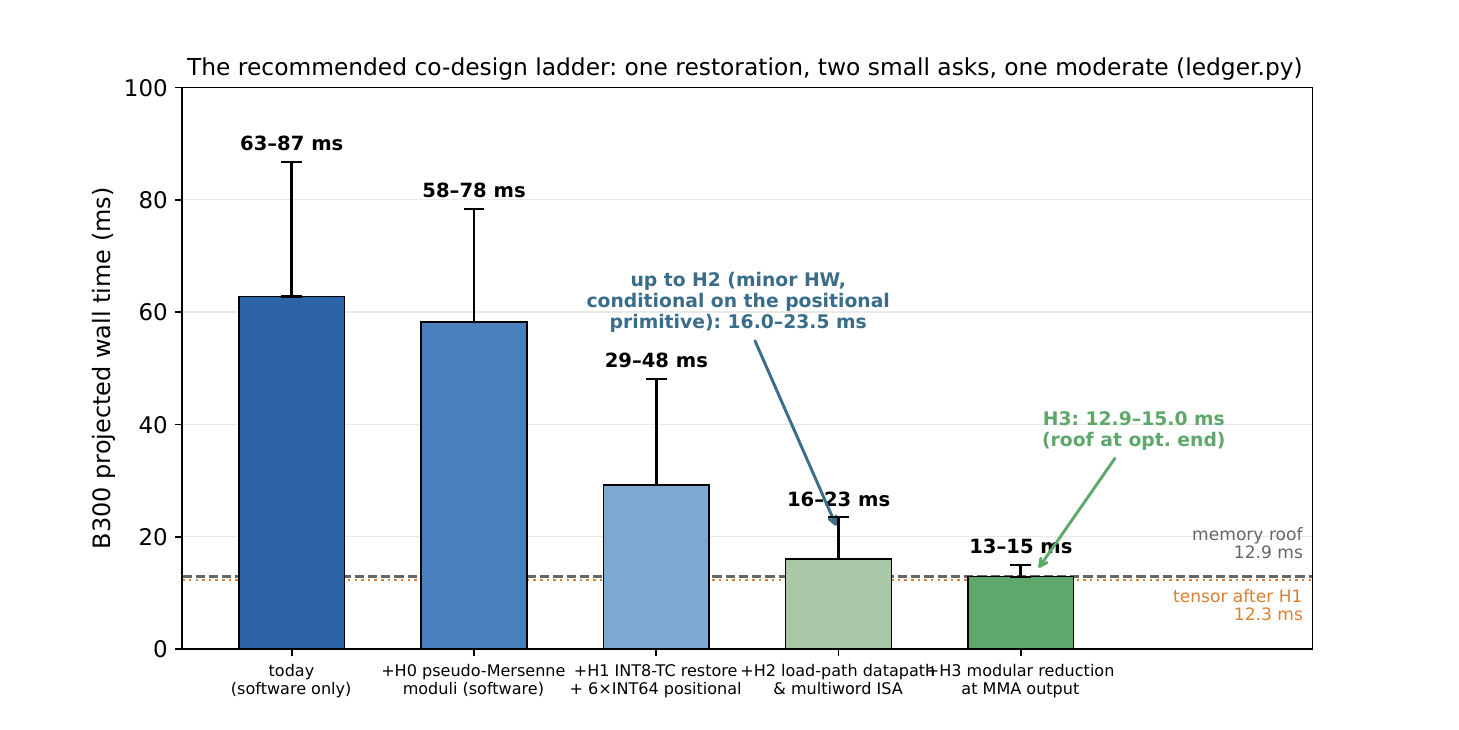}
\caption{\textbf{The recommended co-design ladder
(Table~\ref{tab:inttc-route}, drawn), B300.} Whiskers span the
$c_{\text{epi}}$ sensitivity interval. H0 is software; H1 restores
the INT8 tensor core B200 shipped and adds $6\times$INT64 positional
accumulation; H2 adds the load-path deconstruction datapath and the
multiword-carry/wide-\texttt{cvt} ISA idioms. \emph{Up to H2---the
minor-hardware package---the wall is $16.0$--$23.5$~ms against the
$12.9$~ms roof (dashed), conditional on the position-weighted
accumulation primitive of Eq.~\ref{eq:h1-group}.} The moderate H3
(modular reduction at the MMA output) lands $12.9$--$15.0$~ms: the
roof at the optimistic end of the epilogue interval, $1.17\times$ at
the conservative end. The dotted line is the post-H1 tensor time
($12.3$~ms).}
\label{fig:codesign}
\end{figure}

\paragraph{The ladder.} Table~\ref{tab:inttc-route} (drawn in
Figure~\ref{fig:codesign}) assembles the above into the recommended
ladder. Rung H1 is asks (I)$+$(II) together: the restored INT8-TC at
B200's rate hosting Phase~A, with split wide accumulation
($6\times$INT64: $4$ raw $+$ $2$ fractional groups, $g{=}5$) absorbing
the deposits, the linear combines, and the lift's inner product. \emph{The accumulation H1 needs is not
an ordinary MMA accumulator made wider.} A conventional MMA forms one
independent accumulator per output column, $P_{o,s} = \sum_k
v_{o,k}u_{k,s}$; widening it preserves more bits of each $P_{o,s}$
but never forms $E_o = \sum_s P_{o,s}256^s$, because the $256^s$
weights cross output columns and cannot ride in an \bint{8} operand.
What H1 requires is a \emph{position-weighted cross-column reduction}
in the MMA epilogue, and it must be derived from the workload the
ledger actually times: the \emph{two-limb} \bint{8} formulation. The
residue is split $v_k = v_{0,k} + 256\,v_{1,k}$ with
$v_{0,k}\in[-128,128)$ and $v_{1,k} = (v_k - v_{0,k})/256 \in [0,8]$
(canonical $v_k \le 2038$)---\emph{both} \bint{8}-representable,
unlike the full-width residue itself. Under the shifted-column
identity the high limb multiplies the adjacent slice, giving
$S{+}1 = 16$ raw and $S_{\text{frac}}{+}1 = 7$ fractional columns;
for raw group $j \in \{0,\dots,3\}$ of five columns,
\begin{equation}
G_{o,j} \;=\; \sum_{s=5j}^{\min(5j+4,\,S)} 2^{8(s-5j)}
   \sum_k \big(v_{0,o,k}\,u_{k,s} + v_{1,o,k}\,u_{k,s-1}\big),
\qquad
E_o \;=\; \sum_j G_{o,j}\,2^{40j},
\label{eq:h1-group}
\end{equation}
with out-of-range $u_{k,t} = 0$, and the two fractional groups formed
identically over the seven $w_k$ columns. Column sums are bounded by
$r\cdot 255\cdot(128{+}8) = 416{,}160$ ($20$ signed bits), so a
five-column group needs $52$ bits: the $G_{o,j}$ land in
\emph{six} INT64 registers ($4$ raw $+$ $2$ fractional), assembled by
$3(4{-}1) + 3(2{-}1) = 12$ scalar instructions per output. The
required rate is $6\,n_{\text{out}}$ group-reductions per transform,
i.e.\ $\approx 6\times 10^{12}$/s at the roof. (An earlier version
of this equation accumulated the one-limb $15$-column form with a
full-width $v$---which is not an \bint{8} operand and does not match
the timed workload; the repair adds one group and six assembly
instructions, and its effect is carried through every table below.) \emph{This is a new epilogue primitive, and the ladder's
silicon column says so}: restoring B200-rate INT8-TC is a
restoration; Eq.~\eqref{eq:h1-group} is new hardware, and every rung
below H1 is conditional on it. With H1, the tensor term falls to
$12.3$~ms (under the roof, overshoot retired; ${\approx}14.5$~ms under
the conservative layout count of \S\ref{sec:tme}, in which case the
H3 terminus floors there) and $c_{\text{epi}}$
falls from $203$--$281$ to $94$--$155$ (H0 included): $63$--$87
\rightarrow 29$--$48$~ms. Rung H2 adds the two \emph{small} asks: the load-path
deconstruction datapath (Ozaki 2.5's converter, removing the
streamed half of the deconstruction residual) and the
multiword-carry and wide-\texttt{cvt} ISA idioms (collapsing the
scalar lift and pack to $\approx 8$ instructions). Rung H3 is the one
\emph{moderate} ask: modular reduction to $[0,m_k)$ in the MMA
output path, so canonicalisation never reaches SIMT.

\begin{table}[h]
\centering
\caption{The recommended co-design ladder, cumulative, on B300
(8~TB/s; INT32 at the $41.7$-T~inst/s \emph{optimistic add-equivalent}
endpoint; tensor time after H1 is $12.3$~ms). H1's accumulation is the
\emph{position-weighted} cross-column primitive of
Eq.~\eqref{eq:h1-group}, derived from the two-limb \bint{8} layout
the ledger times ($6\times$INT64 groups, $12$ assembly
inst/output)---a new epilogue operation, not a wider ordinary MMA
register---so every row below H1 is \emph{conditional on that
primitive}. \emph{Up to H2 the package lands $16.0$--$23.5$~ms
against the $12.9$~ms roof} ($1.2$--$1.8\times$); H3 reaches the roof
at the optimistic end of the epilogue interval and $1.17\times$ at
the conservative end. The H0 factor is
derived per modulus (7 of 12 admit fold-and-add; interval
$[0.71,0.81]$, central $0.76$). Generated by \texttt{ledger.py}.}
\label{tab:inttc-route}
\small
\setlength{\tabcolsep}{5pt}
\resizebox{\textwidth}{!}{%
\begin{tabular}{llccl}
\toprule
rung & silicon & $c_{\text{epi}}$ & projected wall (ms) & vs.\ roof \\
\midrule
today (\S\ref{sec:swhw}) & --- & 203--281 & 62.8--86.8 & $4.9$--$6.7\times$ \\
$+$H0 pseudo-Mersenne moduli (7/12 friendly) & \textbf{none} & 188--254 & 58.3--78.4 &
   $4.5$--$6.1\times$ \\
$+$H1 INT8-TC restore $+$ positional $6{\times}$INT64 & restore${}+{}$\emph{new primitive}
   & 94--155 & \textbf{29.2--48.1} & $2.3$--$3.7\times$ \\
$+$H2 load-path datapath \& multiword ISA & small & 52--76 &
   \textbf{16.0--23.5} & $\mathbf{1.2}$--$1.8\times$ \\
$+$H3 modular reduction at MMA output & moderate & 34--49 &
   \textbf{12.9--15.0} & \textbf{at roof (opt.)}--$1.17\times$ \\
\bottomrule
\end{tabular}}
\end{table}

\paragraph{Reading the ladder.} Three points. \emph{First, the
package up to H2 substantially narrows the wall}: $16.0$--$23.5$~ms
against $12.9$, i.e.\ $1.2$--$1.8\times$, conditional on the
positional primitive of Eq.~\eqref{eq:h1-group}. H2's own asks are
small---a converter on a load path that already converts, and
multiword-carry ISA idioms that exist in every CPU ISA---and H0 is
software; the one piece of genuinely new datapath in the package is
Eq.~\eqref{eq:h1-group} itself. \emph{Second}, H3 (modular reduction
at the MMA output) closes most of what remains: $12.9$--$15.0$~ms,
i.e.\ at the roof at the optimistic end of the epilogue interval and
$1.17\times$ above it at the conservative end; the surviving excess
is the deconstruction-residual estimate plus the $12$-instruction
group assembly, both of which an implementation would settle before
any silicon is committed.
\emph{Third}, the ladder needs \emph{no widening of the scalar INT32
pipe}, where the pure-SIMT alternative of \S\ref{sec:codesign}
needed $4.9$--$6.7\times$; the integer tensor core does that work.
All rows are stated at the $41.7$-T~inst/s optimistic add-equivalent
endpoint; the instruction-mix derating of \S\ref{sec:walltime}
applies to every row equally.

\paragraph{And at FP32 precision?} B300 does not need this: native
\fp{32} ($83.3$~TF) clears the $25$-TF \fp{32}-vector floor with
$3.33\times$ headroom (\S\ref{sec:procurement}), so \fp{32} spectral
codes are already memory-bound and emulation buys nothing. The
variant is worth costing only for the case the trend of this paper
implies---a part that collapses \fp{32} as B300 collapsed
\fp{64}---and it is where the narrow accumulator pays. At $k=24$ the
capacity budget \eqref{eq:capacity} falls to $56$ bits, so $r=6$
moduli ($\log_2 M = 62.6$) suffice: $\alpha_{\mathbb C}$ drops
$111\rightarrow 57$, $S$ drops $15\rightarrow 8$, and the monolithic
accumulator drops to $77$ bits (\textbf{INT88}, or \textbf{INT64} in
the $g{=}5$ split). Against a halved $6.4$~ms roof ($Q=48N^3$) the
tensor term is $5.6$~ms and the epilogue $c_{\text{epi}}=109$--$152$,
giving $34$--$47$~ms---\emph{$5.2$--$7.3\times$ its own roof, the same
ratio as \fp{64}}. This is the structural point: halving the output
precision halves the traffic, the tensor work and the accumulator
width, but it does \emph{not} halve $n_{\text{out}}$, so the
per-output epilogue scales down far less than the roof does. The
integer epilogue is not an artefact of \fp{64}; it is an artefact of
reconstructing one scalar at a time.

\subsection{Extension to Rubin-class bandwidth (22~TB/s)}
\label{sec:rubin22}

The same ladder evaluated at Rubin's memory system asks two
questions: how much INT8-TC does Phase~A need against a $4.7$~ms
roof, and where does the epilogue land against Rubin's larger
($65$~T~inst/s derived) scalar pipe.

\paragraph{The INT8-TC requirement scales with bandwidth.} The
Phase-A tensor budget is the roof less the \fp{8} Bailey time:
$4.69 - 2.62 = 2.07$~ms on Rubin, against $3.73$~ms on B300. The
break-even of \eqref{eq:inttc-breakeven} therefore rises to
\begin{equation}
P_{\text{INT8-TC}}^{\text{Rubin}} \;\geq\; \frac{1.42\times 10^{13}}
   {2.07\ \text{ms}} \;=\; 6.9\ \text{POPS},
\label{eq:rubin-inttc}
\end{equation}
$1.53\times$ B200's part---exactly the bandwidth ratio would predict
($6.9/3.8 \approx 22/8 \cdot$ the Bailey correction). A Rubin-class
GPU that restores INT8-TC \emph{scaled as its FP8-TC scaled}
($17.5/4.5 = 3.9\times$ B200 $\Rightarrow \approx 17$~POPS) clears
this with headroom; a flat B200-rate restoration ($4.5$~POPS) leaves
Phase~A at $3.2$~ms against the $2.07$~ms budget, and the tensor term
overshoots to $5.8$~ms---the wall then floors there no matter what
the epilogue does.

\begin{table}[h]
\centering
\caption{The recommended ladder at Rubin-class bandwidth (22~TB/s,
roof $4.69$~ms, INT32 issue $65$~T~inst/s derived). Two INT8-TC
provisions: B200-rate ($4.5$~POPS, tensor $5.78$~ms) and
bandwidth-scaled ($\geq 6.9$~POPS, tensor $4.69$~ms). Native \fp{64}
on the announced Rubin is $4.9$~ms; the emulated route matters on a
Rubin-bandwidth part only if \fp{64} collapses
(\S\ref{sec:swhw}). Generated by \texttt{ledger.py}.}
\label{tab:rubin22}
\footnotesize
\setlength{\tabcolsep}{3.2pt}
\begin{tabular}{lccc}
\toprule
rung & $c_{\text{epi}}$ & wall ($4.5$-POPS int-TC) &
   wall ($\geq 6.9$-POPS) \\
\midrule
today (software only) & 203--281 & 40.2--55.6~ms & 40.2--55.6~ms \\
$+$H0 (software, 7/12 friendly) & 188--254 & 37.3--50.2 & 37.3--50.2 \\
$+$H1 int-TC $+$ positional acc.\ & 94--155 & 18.7--30.8 & 18.7--30.8 \\
$+$H2 datapath \& ISA & 52--76 & 10.3--15.0 & 10.3--15.0 \\
$+$H3 mod-reduce at MMA & 34--49 & \textbf{6.7--9.6} & \textbf{6.7--9.6} \\
\midrule
vs.\ $4.69$~ms roof & & $1.4$--$2.1\times$ & $1.4$--$2.1\times$ \\
\bottomrule
\end{tabular}
\end{table}

\paragraph{What the 22-TB/s ladder (Table~\ref{tab:rubin22}) says.} Three readings.
\emph{First}, the package up to H2 that narrows B300's wall to
$1.2$--$1.8\times$ lands at $10.3$--$15.0$~ms on Rubin
bandwidth---$2.2$--$3.2\times$
off its $4.69$~ms roof. Bandwidth grew $2.75\times$ but the scalar
pipe grew only $1.56\times$, so the epilogue floor
$\etaopt^{\text{epi}} = (c_{\text{epi}}/8)\Bmem$ \emph{rises} with
bandwidth while the ladder's per-rung savings do not. \emph{Second},
even the full ladder lands at $6.7$--$9.6$~ms, $1.4$--$2.1\times$ off:
closing to the roof needs $c_{\text{epi}} \leq 24$, i.e.\ trimming the
surviving deconstruction residual and assembly terms, or a further
$1.4$--$2.0\times$ of scalar issue. The epilogue problem does not go
away at 22~TB/s; it gets relatively harder. \emph{Third}, the
practical conclusion for the announced Rubin is unchanged from
\S\ref{sec:swhw}---its native $4.9$~ms \fp{64} beats every rung
until H3 and ties it there, so emulation remains pointless on that
part. The table's audience is a future part with Rubin's memory
system and B300's \fp{64}: on it, the full ladder turns a
$40$--$56$~ms emulation into $6.7$--$9.6$~ms, within $1.4$--$2.1\times$
of a roof that native arithmetic could not approach at all.

\subsection{Alternative: the pure-SIMT ladder (no integer tensor
core)}
\label{sec:codesign}

For comparison with the recommended route of \S\ref{sec:inttc}, this
subsection prices the ladder we would build if the integer tensor
core were \emph{not} restored: every epilogue term discharged by a
bespoke SIMT-side or output-stage structure instead. Its rungs are
labelled S1--S3 to keep them distinct from the primary ladder's
H-rungs. The comparison's conclusion is stated up front: the
pure-SIMT ladder ends tensor-bound at $13.1$~ms (the FP16-carrier
tile padding survives, since no integer tensor core exists to retire
it), needs the \emph{large} residue-domain output stage to get there,
and is dominated rung for rung by Table~\ref{tab:inttc-route}. It
remains instructive because it prices each epilogue term in
isolation.

Ozaki 2.5 closed with a proposition this paper extends: when a
floor is closed-form, it \emph{names its own hardware escape}. There
the floor was the deconstruction-$\lambda$ ceiling
$R\,P_{\text{int}}/(c_q r)$, and the named asks were cluster reach,
TMEM capacity, and a $c_q$-removing conversion
datapath~\cite{matsuoka2026ozaki25}. Here the floor is the
integer-epilogue floor \eqref{eq:combined-floor},
\[
\begin{aligned}
\etaopt^{\text{epi}} &= \frac{c_{\text{epi}}}{8}\,\Bmem,\\
c_{\text{epi}} &= \underbrace{c_q^{\text{res}}r + c_{\text{dp4a}}
 + c_{\text{canon}}}_{\text{input side}}
 + \underbrace{c_{\text{comb}}}_{\text{combine}}
 + \underbrace{c_{\text{dep}} + c_{\text{lift}}
 + c_{\text{pack}}}_{\text{output side}}
 \;\in\; [203, 281]\ \text{inst/output},
\end{aligned}
\]
and each named term names its mechanism. We propose a four-rung
ladder (H0 plus three silicon rungs), ordered by engineering risk;
Table~\ref{tab:codesign-ladder} computes the cumulative effect (all
values from \texttt{ledger.py}, and the per-rung deltas quoted below
are the ledger's, on top of the H0-reduced counts).

\paragraph{S1 --- the load-path deconstruction datapath
($-c_{\text{dp4a}}$ and the streamed half of $c_q^{\text{res}}r$;
$-16$ to $-33$ inst/output).}
Ozaki 2.5's own hardware ask: an in-flight residue-conversion
datapath (copy-engine- or TMA-integrated) that converts streamed
operands into residue planes as they move, removing the
per-element SIMT residual for the operand half that actually
streams from HBM. It sees only that half: the second-stage operand
$Y'$ is an on-chip intermediate, and the canonicalisation term is
applied to on-chip Bailey MMA outputs, so neither is reached by a
load-path converter---which is why this rung, Part~1's S3C
stream-side converter (Option~C), is worth $16$--$33$ instructions here and not the
whole input side. This rung is not FFT-specific---dense GEMM, SpMV,
and every Ozaki-II kernel shares it---which is precisely why it is
the lowest-risk ask: it already has a constituency.

\paragraph{S2 --- a residue-domain MMA output stage
($-c_{\text{dep}} - c_{\text{comb}} - c_{\text{canon}}$; $-114$ to
$-127$ inst/output).}
This is the large rung, and it bundles three functions at the
tensor-core output path (TMEM on Blackwell-class parts), in
increasing silicon cost: (a)~\emph{wide-accumulate} mode---the
Kulisch deposits are a fixed dataflow, add a bounded integer shifted
by a compile-time amount into a wide accumulator, and tensor cores
already own exactly this pattern at 24-bit width (the FP32
accumulator's exact-integer window), so the ask is $\approx
160$--$192$-bit fixed-point accumulation of the Phase-A column
outputs at their positional offsets, i.e.\ hardware Kulisch
accumulation at the MMA epilogue ($-66$ to $-70$); (b)~the
\emph{linear} residue-domain combine, adds only ($-30$
per output-equivalent); and (c)~\emph{modular reduction} to
$[0,m_k)$ at the output stage, which discharges canonicalisation
($-18$ to $-27$ after H0) and is the expensive part---an array of
modular reducers, discussed below. Function (a) achieves \emph{exactly} what the software
tall-skinny variant of \S\ref{sec:fp8-tallskinny} attempted---and
what its uncosted digitisation destroyed---with no digitisation at
all, because the accumulator, not the operand format, carries the
width. Wide exact accumulators have a long hardware lineage (Kulisch's
original proposal~\cite{kulisch1976,kulisch1986}, reproducible-
summation units~\cite{uguen2017kulisch}); the novelty is only their
placement at the MMA epilogue. Silicon cost is a set of wide adders
in a path that already contains FP32 accumulate hardware.

\paragraph{S3 --- fused \texttt{crtlift} and wide-convert
instructions ($-c_{\text{lift}} - c_{\text{pack}} + 8$; $-37$ to
$-57$ inst/output).}
\begin{sloppypar}
The remaining scalar tail is two fixed-function idioms. (i)
\texttt{crtlift}: given the wide accumulator and the compile-time
constants ($M$ as limbs, the $2^{-f}$ fractional partials), compute
$\widehat z$, subtract $\widehat zM$, and apply the two-sided
centered correction---a small MAC array plus comparators, the
integer analogue of the FP64 divide-free range reduction units
long shipped in DSPs. (ii) a widened \texttt{cvt}: correctly-rounded
conversion from a signed $\le 128$-bit fixed-point register (the
\emph{post-lift centered} result, $|R| < M/2 < 2^{117}$---not the
131-bit pre-lift accumulator) to
\fp{64} with an exponent immediate (the power-of-two scale). CVT
instructions exist; the ask widens their source. We budget the
surviving software residue at $\approx 8$ inst/output.
\end{sloppypar}

\begin{table}[h]
\centering
\caption{The \emph{alternative} pure-SIMT ladder (no integer tensor
core) on B300; cumulative; ranges span the sensitivity intervals;
generated by \texttt{ledger.py}. Tensor time stays $13.1$~ms (the
FP16-carrier tile padding survives without an INT8-TC to retire it),
so this ladder terminates \emph{tensor}-bound at $13.1$~ms and needs
its \emph{large} S2 to do so. Compare Table~\ref{tab:inttc-route},
which reaches $12.9$~ms with no large ask. At the B300-SXM variant's
$37.5$~T~inst/s every integer entry scales by $1.11\times$.}
\label{tab:codesign-ladder}
\small
\setlength{\tabcolsep}{5pt}
\resizebox{\textwidth}{!}{%
\begin{tabular}{lcccc}
\toprule
rung & silicon & int-epilogue (ms) & projected wall (ms) & vs.\ roof \\
\midrule
today (software only) & \emph{none} & 62.8--86.8 & \textbf{62.8--86.8} & $4.9$--$6.7\times$ \\
$+$H0 pseudo-Mersenne moduli & \textbf{none} & 58.3--78.4 & \textbf{58.3--78.4} & $4.5$--$6.1\times$ \\
\midrule
\multicolumn{5}{l}{\emph{$\uparrow$ software only, ships today \quad$|$\quad new silicon required $\downarrow$}}\\
\midrule
$+$S1 load-path decon.\ datapath & small & 53.4--68.3 & 53.4--68.3 & $4.1$--$5.3\times$ \\
$+$S2 residue-domain MMA output stage & \emph{large} & 18.1--28.9 & 18.1--28.9 & $1.4$--$2.2\times$ \\
$+$S3 multiword-carry \& wide-\texttt{cvt} ISA & small & 6.7--11.3 & \textbf{13.1} & \textbf{tensor-bound} \\
\bottomrule
\end{tabular}}
\end{table}

Three observations complete the comparison. \emph{First}, the ladder
is cumulative but separable: S1 alone is worth $\approx 5$--$10$~ms
here and serves every Ozaki-II kernel; S2's wide-accumulate function
alone converts the deposit loop (the single largest output-side
term) and would benefit any wide-accumulation workload,
ReproBLAS-style bitwise-reproducible reductions included
(\S\ref{sec:kulisch-general}). \emph{Second}, we make \emph{no area claim}. A prior formulation asserted that the ladder was cheaper than restoring a 12.5-TF
\fp{64} vector pipe ``($\sim 9\times$ B300's current FP64 area
commitment)''; that $9\times$ was the \emph{throughput} ratio
$12.5/1.4$ relabelled, and we withdraw it. We have no area or power
model, and the honest position is that the relative cost of the
ladder versus FP64 restoration is open. What we can say is
directional: H0 costs nothing, S1 and S3 are small and sit in paths
that already exist, and \emph{S2 is not small}---an array of modular
reducers at the tensor-core output is the same structure that
dominates FHE accelerator area (CraterLake's change-RNS-base unit is
$158.8$~mm$^2$ in 14/12~nm, the largest single functional unit on a
$472$~mm$^2$ die~\cite{samardzic2022craterlake}), and it must run at
$\sim 10^{12}$ outputs/s to keep up with the roof. A reviewer is
right to ask whether simply widening the INT32 pipe is cheaper: from
Table~\ref{tab:codesign-ladder} that alternative needs
$4.9$--$6.7\times$ B300's issue rate, i.e.\ roughly five more copies
of the cheapest array in the SM, with no new pipeline stage, no new
ISA and no new TMEM mode. We do not know which wins, and we no
longer assert that the ladder does. \emph{Third}, on Rubin the same ladder would attack a
$558$--$772$-T-inst/s epilogue requirement against a derived
$65$~T~inst/s issue rate; whether it closes that gap depends on
Rubin's SIMT integer rate, which is not published, and we do not
claim it does. All Rubin figures in this paper are pre-release vendor
announcements and are provisional. In TME terms, either ladder moves the FFT from
the $\gamma$-roof regime back into the $\beta$-roof (memory-bound)
regime where the other four primitives already live---the epilogue
analogue of Ozaki 2.5's $c_q$-removing ask, and, we would argue, the
natural next increment of the same AI-ratchet silicon that created
the problem.

\subsection{Procurement and codesign implications}
\label{sec:procurement}

For HPC system procurement targeting FFT-heavy or spectral
scientific workloads (Quantum Chromodynamics, spectral CFD,
climate emulation, seismic imaging, fusion plasma codes), the
analysis here gives a layered specification. The floors fall into a
primary target, the direct design's scenario, and its integer
fallback---listed in that order, because they make different hardware
demands and rest on different levels of validation:

\begin{itemize}[leftmargin=2em, topsep=2pt]
\item \textbf{Primary target --- native FP64 floor:}
$\eta_{\text{fp64-vec}} \geq 1.56\,\Bmem$. Below this floor, native
FFT becomes compute-bound on the FP64 vector pipe. This is the safe,
validation-free target and is met by H100, B200, and (within 4\%)
Rubin.
\item \textbf{The emulated route (FP64 cut; FP16-carrier Phase~A,
integer-hosted epilogue):} the tensor classes must meet
$\etaopt^{\text{FP8-TC}} = 444\,\Bmem$ and
$\etaopt^{\text{FP16-TC}} = 96\,\Bmem$ (B300 meets both; summed
tensor time $13.1$~ms), and the INT32-SIMT pipe the epilogue floor
$\etaopt^{\text{epi}} = 25$--$35\,\Bmem$ ($203$--$281$~T~inst/s at
8~TB/s---B300's $41.7$~T misses by $4.9$--$6.7\times$). \emph{No shipping
part meets the epilogue floor in software}; B300 lands at
$63$--$87$~ms against the $12.9$~ms roof.
\item \textbf{The wall-closing asks (\S\ref{sec:inttc}):}
either brute-force integer-SIMT provisioning at
$\gtrsim 25$--$35\,\Bmem$ ($4.9$--$6.7\times$ today's ratio), or the
recommended ladder---H1 restore the INT8 tensor core at B200's rate
($\geq 3.8$~POPS; $\geq 6.9$ at 22~TB/s) with $6\times$INT64
positional accumulation ($4$ raw $+$ $2$ fractional groups), H2 the load-path deconstruction datapath and
multiword-carry/wide-\texttt{cvt} ISA idioms, H3 modular reduction
at the MMA output. Up to H2 is minor hardware and lands
$16.0$--$23.5$~ms on B300; H3 lands $12.9$--$15.0$~ms (the roof at
the optimistic end) with today's $41.7$-T-inst/s pipe untouched. We make no area claims; the
ladder is the recommended form because each rung also serves
non-FFT workloads.
\item \textbf{FP32 vector floor (for FP32 spectral codes; corrected
in this revision):} $\eta_{\text{fp32-vec}} \geq 3.125\,\Bmem$
(\S\ref{sec:fp32}---twice the FP64 figure, not equal to it).
\end{itemize}

The native FP64 floor is the safe procurement target; the
FP8-generation tensor floors plus the epilogue provisioning (or the
co-design ladder) form
the emulation-path backup that allows a part to compensate for
sub-floor FP64 silicon. Both H100 and B200 satisfy the native floor
with margin. Rubin meets it within 4\%---essentially
at parity. B300 fails it by $\sim 9\times$ and misses the
integer-epilogue floor by $4.9$--$6.7\times$; no evaluated software
route reaches the roof, and the recovery
the design delivers is at most $1.3$--$1.9\times$ over its collapsed
native path. For codesign of post-Rubin
architectures: prioritise the native FP64 floor; failing that, the
recommended ladder of \S\ref{sec:inttc}---whose up-to-H2 package is
minor hardware---is the most direct roof-recovering package, with
the INT8-TC provision scaled to bandwidth
($\geq 3.8$~POPS at 8~TB/s, $\geq 6.9$ at 22~TB/s). The binding
caution has shifted twice now: first from ``do not cut FP64'' to
``do not starve the integer pipe,'' and---with INT8-TC collapsing
$30$--$70\times$ on B300/Rubin---to its final form: \emph{the
FP8-generation tensor path plus a provisioned (or co-designed)
scalar epilogue is the whole game for exact arithmetic on AI
silicon}.

\subsection{Limitations}
\label{sec:limitations}

Several caveats apply:

\begin{itemize}[leftmargin=2em, topsep=2pt]
\item The per-output integer-epilogue instruction counts
(canonicalisation, combine, deposits, lift, conversion,
deconstruction) are first-order estimates; the stated
$[203,281]$ sensitivity interval bounds but does not
eliminate this uncertainty, and the headline
$63$--$87$~ms range directly depends on them. Likewise the
$c_q^{\text{res,static}}\approx 3$ FFT deconstruction estimate is
ours, not Ozaki 2.5's. The twiddle-fused batched-GEMM schedule of
\S\ref{sec:obfft-cost} likewise awaits an implementation.
\item The double-double Phase~B path of \S\ref{sec:precision} is
projected from arithmetic op counts but has not been measured. Its
actual achievable performance depends on FP32 vector pipeline
efficiency under the dependency pattern of DD additions, which can
degrade throughput by up to $2\times$ in some realisations.
\item The Phase~A formulation requires that the slicing GEMM be
fully fused into the surrounding Bailey kernel; if Phase~A
materialises slice results to shared memory, the bandwidth
multiplier $\beta$ rises and the memory roof shifts. We assume
$\beta = 1$ throughout, consistent with the discipline of
\cite{matsuoka2026tme}, but treat this as an idealisation to be
established by measurement rather than a settled constant; the
register-fusion realism discussion of \S\ref{sec:tme}
($\beta\in[1,(8+2p)/8]$ set by tile shape and register budget, with the
fixed shared-memory DFT$_{32}$ matrix making the FFT case more
favourable than streaming GEMM) and the delivered-$\beta$
measurement of \S\ref{sec:future} qualify this assumption directly.
\item The precision results in Table~\ref{tab:precision} are from
a Python prototype with $16\times 16\times 16$ GEMMs; cumulative
twiddle round-off across $\log_2 N$ Bailey stages may behave
differently at production sizes. Real-hardware measurement is
the natural next step.
\item The FFT op count adopted is the standard $5N\log_2 N$ real
flops per length-$N$ complex 1-D FFT, derived from the radix-2
Cooley-Tukey butterfly (one complex multiply = $4+2=6$ real ops,
plus two complex add/subtract operations = $4$ real ops, giving
$10$ real ops per butterfly $\times\,(N/2)\log_2 N$ butterflies $=
5N\log_2 N$). All quantitative results in the paper---the
operational intensities $\mathrm{OI}_{\text{FP64}}=1.5625$ and
$\mathrm{OI}_{\text{FP32}}=3.125$~flops/byte, the
parity floors, and the architecture comparisons---are
derived from this convention.\footnote{Higher-radix algorithms
(split-radix, mixed-radix, prime-factor) reduce the constant
slightly~\cite{frigo2005fftw}, but the $5N\log_2 N$ count is the
standard benchmark for radix-2 Cooley-Tukey and is the convention
used by all of the FFT literature we cite. We are grateful to
Daisuke Takahashi (private communication) for pointing out an
earlier draft's incorrect alternative claim.}
\item \textbf{Operand-dependent precision selection is a real,
unmodelled cost.} Fixed-point and integer-modular arithmetic carry
no exponent, so the number of moduli $r$ (equivalently, the slice
count and the integer scale factors $s_A,s_B$) that guarantees full
\fp{64} accuracy is not a universal constant: it depends on the
dynamic range of the actual operands. Before each emulated GEMM the
operands must be inspected---at minimum a max-abs reduction to set
$s_A,s_B$ and avoid integer overflow, and in the adaptive case a
per-row/per-block exponent-span analysis to choose $r$. This
analysis is not free; it is INT32/SIMT work that competes with the
reconstruction pipe, and because it can raise $r$ above the nominal
$r=12$ used in our projections, both the compute cost ($\alpha=3r+1$)
and the residue-storage footprint (hence $\beta$) grow with it. Our
projections fix $r=12$, the value recommended for \fp{64}-equivalent
Ozaki-2/FP8~\cite{uchino2026fp8}; on adversarial inputs the
exponent-span estimators of Schwarz et al.\ (Automatic Dynamic
Precision / Exponent-Span Capacity~\cite{schwarz2025dgemm}) may select
a larger $r$, with that work reporting under $10\%$ overhead on
realistic matrices but no guarantee for arbitrary inputs. Two points
bound the exposure for FFT specifically. First, the cost falls
asymmetrically: it attaches to the Phase~A \emph{input scaling}
(shared with Ozaki-II GEMM, Part~1), \emph{not} to the Kulisch
Phase~B accumulator, whose width is fixed at compile time by the
modulus product and is therefore independent of operand magnitude.
Second, FFT inputs from physical-simulation fields are typically
well-conditioned, so the static $r=12$ is expected to hold for the
target workloads; quantum-chemistry and other extreme-dynamic-range
spectra are the regime where adaptive $r$, and an accompanying
fallback to native \fp{64} where the adaptive cost is prohibitive,
must be measured rather than assumed. Quantifying the delivered $r$
distribution and ADP-fallback frequency on real application inputs is
an explicit task of the validation programme (\S\ref{sec:future}).
\end{itemize}

\section{Future Work}
\label{sec:future}

The substantive next step is constructive rather than analytic.
The headline finding---that the direct design eliminates \fp{64}
entirely but walls on the integer epilogue $4.9$--$6.7\times$ above the memory roof on B300---rests on projected instruction counts that
must be measured on real hardware.

\paragraph{(0) Correctness reference first.}
Before any performance work: a bit-exact arbitrary-precision
reference of the direct CRT. \emph{Delivered with this revision}:
\texttt{ledger.py} generates every constant, and \texttt{verify.py}
unit-tests the construction in exact arithmetic---CRT idempotents,
the shifted-column two-limb identity, the biased signed-deposit
accumulator (including the $P_0=-1$ counterexample), the two-sided
centered lift (including $C=\pm(M/2{-}1)$ and boundary sweeps), and
end-to-end reconstruct-and-pack against Python big integers on
random and adversarial inputs. Every hardware kernel below must
reproduce this reference bit-for-bit; the reviews' counterexamples
are its permanent regression tests.

\paragraph{(1) Implement and measure Kulisch
Phase~B on B300/Rubin.}
The principal experimental task is to implement
Algorithm~\ref{alg:obfft} with a Kulisch fixed-point Phase~B in CUDA
and measure on B200, B300, and (when available) Rubin hardware.
(No CUDA implementation yet exists; a prior reference to a
supplied skeleton was inaccurate and is withdrawn---the artifact
currently contains the analysis scripts only.) The benchmark
protocol:
\begin{enumerate}[leftmargin=2em, topsep=2pt, itemsep=0pt]
\item \emph{Baseline}: measured cuFFT on B200 at $1024^3$ FP64
(expected $15\text{--}25$~ms based on prior reports, against the
$12.9$~ms theoretical memory roof).
\item \emph{B200 \OBFFT\ paths}: implement and measure the
recursive-Garner, direct-CRT + fp64 sum, and direct-CRT + Kulisch
paths on B200 as well, to verify the model in a regime where
native FFT is bandwidth-bound and the comparison is clean.
\item \emph{B300 native cuFFT}: measure the FP64-collapsed
native path (expected $\sim 115\text{--}140$~ms compute-bound).
\item \emph{B300 \OBFFT\ + Kulisch (the specified design)}: the
headline measurement, and the direct test of the epilogue
accounting. The design projects to
$\geq 63$--$87$~ms at peak integer issue (stretching at
realised efficiency), the range set by the deconstruction residual
and the per-term epilogue instruction intervals. The measured SASS
counts against the $[203,281]$ ledger, and where within $[3.0,6.3]$
the FFT's static-exponent deconstruction
actually lands, are the two numbers the whole model turns on.
\item \emph{B300 \OBFFT\ + fp32+Kahan}: cross-check the reduced-precision
path ($\sim 14$~ms projected, fp32-class $\sim 10$--$18$-bit precision).
\item \emph{Rubin native cuFFT}: validate the native-parity claim
($\sim 5\text{--}8$~ms projected against $4.7$~ms memory roof).
\end{enumerate}
The comparison across these six paths will validate (or
refute) the epilogue accounting and the bandwidth-parity floors of
\S\ref{sec:parity}.

\paragraph{(1a) End-to-end accuracy battery.}
``Full \fp{64}'' must be demonstrated end-to-end, not only for the
reduction step. The battery: forward and inverse transforms at
multiple sizes ($256^3$--$2048^3$); input classes covering random,
impulse, single-frequency, cancellation-heavy, signed, and
wide-exponent-range data; comparison against a higher-precision
(MPFR) reference at small sizes and against cuFFT/FFTW at production
sizes; normwise forward error, backward error, inverse-transform
residual, and Parseval error; and the observed distribution of
selected $r$, overflow-guard trips, and adaptive-precision fallback
frequency.

\paragraph{(1b) Measuring the \emph{delivered} cost model, not just
the synthetic kernel.}
The six-path benchmark above measures a single, well-conditioned
$1024^3$ transform. Establishing the thesis requires closing the gap
between the analytic TME parameters and what hardware actually
delivers, on inputs that are not hand-picked. Three quantities must be
measured rather than assumed: (i)~the \emph{delivered $\alpha$},
including the pre-computation (operand scaling, rounding-to-integer,
per-modulus reduction) that the throughput model charges only as
tensor-core MMAs---i.e.\ the fraction of the $(3r+1)$ budget that is
genuinely hidden behind the MMA stream versus exposed on the SIMT
pipe; (ii)~the \emph{delivered $\beta$} as a function of tile shape
and register budget, to locate where on the $[1,(8+2p)/8]$ interval a real
fused kernel lands and at what tile size it begins to spill; and
(iii)~the \emph{ADP-fallback frequency}---how often the
exponent-span estimator~\cite{schwarz2025dgemm} must raise $r$ above
$12$, or fall back to native \fp{64}, when driven by real
operand distributions rather than synthetic well-conditioned data.
Only once these three are characterised can the projected wall times
be read as predictions rather than ideal lower bounds on time.

\paragraph{(1c) Validation on real spectral applications.}
The decisive test is end-to-end, on production scientific codes whose
performance is dominated by 3-D FFTs, not on a standalone transform.
Candidate drivers include spectral-element and pseudospectral CFD
(homogeneous-turbulence DNS), planewave DFT (\textsc{Quantum
ESPRESSO}, \textsc{VASP}-class FFT kernels), particle-mesh Ewald and
PIC electrostatics, and ab-initio molecular dynamics---workloads that
span the favourable (well-conditioned, physical-field) and the
stressful (wide-dynamic-range, quantum-chemistry) ends of the input
spectrum identified in the Limitations (\S\ref{sec:limitations}). For each, the measurement of
interest is not the kernel microbenchmark but the change in
application time-to-solution and end-to-end accuracy when the native
\fp{64} FFT is replaced by the Ozaki-Bailey-Kulisch path on a
\fp{64}-collapsed GPU, including the cost of any
\fp{64} fallback the input distribution forces. This validation is the
natural locus of a joint measurement programme with hardware-vendor
and laboratory partners, and is where the present analytic projections
are either confirmed as a basis for codesign or revised; it is folded
into the broader FP64-transition study from which this paper's kernel
analysis is drawn.

\paragraph{(1d) Verifying the integer-hosted Phase~B and its
narrow-width variants --- the highest-leverage experiment.}
\begin{sloppypar}
Of all the items here, confirming the Phase~B substrate results of
\S\ref{sec:design} and \S\ref{sec:noint32} is the most consequential,
because it determines whether the title thesis holds in its strong
form and whether the architectural recommendation survives the next
hardware generation. Two measurements, in order of importance.
\end{sloppypar}

First, the \textbf{epilogue instruction ledger}, on which the
headline range directly rests: compile the Algorithm-\ref{alg:kulisch}
kernel (biased carry-save deposits, two-sided lift, correctly-rounded
pack) plus canonicalisation and combines, report SASS instruction
counts against the stated $[203,281]$ sensitivity interval, and
validate bit-exactness against \texttt{verify.py}'s reference
(whose adversarial cases---$C=\pm(M/2{-}1)$, negative deposits,
rounding ties---are permanent unit tests). Second in the same
family: the FP16-carrier Phase~A exact-integer accumulation contract
(FP32 accumulators, integer operands) on real silicon, with the
E4M3-plane variant of Appendix~\ref{app:garner-detail} as fallback;
and the deconstruction-residual measurement (where in $[3.0,6.3]$
the FFT's static-scale case lands). The demoted tall-skinny deposit
route needs a digitisation ledger before it re-enters the
quantitative comparison.

Second, the \textbf{narrow-width hostings (INT16/INT8)}.
These should be implemented and measured as the design's
insurance against the next AI-ratchet turn: the integer-hosted
Phase~B is the design, and its narrow-width variants remove any
dependence on a wide INT32 pipe, the one accidental hardware feature a pure-integer
realisation would lean on (\S\ref{sec:noint32}). The measurement is
the realised word-level deposit and carry-resolve
throughput at $w=16$ and $w=8$---an ISA-specific instruction ledger
this paper deliberately does not predict
(\S\ref{sec:width-subfloor}): narrow scalar types do not
automatically provide independent carry-chain lanes, and no public
issue model settles the question. Establishing where INT16/INT8
land relative to INT32 on real silicon would turn the
width-independence \emph{correctness} property into a
\emph{performance} statement, hardening the emulated procurement
route of
\S\ref{sec:procurement} against any narrowing of INT32. Both route
families share the \texttt{libKulisch} harness of
\S\ref{sec:kulisch-general} and the hardware platforms of item~(1).

\paragraph{(2) Hardening the FP64 degradation floors as empirical
codesign metrics.}
The architectural finding---that
$\eta_{\text{fp64-vec}} \geq 1.56\,\Bmem$ is the floor for native FFT
to remain memory-bound, $\eta_{\text{int-vec}} \geq
25$--$35\,\Bmem$ is the integer-epilogue floor, and
$444\,\Bmem$ (FP8-TC) / $96\,\Bmem$ (FP16-TC) are the per-class
tensor floors---deserves to be hardened from a predicted
floor into an empirically established one. This is the bottom
line for FP64 vector degradation that future GPU generations should not
cross, if FFT-heavy scientific codes are to retain bandwidth-bound
execution. The model predicts the achieved-to-peak bandwidth ratio
drops sharply at each floor and stays flat above it. Hardening all
three---including confirming that the integer floor's correctness is
substrate-width-independent, and measuring the epilogue instruction
counts that set its coefficient---would establish the full set as
procurement specifications, and give the FugakuNEXT,
Doudna~\cite{nersc2026doudna}, and Blue
Lion communities a clean rule for evaluating post-Rubin GPU designs,
covering both the case where an integer pipe is retained and the case
where \fp{8} tensor cores carry the deposits.

\paragraph{(3) Building the generalised \texttt{libKulisch} library.}
The Kulisch rescue applies more broadly than Ozaki-Bailey FFT
(\S\ref{sec:kulisch-general}). The natural medium-term workstream is
to implement the templated wide-accumulator primitive described in
\S\ref{sec:kulisch-general}, instrument it with the benchmark harness
proposed there, and run the initial three-kernel benchmark campaign
(Ozaki-Bailey FFT, Ozaki-II DGEMM, CG dot products). If the
broader hypothesis is confirmed---that Kulisch provides meaningful
end-to-end speedup whenever the reduction-phase Amdahl fraction
$f \gtrsim 0.5$---then \texttt{libKulisch} graduates from a research
prototype to a production library worth integrating into the cuBLAS
emulation stack~\cite{nvidia2025cublas} and the broader HPC stack. This is a $\sim$6--12 month engineering
effort with potentially broad downstream impact.

\paragraph{(4) Refining the Phase~B path beyond Kulisch.}
Three further open problems on the reduction phase:
(i)~empirical measurement of the double-double Phase~B path described
in \S\ref{sec:precision} to validate the projected $\sim 30$~ms /
$\sim 48$-bit point;
(ii)~recursive Ozaki-II of Phase~B itself, where the per-output
reduction is replaced by a smaller-$r$ Ozaki-II emulation on tensor
cores;
(iii)~whether the Phase~B \fp{64} reduction can be expressed as a
structured tall-skinny GEMM amenable to BLAS optimisation. None of
these is expected to surpass Kulisch on B300, but they may matter on
architectures with different pipe ratios.

\paragraph{(5) FP32 parallel implementation.}
The analysis of \S\ref{sec:fp32} predicts that current architectures
comfortably exceed the FP32 floor for spectral codes. A direct
demonstration---running cuFFT in single precision on B300 at
$1024^3$ and measuring achieved bandwidth---would confirm this and
would establish the cross-precision design metric for post-Rubin GPUs.

\paragraph{(6) Other FFT-class kernels and decompositions.}
The Bailey factorisation is one of several FFT decompositions
(four-step, six-step, Stockham, Pease) all of which admit tensor-core
acceleration in principle. A systematic comparison of which
decomposition minimises $\gamma$-overhead under \OII\ emulation,
particularly with Kulisch Phase~B, would close out the FFT story;
preliminary inspection suggests Bailey six-step is competitive but
not uniquely optimal.

\paragraph{(7) The Bluestein--NTT route on integer tensor cores.}
Kawakami and Takahashi~\cite{kawakami2026bluestein} have recently
shown that the Ozaki scheme can be applied to FFT via a Bluestein
reduction to cyclic convolution, with the split component
convolutions computed exactly using NTTs (see
\S\ref{para:kawakami}). Their CPU implementation reports
$107$--$1315\times$ FFTW double-precision slowdown---a baseline
result on a hardware regime where FP64 is natively supported.
Translating this approach to GPUs with strong integer tensor-core
throughput---B200-class parts today, or any part on which the INT8
tensor core is restored (rung H1), since B300 and Rubin collapsed it
$30$--$70\times$---is a natural extension. The
bandwidth-parity floor framework of~\S\ref{sec:parity} would carry
over with the appropriate substitutions: $\Plow$ replaced by integer
tensor-core (INT8 operand, INT32 accumulate) throughput, and the Garner step replaced by the CRT
recombination of the NTT outputs. A direct comparison between the
Ozaki-Bailey-Kulisch route (FP8 tensor cores + INT32 SIMT
Kulisch) and the Bluestein-NTT route (integer tensor cores + FP64
output reconstruction) on the same hardware would
identify which substrate---FP8 floating-point or integer
tensor cores---is the more efficient base for emulated full-FP64 FFT,
and may suggest hybrid combinations.

\section{Conclusion}

The result of this paper is a single design, stated and derived
directly: a full-\fp{64} 3-D FFT with \emph{no \fp{64} arithmetic
anywhere in its kernel}---Bailey GEMMs emulated by \OII\ on \fp{8}
tensor cores, the twiddle Hadamard fused into the second-stage DFT
operands, Karatsuba combines in the residue domain, CRT Phase~A on
the FP16 path of the same tensor cores (forced there by the
$30$--$70\times$ INT8-tensor collapse on B300/Rubin), an exact
fixed-point (Kulisch) reduction in biased carry-save form with a
two-sided modulo-$M$ centered lift---machine-generated from the
Rev-34 modulus set and unit-tested bit-exactly
(\texttt{verify.py})---and power-of-two
scaling folded into the final conversion. On B300 the
service-maximum lower bound is $63$--$87$~ms for $1024^3$
against the $12.9$~ms memory roof---\emph{at most}
$1.3$--$1.9\times$
the $116$~ms native path if attained, with a worst-case reduction
bound stronger than naive \fp{64} summation, and $4.9$--$6.7\times$
short of the roof itself. The binding resource is neither \fp{64}
(eliminated) nor the tensor classes (at the roof at $13.1$~ms,
subject to the layout caveat of \S\ref{sec:tme})
but the \emph{integer-SIMT
epilogue}: the deconstruction residual, canonicalisation, combines,
deposits, lift, and conversion that no evaluated substrate
reassignment
removes---and whose closed-form floor names the hardware
ladder (\S\ref{sec:inttc}) that takes B300-class silicon to the
roof. This completes, for the fifth canonical
primitive, the canonical-kernel coverage of the post-FP64 stack begun
in \cite{matsuoka2026tme}---and it revises that coverage's headline
for FFT from ``at the roof'' to ``at the integer-epilogue wall,
with the wall's removal specified.''

The design is forced---within the substrate assignments evaluated
here---by the cost model. Bailey's small inner factor
$k\approx\sqrt N\approx 32$
pushes the kernel into the regime $k\ll r^2\!\sim\!144$ where the TME
parameter $\gamma$ (reconstruction cost) no longer amortises and
binds---$\sim 111$~ms under recursive Garner on B300, $\sim 8.6\times$
the roof---while, in the four-parameter $(\alpha,\beta,\gamma,c_q)$
model adopted after the NVIDIA technical
review~\cite{bayraktar2026tme},
the input-side deconstruction cost $c_q$ simultaneously claims the
integer SIMT pipe. Both conversion terms bind at once: this is the
distinctive feature of the FFT case among the five primitives. The
ablation ladder confirms the assignment: recursive Garner
($\sim 111$~ms), direct CRT with naive \fp{64} summation
($\sim 278$~ms on the collapsed vector pipe), INT8-tensor Phase~A
($\sim 85$~ms, dead)---each undone ingredient lands
further from the roof.

The principal architectural results are the closed-form
bandwidth-parity floors, all regenerated in this revision from the
actual modulus set and a single stated operation-counting convention
(and all reproducible from the supplied \texttt{ledger.py}). The
native FFT floor
$\etaopt^{\text{FP64}} \approx 1.56\,\Bmem$ is the reference: at
any $\eta_{\text{FP64-vec}}\geq 1.56\,\Bmem$, native fp64 FFT reaches
the memory roof, and \OII\ for FFT is unnecessary. H100 and B200
satisfy it with margin; Rubin sits within 4\% (essentially at
parity); B300 misses it by $\sim 9\times$. For the emulated route the
floors are, per instruction class: FP8-TC $444\,\Bmem$ and FP16-TC
$96\,\Bmem$ (both met on B300; summed tensor time $13.1$~ms) and
integer epilogue $25$--$35\,\Bmem$,
which B300's $5.2\,\Bmem$ misses by $4.9$--$6.7\times$.
\emph{No shipping part reaches the emulated
memory roof in software}; what B300 gains from the design is at
most a $1.3$--$1.9\times$ recovery over its collapsed native path,
and the recommended co-design ladder of \S\ref{sec:inttc} recovers
the rest: minor hardware (up to H2) to $16.0$--$23.5$~ms, one
moderate ask to $12.9$--$15.0$~ms---the $12.9$~ms roof at the
optimistic end of the epilogue interval.

The design carries its insurance at a stated confidence gradient
(\S\ref{sec:noint32}). The width-parameterised Kulisch accumulator
runs the
identical exact reduction on any integer pipe of width
$\geq 8$~bits---correctness certain, classical reduced-radix
arithmetic; cross-width \emph{performance} deliberately deferred to
an ISA-specific ledger. The width-independence matters because
B300's wide integer pipe is an \emph{accident} of the AI design
pivot, not a guaranteed feature. What this revision withdraws, after
two independent reviews, is every residue of the stronger earlier
claims: the integer pipe does \emph{not} ``drop out entirely''
(Ozaki 2.5's residual is elementwise-nonlinear and not
tensor-migratable; the epilogue is per-output scalar work with no
known software tensor formulation), and the tensor-hosted deposit
variant is qualitative until its digitisation is costed.
\fp{8}-generation tensor cores are
all you need \emph{in place of \fp{64}}---no branch of the design
touches \fp{64} silicon, and the collapsing INT8-TC path is routed
around too---but \fp{8}-plus-nothing-else is not
established, and on shipping parts the integer pipe is the wall.

A GPU meets memory-roof FFT parity at full \fp{64} if it satisfies
\emph{either} the native FP64 floor, \emph{or} the route-complete
emulated set: FP8-TC $444\,\Bmem$, FP16-TC $96\,\Bmem$, \emph{and}
the integer-epilogue floor $25$--$35\,\Bmem$. H100 and B200 satisfy
the
native floor with margin; Rubin within 4\%; B300 fails it by
$\sim 9\times$ and fails the epilogue floor---the honest summary
is that B300 buys back at most $1.3$--$1.9\times$ of its collapse in
software
and leaves the last $4.9$--$6.7\times$ to hardware. A hypothetical
B300+ at 12.5~TFLOPS FP64
vector would restore native parity at 8~TB/s and obviate the emulated
workaround; so, at a silicon cost we do not model, would the recommended
ladder of \S\ref{sec:inttc}.

The parallel FP32 analysis, corrected in this revision (the floor is
$3.125\,\Bmem$, not $1.56$: halved traffic doubles operational
intensity), shows FP32 FFT is adequately provisioned on B300-class
bandwidth ($3.33\times$ headroom) but nearly consumed on Rubin-class
bandwidth ($1.89\times$ at Rubin's published 130~TF)---so FP32
vector silicon is \emph{not} free
to cut on high-bandwidth parts, a reversal of the prior
conclusion. The asymmetry between FP64 (critically
under-provisioned on B300) and FP32 (adequately provisioned today,
thinning as bandwidth grows) remains the clearest signal of how the
architectural pivot toward AI4S has re-shaped the cost--benefit
calculation for native floating-point units.

The author's view, on the basis of this analysis, is the following.
\emph{Software alone reaches $63$--$87$~ms on B300 and is a net loss on Rubin (\S\ref{sec:swhw}); everything below is a hardware ask.} \emph{The safe codesign target for any GPU intended to serve
spectral or FFT-heavy scientific workloads is FP64 vector
throughput $\geq 1.56\,\Bmem$, with FP32 vector at
$\geq 3.125\,\Bmem$.} If FP64 silicon is cut, the emulated route
requires the route-complete floors of \S\ref{sec:parity}: FP8-TC
$444\,\Bmem$, FP16-TC $96\,\Bmem$, \emph{and} integer epilogue
$25$--$35\,\Bmem$---the last of which no shipping part
meets, and which is therefore the concrete procurement ask. Its
recommended form is the ladder of \S\ref{sec:inttc}: restore the
INT8 tensor core B200 shipped ($\geq 3.8$~POPS at 8~TB/s,
$\geq 6.9$ at 22~TB/s) with $6\times$INT64 positional accumulation
(H1); add the load-path deconstruction datapath and
multiword-carry/wide-\texttt{cvt} ISA idioms (H2); and, to close the
sensitivity interval, modular reduction at the MMA output (H3).
Up to H2 is minor hardware and takes the design from $63$--$87$~ms
to $16.0$--$23.5$~ms; H3 lands $12.9$--$15.0$~ms (the roof at the
optimistic end) on today's $41.7$-T-inst/s integer pipe. That a
double-precision spectral workload can run to
full \fp{64} accuracy with no \fp{64} arithmetic at all is
established here by construction; making it run \emph{at the memory
roof} is, on the evidence assembled, a four-rung hardware ladder
away---and every rung pays for itself outside FFT.

Building and measuring the kernel is the immediate next step
(\S\ref{sec:future}): the bit-exact CRT reference first, then the
six-path benchmark and the accuracy battery, then the epilogue
instruction ledger on which this paper's headline range directly
rests. For the
FugakuNEXT codesign and the broader 2028--2030 procurement window,
the layered specification stands as: \emph{safe}
($\eta_{\text{FP64-vec}} \geq 1.56\,\Bmem$,
$\eta_{\text{FP32-vec}} \geq 3.125\,\Bmem$); \emph{emulated-fallback}
(the route-complete tensor $+$ integer-epilogue floors above, on any
integer width $\geq 8$~bits); and \emph{aspirational} (the recommended ladder of
\S\ref{sec:inttc}, which discharges the epilogue and makes the
roof reachable with no \fp{64} and today's integer pipe). The
sharpest single sentence this analysis supports is no longer ``FP8
and memory alone,'' but something more useful to a hardware
architect: \emph{the future of full-precision FFT on AI silicon is
decided not by floating-point units at all, but by the integer
epilogue---and the epilogue, unlike the FP64 pipe it replaced, is
three small, reusable, named pieces of silicon away from free.}

\section*{Acknowledgements}

The author thanks Katsuhisa Ozaki, Yuki Uchino, Toshiyuki Imamura,
and Daichi Mukunoki for the body of work on which this analysis
rests; any errors in interpretation are the author's. The author is
particularly grateful to Toshiyuki Imamura for clarifying the
Ozaki-2/FP8 moduli count and MMA-cost structure (recommending
$r=12$ with $(3r+1)$ Matmuls per Bailey GEMM due to the internal
Karatsuba emulation of signed int9), and to Daisuke Takahashi for
correcting the FFT operation-count discussion in an earlier draft
and for sharing the Kawakami--Takahashi~\cite{kawakami2026bluestein}
Bluestein--NTT Ozaki-FFT work that informs the comparison in
\S\ref{para:kawakami}; both sets of corrections are incorporated into
the present version. The author also thanks the broader RIKEN R-CCS
team, the Institute of Science Tokyo faculty, and the NVIDIA
cuBLAS team for technical discussions that shaped this paper. The
author is grateful to Dan Ernst and the NVIDIA libraries and DevTech
teams for a detailed technical review of an earlier draft of this work
and its companion~\cite{matsuoka2026tme}; their feedback on the
practical realisability of register-level fusion, on operand-dependent
precision selection, and on the gap between throughput models and
delivered performance directly shaped the implementation-realism
discussion of \S\ref{sec:obfft} and the validation programme of
\S\ref{sec:future}. The author further thanks Jack Dongarra for a
careful reading of an earlier draft, whose comments sharpened the
distinction between projected and measured results and the precise
use of the terms ``precision'' and ``accuracy'' throughout. This
work was undertaken as part of the FugakuNEXT project and related
R-CCS initiatives on AI for Science.

\section*{Disclosure of AI-assisted investigation and writing}

This manuscript and the underlying technical investigation---algorithm
design (\OBFFT, tc-CRT reformulation, Kulisch Phase~B), performance
modelling, parity-floor derivation, prototype implementation, figures,
and initial \LaTeX{} drafting---were carried out in close collaboration
with Anthropic's Claude (Opus~4.7), under the author's direction.
Gemini~3 was used for copy-editing. The author posed the research
questions, challenged Claude's analyses, identified inconsistencies in
earlier drafts (notably the misleading inclusion of reduced-precision
paths in the wall-time chart, leading to the corrected identification
of $1.56\,\Bmem$ as the binding floor), and takes full responsibility
for all scientific content. The present revision additionally
incorporates the findings of two independent technical reviews of
the July~29--30 drafts. The first identified the stale CRT slice
count, the missing modulo-$M$ reduction, the conflated Karatsuba
layers, the uncounted twiddle Hadamard, and the FLOP-convention and
FP32 operational-intensity errors; the second identified the
INT8-versus-FP8 tensor-class rate error, the two-limb radix defect,
the invalid signed-deposit and one-sided-centering pseudocode (with
exact counterexamples now enshrined in \texttt{verify.py}), the
Barrett double-count, and the unsubstantiated epilogue instruction
counts now carried as sensitivity intervals. Each correction is
acknowledged at the point of the fix. The two later revisions were
likewise AI-assisted under the author's direction: Revision~9
(the August~19 cross-paper reconciliation) with Claude (Opus~4.8 and
Fable~5), and Revision~10 (the September~3 final pass, which added the
layout caveat of \S\ref{sec:tme}, reconciled the pure-SIMT ladder prose
with the ledger, and repaired the stale constants listed in
\S\ref{sec:changelog}) with Claude (Fable~5.1). Empirical hardware
measurement is the explicit next step.


\appendix

\section{Derivation of the FP64 Bandwidth-Parity Formula}
\label{app:parity}

We derive the parity formula $\etaopt = \mathrm{OI}\cdot \Bmem$
and its FFT-specific instantiation $\etaopt \approx 1.56\,\Bmem$
(native) and $\etaopt = 3.75\,\Bmem$ (naive \OBFFT, $r=12$, $S=15$,
FMA $=2$).

\paragraph{Setup.}
Consider a 3-D FFT of size $N^3$ with $N=pq$ (Bailey factorisation
with $p\approx q\approx \sqrt N$). The total work in complex-flop
units is
\begin{equation}
W_{\text{FFT}} = 3 \cdot 5 N^3 \log_2 N
\end{equation}
flops (counted as multiply-adds), where the factor of 3 accounts
for the three axes and the $5N\log_2 N$ per 1-D FFT is the standard
radix-2 count: one complex multiply ($6$ real ops) plus two complex
add/subtracts ($4$ real ops) per butterfly, i.e.\ $10$ real ops
$\times\,(N/2)\log_2 N$ butterflies---no rounding involved.

The total HBM traffic is
\begin{equation}
Q_{\text{FFT}} = 6 N^3 \cdot 16 \text{ bytes}
              = 96 N^3 \text{ bytes},
\end{equation}
counting one read and one write per axis (3 axes), each of $N^3$
complex-fp64 elements (16~bytes per element).

Hence the operational intensity is
\begin{equation}
\mathrm{OI}_{\text{FFT}} = \frac{W_{\text{FFT}}}{Q_{\text{FFT}}}
= \frac{15 N^3 \log_2 N}{96 N^3}
= \frac{15 \log_2 N}{96}.
\end{equation}
For $N = 1024$, $\log_2 N = 10$, giving
$\mathrm{OI}_{\text{FFT}} = 150/96 = 1.5625$~FLOPS/byte,
which we round to $1.56$ in the main text.

\paragraph{Native parity.}
The native execution time is bounded below by the larger of the
compute-bound and memory-bound times:
\begin{equation}
T_{\text{nat}} = \max\!\left(\frac{W_{\text{FFT}}}{\eta},\ \frac{Q_{\text{FFT}}}{\Bmem}\right).
\end{equation}
Setting these equal gives the parity point:
\begin{equation}
\frac{W_{\text{FFT}}}{\eta} = \frac{Q_{\text{FFT}}}{\Bmem}
\quad\Longleftrightarrow\quad
\eta = \frac{W_{\text{FFT}}}{Q_{\text{FFT}}}\cdot \Bmem = \mathrm{OI}\cdot \Bmem.
\end{equation}
Thus $\etaopt^{\text{nat}} = 1.56\,\Bmem$. \emph{This is the binding
floor.}

\paragraph{\OBFFT\ parity (Phase~B-bound).}
For the \OBFFT\ kernel with the naive fp64 reduction, the Phase~B
time is $2\,N_{\text{out}}^{\text{total}} \cdot S$ FLOPs (FMA $=2$,
matching the TFLOP/s convention of the native formula above) divided
by the FP64 vector throughput $\eta$, where
$N_{\text{out}}^{\text{total}}$ is the number of scalar CRT
reconstructions.

With the residue-domain combine of \S\ref{sec:obfft-cost}
(Algorithm~\ref{alg:obfft}), only two components (Re, Im) per
complex output are reconstructed:
\begin{equation}
N_{\text{out}}^{\text{total}} = 3\ \text{axes} \cdot 2\ \text{stage
GEMMs} \cdot 2\ \text{components}\cdot N^3 = 12 N^3
\end{equation}
(an earlier draft used $18N^3$---three separate real-product
reconstructions---while its algorithm combined first; the mismatch
was flagged by independent review and the dataflow is now fixed as
combine-first).

Phase B time is $2\cdot 12 N^3 \cdot S / \eta$. Setting equal to
memory time:
\begin{equation}
\frac{2\cdot 12 N^3 \cdot S}{\eta} = \frac{96 N^3}{\Bmem}
\quad\Longleftrightarrow\quad
\eta = \frac{24 S}{96}\,\Bmem = \frac{S}{4}\,\Bmem.
\end{equation}
For $r=12$, $S = 15$ (Rev-34 set),
$\etaopt^{\OBFFT} = 3.75\,\Bmem$. \emph{This is informational rather
than binding}, because
at the native floor $1.56\,\Bmem$ the native FFT is already memory-bound
and \OBFFT\ is not required.

\paragraph{Why the native floor binds and naive \OBFFT\ does not rescue.}
The key inequality is $\etaopt^{\OBFFT} > \etaopt^{\text{nat}}$, i.e.,
$3.75 > 1.56$. This means the naive Ozaki path requires \emph{more} FP64
vector throughput per byte of memory traffic than the native path does.
Equivalently: native FFT does $\frac{15\log_2 N}{96} \approx 1.56$
\fp{64} flops per byte; Phase B does $\frac{24 S}{96} = 3.75$
\fp{64} flops per byte. The structural reason naive
\OII\ cannot beat native FFT on the FP64-vector dimension is that
Phase~B must scalar-process each output through the same arithmetic
pipe with no opportunity for tensor-core acceleration (the inner
dimension is $r$, far too small for tensor-core efficiency). Hence:
any GPU that fails the native floor also fails the naive Ozaki
floor on the FP64 vector pipe, by a strictly larger margin. The
Kulisch sub-floor of the next paragraph addresses this by routing
Phase~B onto a different pipe entirely.

\paragraph{The Kulisch INT32 sub-floor.}
\label{app:kulisch-floor}
For the Kulisch fixed-point Phase~B of \S\ref{sec:classical-phasB},
the per-output instruction interval is $[66,70]$ for the biased
carry-save deposits ($S_{\text{dep}}=15$ shifted columns, plus the
guard-bit normalisation; \texttt{verify.py} T3), $[25,35]$ for the
two-sided modulo-$M$ lift, and $[20,30]$
for the correctly-rounded pack with the power-of-two scale folded
into the exponent---$c_{\text{rec}}\in[111,135]$ INT32 instructions.

Total Phase~B work across all $12 N^3$ outputs is
$12 N^3\, c_{\text{rec}}$ INT32 instructions. Setting the Kulisch
time equal to the memory-roof time $Q/\Bmem = 96 N^3/\Bmem$:
\begin{equation}
\frac{12 N^3\, c_{\text{rec}}}{\eta_{\text{INT32}}}
   = \frac{96 N^3}{\Bmem}
\;\;\Longleftrightarrow\;\;
\eta_{\text{INT32}}^{\mathrm{Kulisch}}
   = \frac{c_{\text{rec}}}{8}\,\Bmem
   = 14\text{--}17\,\Bmem.
\label{eq:app-kulisch}
\end{equation}
At $\Bmem=8$~TB/s this is $111$--$135$~T~inst/s---\emph{well above}
B300's $41.7$~T~inst/s before deconstruction is even charged (the
July-29 draft's ``79 inst / $14\%$ margin'' descended from the
broken two-word signed deposit and the unsubstantiated 15/8-inst
lift/pack counts). At $\Bmem=22$~TB/s it is $306$--$372$~T~inst/s---but
Rubin meets the native FP64 floor and does not need the emulated
route.

\paragraph{Adding the rest of the epilogue.}
Equation~\eqref{eq:app-kulisch} charges reconstruction only. The same
pipe carries the per-input residue deconstruction
(\S\ref{sec:deconstruction}): $W_{\text{dec}} = n_{\text{dec}}\,
c_q^{\text{res}}\, r$ integer instructions over
$n_{\text{dec}} = 12N^3$ scalar
decompositions. Setting $W_{\text{dec}}/\eta = Q/\Bmem$ gives the
deconstruction sub-floor
\begin{equation}
\etaopt^{\text{dec}} = \frac{12\, c_q^{\text{res}}\, r}{96}\,\Bmem
   = \frac{c_q^{\text{res}}\, r}{8}\,\Bmem
   \approx 4.5\text{--}9.5\,\Bmem \quad (c_q^{\text{res}}\!\in\![3.0,6.3],\ r\!=\!12),
\label{eq:app-decon}
\end{equation}
plus the dp4a bulk, canonicalisation, and residue combines
(\S\ref{sec:deconstruction}). The total is
\begin{equation}
\etaopt^{\text{epi}}
  = \frac{c_{\text{epi}}}{8}\,\Bmem
  = 25.4\text{--}35.1\,\Bmem
  \qquad (c_{\text{epi}}\in[203,281]),
\label{eq:app-combined}
\end{equation}
which B300's $41.7$~T~inst/s ($=5.2\,\Bmem$) misses by
$4.9$--$6.7\times$.
Like every floor in this appendix, \eqref{eq:app-combined} is derived
against the memory-roof time and is therefore already an ideal-overlap
quantity; no scheduling relief applies on top of it
(Appendix~\ref{app:amdahl}).

\paragraph{The Kulisch coefficient $K \equiv c_{\text{rec}}/8$.}
The Kulisch coefficient depends on the Ozaki-II configuration through
$S$ and on the implementation through the deposit, lift, and pack
counts---every one an interval pending SASS measurement. The
range $K \in [14, 17]$ spans the specified implementation---well
\emph{above} both FP64-pipe coefficients
($1.56$ native, $3.75$ naive-Ozaki), which is precisely the point: the
Kulisch path is viable only because it routes that heavier per-output
work onto a pipe $30\times$ wider than B300's \fp{64} vector unit.
The constructive content of~\eqref{eq:app-kulisch} is
that reconstruction alone costs integer throughput of order
$14$--$17\,\Bmem$; with the full epilogue
(\eqref{eq:app-combined}), the integer pipe must supply
$25$--$35\,\Bmem$
to retain memory-roof FFT parity through the emulated path,
regardless of its FP64 vector spec---which is why B300, at
$5.2\,\Bmem$, walls above the roof on every evaluated software
route, and why the co-design ladder of \S\ref{sec:inttc} is the
constructive answer.

\section{The Bernstein Fractional CRT Reconstruction}
\label{app:bernstein}

The forward CRT formula~\eqref{eq:fcrt} can be written in
fractional form---\emph{with termwise modular reduction}, which
earlier drafts omitted: without it, an unreduced term
$v'_k y_k/m_k$ is of order $m_k$, and the range bound below is
false. Reduce each numerator first,
\begin{equation}
t_k = \big(v'_k\, y_k\big) \bmod m_k
\quad\text{(one Barrett or fold-and-add step per modulus)},
\end{equation}
so that $t_k/m_k \in [0,1)$ and
\begin{equation}
C \equiv \sum_k t_k \cdot \frac{M}{m_k} \pmod M
= M\cdot \left(\sum_k \frac{t_k}{m_k} \bmod 1\right).
\end{equation}
The sum $s = \sum_k t_k/m_k \in [0, r)$ now has the bounded range
the estimate below needs, so its fractional part
$f = s - \lfloor s\rfloor \in [0, 1)$ can be computed in fp
arithmetic (the centered representative uses the same two-sided
convention as Algorithm~\ref{alg:kulisch}). The reconstructed value
is
\begin{equation}
C / (s_A s_B) = (M / (s_A s_B))\cdot f.
\end{equation}
This repaired formulation is unit-tested (\texttt{verify.py}, T8),
including a guard confirming the unreduced form violates $[0,r)$.

\paragraph{Precision analysis.}
We carry the analysis at the prototype configuration of
Table~\ref{tab:precision}: $r=16$ moduli of $\approx 7$~bits each,
the configuration the Python prototype actually ran (the shipping
Rev-34 set is $r=12$ moduli of $9$--$11$ bits with
$\log_2 M = 117.88$---nearly the same total CRT range, so the
headroom structure below carries over; we flag explicitly that this
subsection's arithmetic is the $r=16$ prototype's, not the Rev-34
set's). With $r=16$,
$M \approx 2^{112}$ and
$M / (s_A s_B)\approx 64$ (since $s_A s_B \approx 2^{106}$).

\emph{The following error estimate is heuristic, and we label it as
such.} (A review of this passage correctly noted that a rigorous
bound must track absolute versus relative error per term---the
output needs $\approx 2^{-53}$ \emph{relative} accuracy, i.e.\
$\approx 2^{-59}$ absolute accuracy in $f$ when the scale factor is
$64$---and must weight the $\gamma_{r-1}$ summation bound by the
magnitude of the summed terms, which can be of order $r$.) The
heuristic sketch: the sum $s \in [0,16)$ carries $\log_2 r = 4$
integer bits, fp64 summation of $16$ terms loses a few ulps at that
scale, and the multiplication by $M/(s_As_B) \approx 64$ amplifies
the absolute error by $6$ more bits. This predicts an output several
bits short of full \fp{64}, in the direction and rough magnitude of
the $\sim 9$-bit loss \emph{measured empirically} in
Table~\ref{tab:precision} ($\sim 47$-bit observed accuracy). We
make no exact bit-loss claim for this legacy ablation; the adopted
kernel avoids the issue entirely by performing the lift in exact
integer arithmetic (Algorithm~\ref{alg:kulisch}), where the error
budget is the proven $\tfrac14$-ulp fixed-point bound.

\paragraph{In fp32+Kahan compensated sum.}
With fp32 (24-bit mantissa), $s$ has relative precision $r\cdot 2^{-24}
\approx 2^{-20}$. Kahan compensation gives $\sim 2^{-44}$ for
benign sums, but the fractional-part extraction amplifies the
absolute error by $M / (s_A s_B) \approx 64$, yielding result
precision $\sim 2^{-38}\approx 10^{-12}$ in the best case. The
measured $10^{-3}$ in Table~\ref{tab:precision} indicates
non-benign accumulation (terms are not aligned in magnitude),
giving only $\sim 10$-bit precision. Iterative refinement closes
this gap at $\sim 3\times$ runtime.

\section{Garner Reconstruction: Cost and Precision Analysis}
\label{app:garner-detail}
\label{app:slicing}

This appendix documents both the per-formulation cost analysis
(referenced from \S\ref{sec:tcgarner}) and the per-scheme precision
analysis (referenced from \S\ref{sec:precision}). The two are
interleaved because each Garner formulation pairs naturally with a
specific subset of Phase~B reduction schemes.

\subsection{Per-formulation cost analysis}

We tabulate the operation counts for the three Garner formulations
(Recursive, Slicing, Bernstein) and, within the Slicing formulation,
the four Phase~B reduction choices (fp64 sum, Sum2/SumK, DD on fp32,
Kulisch fixed-point, and the reduced-precision fp32+Kahan path) used
in the main text.

\paragraph{Recursive Garner (legacy).}
The mixed-radix recursion
$v_k = (v'_k - \sum_{j<k} v_j \cdot M_{jk})\cdot m_k^{-1} \bmod m_k$
requires, per output element:
\begin{itemize}[topsep=2pt, leftmargin=2em]
\item Inner loop: $k(k-1)/2$ iterations
\item Per iteration: 1 multiply + 1 mod + 1 add + 1 multiply + 1 mod (for prefix update)
\item Total: $\sim 2.5 r^2$ integer ops, dominated by modular reduction
\end{itemize}
For $r=12$: $\sim 360$ INT32 ops per output. Modular reduction on
B300 SIMT is charged once, at the issued-instruction rate; earlier drafts divided again by an ``effective 25~TOPS'', charging Barrett twice, on a $41.7$~T~inst/s nominal
INT32 pipe.

\paragraph{Slicing-based direct CRT.}
Per output:
\begin{itemize}[topsep=2pt, leftmargin=2em]
\item Phase A: one canonical residue operand per modulus; the $2$ in $2r(S{+}S_{\text{frac}})$ counts MAC${}={}2$ $\times\ (S{+}S_{\text{frac}})$
   \bint{8} multiply-adds ($S = 15$ from the Rev-34
   $\log_2 M = 117.88$; $S_{\text{frac}}=6$ lift columns)
\item Phase B (choice-dependent): see options below; the exact
   variants additionally carry the modulo-$M$ lift and
   normalise-pack conversion ($\approx 23$ INT32 inst/output)
\end{itemize}
Total Phase~A: $2r(S{+}S_{\text{frac}}) = 504$ \fp{16} tensor ops ($252$ MACs) per output ($\approx 4.0$~ms on B300's 2.5-PF FP16-TC with tile padding charged, MAC $=2$; \S\ref{sec:classical-phasB}).

The Phase~B options (FMA counted as 2 FLOPs against FP peaks):
\begin{itemize}[topsep=2pt, leftmargin=2em]
\item \emph{fp64 sum} (naive): $2S$ FLOPs per output on FP64 vector
pipe (and no modulo-$M$ lift---it is the accuracy baseline, not a
correct CRT realisation).
\item \emph{Sum2/SumK} \cite{ogita2005}: $\sim 6\times$ the naive
fp64 cost on the FP64 pipe (full fp64 with logarithmic error growth).
\item \emph{DD on fp32}: $\sim 7S$ fp32 ops on FP32 vector pipe
($\sim 48$-bit effective precision).
\item \emph{Kulisch fixed-point}: $c_{\text{rec}}\in[111,135]$ INT32 instructions on
the INT32 SIMT pipe (deposits $+$ lift $+$ conversion), full fp64
with single final rounding.
\item \emph{fp32+Kahan}: $4S$ fp32 ops on FP32 vector pipe (reduced
precision, $\sim 10$--$18$~bits in our specific Phase~B setting).
\end{itemize}

\paragraph{Bernstein fractional CRT.}
Per output:
\begin{itemize}[topsep=2pt, leftmargin=2em]
\item Phase A: $r$ fp multiply-adds (Bernstein sum)
\item Phase B: 1 floor, 1 subtract, 1 multiply
\end{itemize}
Total: $r=12$ fp MAC + constant overhead per output. Phase~A here
is an inner product across the $r$ residues; if done in fp64 it
maps onto the FP64 vector pipe. The small inner dimension $r=12$
keeps it on vector pipes rather than tensor cores.

\paragraph{Net comparison on B300 for $1024^3$ FFT.}
\begin{center}\footnotesize
\setlength{\tabcolsep}{4pt}
\resizebox{\textwidth}{!}{%
\begin{tabular}{lrrll}
\toprule
Formulation & Phase~A & Phase~B $+$ epilogue & Bottleneck pipe & Precision\\
\midrule
Recursive Garner             & ---     & ---            & INT32 SIMT $\gamma$ at 111 ms    & full fp64 \\
Slicing + fp64 sum            & 3.96 ms & 278 ms (fp64)  & FP64 vec at 278 ms                & full fp64 (no lift) \\
Slicing + Sum2/SumK (est.)    & 3.96 ms & $\sim$1550 ms (fp64) & FP64 vec at $\sim$1550 ms          & full fp64 \\
Slicing + DD on fp32 (est.)   & 3.96 ms & 30 ms (fp32)   & FP32 vec at 30 ms                 & $\sim 48$ bits \\
Slicing on INT8-TC            & 85 ms  & ---            & INT8-TC at 167 TOPS (dead)        & --- \\
\textbf{Slicing (FP16-TC) + Kulisch, int-hosted}
                              & 3.96 ms & 63--87 ms (int32) & \textbf{int epilogue} & \textbf{full fp64} \\
Slicing + fp32+Kahan          & 3.96 ms & 14 ms (fp32)   & FP32 vec / mem.\ roof             & $\sim 10$--$18$ bits \\
Bernstein + fp64 sum          & 3.96 ms & $\sim$220 ms (fp64)  & FP64 vec               & full fp64 \\
\bottomrule
\end{tabular}}
\end{center}

\paragraph{Reading the table.}
At full \fp{64} precision:
\begin{itemize}[topsep=2pt, leftmargin=2em]
\item \emph{Recursive Garner} ($\sim 111$~ms) and \emph{Slicing + fp64
sum} ($\sim 278$~ms) are bottlenecked on the $\gamma$ SIMT path
and FP64 vector pipe respectively. These hit the FP64 floor of
$1.56\,\Bmem$
(\S\ref{sec:parity}), which B300 fails by $\sim 9\times$;
INT8-TC hosting dies on the 167-TOPS class rate.
\item \emph{Slicing (FP16-TC) + Kulisch} is bottlenecked on
the integer-SIMT \emph{epilogue} at $63$--$87$~ms against the
$12.9$~ms roof; no software hosting reaches the roof on B300
(\S\ref{sec:inttc} for the hardware that would).
\end{itemize}
The fp32+Kahan path is projected to reach the memory roof at 14~ms but
only at fp32-class ($\sim 10$--$18$-bit) precision, not full \fp{64}. DD on fp32 projects to $\sim 30$~ms
at $\sim 48$~bits---an intermediate option for codes that can tolerate
half a dozen bits below \fp{64}.

The architectural conclusion is therefore not that ``the floor is set
by FP64 vector throughput'' simpliciter, but that \emph{the floors are
set by whichever pipes the chosen conversion assignment lands on,
input side included}. For naive Phase~B paths the FP64 vector pipe
binds; for the exact Kulisch hosting the integer SIMT pipe binds at
the \emph{epilogue} floor (which B300 misses); and the floor rule
of \S\ref{sec:parity} states that a GPU meets memory-roof FFT parity at
full \fp{64} if it satisfies one of the \emph{route-complete} floor
sets---native, or the emulated tensor-plus-epilogue set. The
epilogue provisioning---or the co-design ladder of
\S\ref{sec:inttc}---is what decides roof attainment.

\subsection{Per-scheme precision analysis}

We measured reconstruction error on a Python prototype with $r=10$
(fp32-equivalent target) and $r=16$ (the prototype's fp64-equivalent
configuration; the recommended FP8-substrate set $r=12$ sits between
the two and shares the $r=16$ row's qualitative behaviour) at
$16\times 16\times 16$ matrix sizes. Table~\ref{tab:precision} reports
the maximum relative error against the true \fp{64} matrix product.

\begin{table}[h]
\centering
\caption{Max relative error vs.\ true \fp{64} product, by reconstruction
method, on the prototype configurations $r=10$ and $r=16$.
Slicing/Bernstein with \fp{64} sum retains precision at $r=10$
but loses $\sim 9$ bits at the fp64-equivalent configuration because
the unweighted sum has dynamic
range exceeding fp64's 53-bit mantissa. The Bernstein fractional CRT
amplifies absolute error by $M/(s_A s_B) \approx 64$ when extracting the
fractional part, which makes the \fp{32}+Kahan variant lose precision
sharply for non-benign sums.}\label{tab:precision}
\small
\begin{tabular}{lcccccc}
\toprule
$r$ & recursive & slice/fp64 & slice/fp32+K & bern/fp64 & bern/fp32+K & bern/fp32 \\
\midrule
10 & $1.3\times 10^{-9}$ & $1.3\times 10^{-9}$ & $1.6\times 10^{-3}$ & $1.3\times 10^{-9}$ & $1.7\times 10^{-3}$ & $3.6\times 10^{-3}$\\
16 & $1.3\times 10^{-14}$ & $6.4\times 10^{-12}$ & $3.4\times 10^{-3}$ & $4.7\times 10^{-12}$ & $1.3\times 10^{-3}$ & $2.5\times 10^{-3}$\\
\bottomrule
\end{tabular}
\end{table}

Three regimes emerge in the B300 picture (Figure~\ref{fig:precision}):
\begin{itemize}[leftmargin=2em, topsep=2pt]
\item \emph{Full \fp{64}} (53-bit mantissa): \fp{64} sum Phase~B at
$278$~ms, $\sim 20\times$ over the memory roof. Slicing direct-CRT
with an fp64 sum loses $\sim 9$ bits at the fp64-equivalent
configuration because the unweighted sum has $\gtrsim 100$-bit
dynamic range and \fp{64} mantissa is only 53 bits; recovering full
\fp{64} requires either remaining with the slow recursive Garner
($\sim 111$~ms), accepting the $\sim 6\times 10^{-12}$ error of the
fp64-sum slicing variant, or the exact Kulisch accumulator of the
main text (which is the design's answer).
\item \emph{Double-fp32 (DD), $\sim 48$-bit} (estimated, not measured):
representing each accumulator as two \fp{32} words gives $\sim 48$~bits
of precision at $\sim 4\times$ the \fp{32} cost. On B300 this projects
to $\sim 30$~ms for Phase~B---above the memory roof but $\sim 7\times$
faster than full \fp{64} sum, and the natural fallback for applications
that can tolerate $\sim 48$-bit precision.
\item \emph{\fp{32}+Kahan, fp32-class $\sim 10$--$18$-bit} (measured): Phase~B in 14~ms,
matching the memory roof, but the precision is fp32-class or below
rather than fp64. For codes that need full \fp{64} accuracy this is
unusable; for codes already running at \fp{32}, it is the natural
choice but then \OII\ is also unnecessary (see \S\ref{sec:fp32}).
\end{itemize}

\paragraph{Classical-scheme survey, with Kulisch comparison.}
The non-Kulisch classical schemes mentioned in \S\ref{sec:classical-phasB}:
\begin{itemize}[topsep=2pt, leftmargin=2em]
\item \emph{Compensated summation} (Kahan~\cite{kahan1965},
Neumaier~\cite{neumaier1974},
Ogita-Rump-Oishi Sum2/SumK~\cite{ogita2005}): all variants run on the
FP64 vector pipe and cost $4$--$6\times$ naive fp64 sum---$\sim
1030$--$1550$~ms for Sum2 at full fp64 substrate.
\item \emph{Multi-double} (Dekker DD~\cite{dekker1971},
Bailey-Hida-Li QD~\cite{baileyhidaQD2001}): on fp64 substrate, DD
costs $6$--$10\times$ a single fp64 op---worse than naive on B300. On
fp32 substrate, DD-32 yields $\sim 48$~bits at fp32 vector throughput;
QD-32 reaches $\sim 72$~bits at $\sim 60$~ms.
\item \emph{Reproducible BLAS}
(Demmel-Hida-Nguyen~\cite{demmel2008,demmel2013}): bin-based
signed-integer accumulators carry $\sim 5\times$ overhead in practice;
on B300, the dominant pipe is still fp64 vector (for the bin floats).
\item \emph{Tall-skinny DGEMV} on cuBLAS: same fp64-vector bottleneck
since cuBLAS does not use tensor cores for fp64 reductions on
Blackwell/Rubin generations. $\sim 210$~ms.
\end{itemize}
The Kulisch scheme is the unique entry in this catalogue that routes
Phase~B onto a different pipe (INT32 SIMT) and therefore the unique
escape from the FP64-vector bottleneck. The main-text
\S\ref{sec:classical-phasB} contains the projected wall time and
precision analysis; the integrated comparison table earlier in this
appendix shows where each scheme places in the cost/precision plane.

\section{Generalised Kulisch: Beneficiaries, Amdahl, and Library Notes}
\label{app:beneficiary-analysis}
\label{app:kulisch-general-details}

This appendix consolidates the supporting material for the generalised
Kulisch rescue: the per-algorithm beneficiary analysis behind
Table~\ref{tab:kulisch-candidates} (\S\ref{app:beneficiary-analysis-detail}),
the Amdahl computation and the overlap discussion referenced from
\S\ref{sec:kulisch-general} (\S\ref{app:amdahl}), and the implementation
notes for \texttt{libKulisch} (\S\ref{app:libkulisch}).

After Phase~A, many kernels need a per-output reduction of the form
$y_i = \sum_{j=1}^{W} P_{ij}\cdot w_j$, where each $P_{ij}$ is a small
integer (bounded number of bits) and $w_j$ is a precomputed \fp{64}
positional weight. The Kulisch sub-floor on integer-vector throughput is
\begin{equation}
\boxed{\;\etaopt^{\text{Kulisch}}
   = \frac{c\,W}{Q_{\text{out}}}\,\Bmem
   = c \cdot \mathrm{OI}_{\text{red}} \cdot \Bmem,\;}
\label{eq:kulisch-general}
\end{equation}
where $c$ is the average INT32 instruction count per weighted term
(deposit plus the amortised lift/convert epilogue; $c_{\text{rec}}/S
\in [111,135]/15 \approx 7.4$--$9$ for the FFT configuration),
$Q_{\text{out}}$ is bytes of HBM traffic per output, and
$\mathrm{OI}_{\text{red}} = W/Q_{\text{out}}$ is the reduction-phase
operational intensity. For Ozaki-Bailey FFT this recovers~\eqref{eq:parity-kulisch}.

\begin{table}[h]
\centering
\caption{Algorithms with at least moderate Kulisch benefit. Speedup
column assumes $k=7\times$ Kulisch-vs-naive on the reduction itself
($k\approx 6$--$8$ is the FFT's reconstruction-only ratio, $278$~ms
against $34$--$42$~ms; with the full integer epilogue charged the
FFT's own ratio is $3.2$--$4.4$, so its row is an upper bound and the
epilogue-charged value is given in parentheses; workloads with
lighter epilogues land nearer $k$). Rejected candidates and per-algorithm rationale are in
\S\ref{app:beneficiary-analysis-detail}.}\label{tab:kulisch-candidates}
\small
\setlength{\tabcolsep}{5pt}
\begin{tabular}{p{0.31\textwidth}p{0.28\textwidth}p{0.13\textwidth}l}
\toprule
Algorithm & Common context & Amdahl $f$ & Speedup ($k=7$) \\
\midrule
Ozaki-Bailey FFT (this work) & Spectral CFD, QCD, climate &
0.95--0.98 & $\sim 5$--$6\times$ ($2.9$--$4.1$) \\
Ozaki-II SpMV & PDE discretisations, irreg.\ meshes &
0.70--0.90 & 2.5--4$\times$ \\
ReproBLAS replacement & Bitwise-reproducible HPC codes &
0.50--0.90 & 1.8--4$\times$ \\
Bandwidth-bound multi-term stencils & High-order finite-difference codes &
0.50--0.80 & 1.7--3$\times$ \\
Ozaki-II batched-small DGEMV & Batched factorisations, ML inference &
0.30--0.60 & 1.3--1.9$\times$ \\
\bottomrule
\end{tabular}
\end{table}

\subsection{Beneficiary-algorithm analysis}
\label{app:beneficiary-analysis-detail}

Each candidate is evaluated against the four criteria of
\S\ref{sec:kulisch-general}: (i)~per-output fp64 reduction is the
binding bottleneck; (ii)~the bottleneck is on the FP64 vector pipe;
(iii)~operands have bounded dynamic range fixed at compile time;
(iv)~the Amdahl fraction $f$ is large enough to matter.

\subsubsection{The Ozaki-II family, by operational intensity}

Within the Ozaki-II family, the Phase~B Amdahl fraction depends on
the kernel's operational intensity (OI). High-OI kernels (dense GEMM)
are dominated by tensor-core Phase~A; low-OI kernels (FFT, SpMV) by
fp64-vector Phase~B. Concretely on B300:
\begin{itemize}[topsep=2pt, leftmargin=2em]
\item Dense DGEMM via Ozaki-II ($M=N=K=8192$, real): Phase~A on fp8
tensor cores at 5~PF ($\alpha=3r{+}1=37$; real GEMM, so no complex
factor) takes $\sim 8$~ms; Phase~B ($S=15$ deposits per
output, $MN$ outputs on the integer pipe) takes $\sim 0.1$~ms. Amdahl
$f\approx 0.01$, Kulisch speedup negligible. Not a
Kulisch beneficiary.
\item Ozaki-II batched-small DGEMV (inner length $k\sim 10$--$100$):
Phase~B work scales inversely with $k$; for small $k$ Phase~B
dominates. $f\sim 0.3$--$0.6$, speedup $1.3\text{--}1.9\times$ at $k=7$.
Moderate.
\item Ozaki-II SpMV: per-row CRT-encoded reductions over
sparsely-populated columns. Phase~B dominates per-row work,
$f\sim 0.7$--$0.9$, speedup $2.5\text{--}4\times$ at $k=7$. Strong.
\item Ozaki-Bailey FFT (this work): OI $\approx 1.56$~FLOPS/B intrinsic
to FFT structure; Phase~A small, Phase~B dominant. $f\sim 0.95$--$0.98$,
speedup $\sim 5$--$6\times$ at $k=7$ ($2.9$--$4.1\times$ with the full
epilogue charged). Principal beneficiary.
\end{itemize}
The qualitative rule: within Ozaki-II, Kulisch benefits scale
inversely with operational intensity. High-OI kernels are already
served by tensor-core Phase~A; low-OI kernels need Kulisch to escape
the FP64-vector Phase~B.

\subsubsection{Iterative solvers and Krylov dot products: the bottleneck
is elsewhere}

A natural expectation is that CG, GMRES, and other Krylov solvers
benefit from Kulisch because they involve dot products. They typically
do not. In a CG iteration the cost is dominated by one SpMV (which is
memory-bound for typical sparse matrices) plus two dot products
$\langle r, r\rangle$ and $\langle r, Ap\rangle$ and three AXPYs---all
memory-bound, not FP64-vector-bound. The dot products' arithmetic
intensity is $\sim 1$~FLOP/B; on B300 the collapsed 1.4~TFLOPS FP64
vector handles them in $\sim 0.03$~ms for $n=10^7$, while the SpMV
takes $\sim 10$--$50$~ms. Amdahl $f$ for the reductions is
$\sim 0.05$--$0.15$ in a well-tuned CG; max Kulisch speedup
$1.05$--$1.18\times$.

The SpMV \emph{itself} could be accelerated by a Kulisch-style
accumulator on its per-row dot products, but only if the SpMV is
FP64-vector-bound rather than memory-bound. For most PDE
discretisations this is not the case. For unusually compute-dense rows
(block-sparse formats with large blocks, structured stencil operators
implemented as banded GEMV with $k$ in the hundreds) the SpMV may
become FP64-vector-bound, in which case the Kulisch rescue applies:
this is the ``bandwidth-bound multi-term stencils'' case in
Table~\ref{tab:kulisch-candidates}.

The earlier draft of this paper double-counted by listing ``iterative
solvers'' and ``Krylov subspace dot products'' as separate
beneficiaries. They refer to the same underlying reductions and share
the same low Amdahl fraction; both are excluded from the recommended
set.

\subsubsection{Other candidate algorithms}

\begin{itemize}[topsep=2pt, leftmargin=2em]
\item \textbf{N-body Coulomb sums}, $\sum_j q_iq_j/r_{ij}$: the sum is
the algorithm; no separable Phase~A/B structure. Each term involves
an inline divide. Speculative fit: applies only if $q_j/r_{ij}$ values
are pre-tabulated and final summation dominates, which is unusual.
\item \textbf{Quadrature / numerical integration},
$\sum_k w_k f(x_k)$: wall time dominated by function evaluation
$f(x_k)$, not the sum. Limited fit unless $f$ is a single fp64
multiply.
\item \textbf{Convolution-as-GEMM}: structured GEMM, compute-bound on
tensor cores, analogous to dense DGEMM. Limited fit.
\item \textbf{Polynomial evaluation (Horner)}: sequential dependency
chain, no Phase A/B split. For typical degrees 10--100, the wall time
is microseconds and Kulisch setup overhead exceeds savings. Not useful.
\item \textbf{Reproducible BLAS} (ReproBLAS by Demmel, Hida and
Nguyen~\cite{demmel2008,demmel2013}): already uses signed-integer
accumulator bins for the explicit purpose of reproducible
summation; this is essentially a different Kulisch variant.
Strong fit as a replacement ReproBLAS implementation, with the 6-word INT32 wide accumulator a cleaner alternative to ReproBLAS's
bin structure. Included in Table~\ref{tab:kulisch-candidates}.
\item \textbf{Bandwidth-bound stencils with multi-term updates}: when
a stencil involves $> 5$ contributions per output and the kernel is
FP64-vector-bound, Kulisch can move the update onto the INT32 pipe.
Strong fit for high-order stencils with bounded coefficients (e.g.,
$\sim 13$-term Laplacian variants on non-uniform grids). Included in
Table~\ref{tab:kulisch-candidates}.
\item \textbf{Polynomial-kernel methods in ML inference} (kernel SVMs,
GP regression): involve dot products of feature vectors followed by
polynomial evaluation. Speculative fit if per-query dot products
dominate and feature range is bounded; benchmarking would clarify.
\item \textbf{HPL panel updates}: rank-$k$ trailing-matrix updates,
firmly compute-bound on tensor cores. Not useful.
\end{itemize}

\subsection{Amdahl analysis, and overlap as a model assumption}
\label{app:amdahl}

This subsection details the Amdahl computation summarised in
\S\ref{sec:kulisch-general} and clarifies the role of overlap: the
floors already assume it, so it is an attainment condition, not a
relief.

\subsubsection{Comparison to the FP64-vector floor for the same reduction}

Without Kulisch, the same per-output reduction
$y_i = \sum_j P_{ij} w_j$ from \S\ref{sec:kulisch-general} runs on
the FP64 vector pipe at $W$ multiply-adds per output, giving
$\etaopt^{\text{FP64-vec}} = \mathrm{OI}_{\text{red}}\cdot \Bmem$.
Kulisch therefore trades a factor of $c_{\text{rec}}/S \approx 7.4$--$9$ in operation
count (deposits plus amortised lift/convert)
for the ability to use the INT32 SIMT pipe instead of FP64 vector. On
B300 the INT32 pipe has $\sim 30\times$ the throughput of the FP64
vector pipe ($41.7$~T~inst/s instruction issue against 1.39~TFLOP/s, i.e.\
$\sim 27\times$ in FMA units); net, Kulisch buys an effective
$\sim 6.6$--$8.2\times$ speedup on the reconstruction itself relative
to the naive fp64 sum at $c_{\text{rec}}/S\approx 7.4$--$9$
($278$~ms against $34$--$42$~ms), provided the INT32 sub-floor is met.

\subsubsection{Serial Amdahl}

Let $T_{\text{reduce}}$ and $T_{\text{rest}}$ be the wall times of the
reduction and non-reduction phases in the current implementation, and
let $f = T_{\text{reduce}}/(T_{\text{reduce}} + T_{\text{rest}})$ be
the reduction-phase fraction. Suppose Kulisch reduces the reduction
time by a factor $k$ (on FP64-collapsed B300, $k\approx 6$--$8$
relative to naive fp64 Phase~B once the full epilogue is charged).
The total speedup under serial
execution is
\begin{equation}
S_{\text{serial}}(f, k) = \frac{1}{(1-f) + f/k}.
\label{eq:app-amdahl}
\end{equation}
At $k=7$:
\begin{center}\small
\begin{tabular}{lll}
\toprule
$f$ & $S$ & Verdict \\
\midrule
0.98 (\OBFFT) & $6.2\times$ & strong speedup; principal beneficiary \\
0.80 (Ozaki-II SpMV) & $3.2\times$ & strong speedup \\
0.50 & $1.75\times$ & moderate \\
0.20 & $1.21\times$ & marginal; Kulisch not worthwhile \\
\bottomrule
\end{tabular}
\end{center}
The practical threshold for ``Kulisch is worth implementing'' is
$f \gtrsim 0.5$.

\subsubsection{Overlap: already assumed by the floors, not a further
relief}
\label{eq:app-overlap-floor}

An earlier draft derived a factor-of-2 ``overlap loosening'' of the
Kulisch sub-floor by comparing a serial-sum wall time
$T_{\text{Kulisch}} + T_{\text{memory}}$ against the overlapped
$\max(T_{\text{Kulisch}}, T_{\text{memory}})$. That derivation
double-counted: every bandwidth-parity floor in this paper
(Appendix~\ref{app:parity} and \eqref{eq:kulisch-general}) is obtained
by setting the pipe's compute time \emph{equal to the memory-roof
time}, i.e.\ under the service-maximum (ideal-overlap) model of
\eqref{eq:emuroofline}. The floors are therefore already
ideal-overlap quantities, and no scheduling discipline lowers them
further; we retract the loosened figure. What double-buffering
determines is whether the ideal-overlap bound is \emph{attained}: a
kernel that fails to overlap integer work with HBM traffic runs
\emph{above} the model's projection, never below the floor.

Attaining the bound is plausible for the Ozaki-Bailey kernel because
Phase~A (tensor-core GEMMs), the integer conversions, and HBM traffic
use disjoint hardware pipes and disjoint register-bank classes,
allowing one tile's conversions to run under the next tile's loads.
The benchmark of \S\ref{sec:future}~(1) measures the realised overlap
factor empirically. For algorithms beyond Ozaki-Bailey: fully
data-parallel reductions (per-row \texttt{spmv}, batched quadrature)
should approach the ideal-overlap bound; algorithms with sequential
dependency chains (Horner polynomial evaluation, recursive sums) will
run above it. The service-maximum model therefore applies to most of
the Table~\ref{tab:kulisch-candidates} beneficiaries but not
uniformly.

\subsection{Implementation notes for libKulisch}
\label{app:libkulisch}

This subsection documents the implementation specifics referenced
in~\S\ref{sec:kulisch-general}.

\subsubsection{Library components}

\texttt{libKulisch} offers four components:
\begin{enumerate}[topsep=2pt, leftmargin=2em]
\item A templated wide-accumulator primitive parameterised on output
bit-width (typically 128--256 bits, packed into 4--8 INT32 words) and
on the source operand format (INT8/INT16/INT32, with compile-time
positional weights).
\item Warp-level shuffle primitives for carry propagation across
word boundaries, optimised for 32-thread warps (Blackwell) and
for wider-SIMT parts, if any appear.
\item Drop-in replacements for the per-output reduction loops in
GEMMul8~\cite{ozaki2025ii}, cuBLAS Ozaki, SpMV libraries (cuSPARSE,
PETSc), and ReproBLAS---matching the beneficiary list of
Table~\ref{tab:kulisch-candidates}.
\item A benchmark harness measuring wall time and achieved INT32
throughput across $\Bmem$-bound and $\eta_{\text{INT32}}$-bound
regimes, validating~\eqref{eq:kulisch-general} empirically.
\end{enumerate}

The goal of \texttt{libKulisch} is to deliver an empirical answer to a
single question: within the narrowed beneficiary class of
\S\ref{sec:kulisch-general}, how often does Kulisch actually provide
meaningful end-to-end speedup, and how often is the predicted Amdahl
fraction $f$ accurate to within a factor of 2? The model predicts
$1.3$--$6\times$ end-to-end speedup across the five beneficiary classes
of Table~\ref{tab:kulisch-candidates}; empirical confirmation would
establish Kulisch as a standard tool in the post-FP64-collapse
arsenal, complementary to the tensor-core Phase~A acceleration that
already serves the high-OI half of the Ozaki-II family.

\subsubsection{Caveats and limitations}

Three implementation caveats apply:
\begin{enumerate}[topsep=2pt, leftmargin=2em]
\item Kulisch requires bounded dynamic range. Reductions with truly
unbounded inputs (e.g., quantum-chemistry codes with extreme condition
numbers) require either application-level rescaling or adaptive-precision
extensions~\cite{schwarz2025dgemm}.
\item The wide accumulator consumes register pressure ($\sim 6$ INT32 registers per output for the Ozaki-Bailey case). Kernels already at
the register-pressure limit may need to spill, degrading throughput.
The benchmark harness reports per-kernel register use to flag
spill-prone cases.
\item The achievable overlap factor (Appendix~\ref{app:amdahl}) depends
on the surrounding kernel's dependency structure; algorithms with
sequential reductions cannot attain the ideal-overlap bound the
floors assume, and run above the model's projection. This is automatically detected by the benchmark harness.
\end{enumerate}

\end{document}